\documentclass[12pt, a4paper]{article}
\usepackage{amsthm,amsmath,amsfonts,amssymb,cases}
\usepackage{xcolor}
\usepackage{enumerate}
\usepackage{esint}
\usepackage{comment}

\newcommand{\R}{{\mathord{\mathbb R}}}
\newcommand{\Z}{{\mathord{\mathbb Z}}}
\newcommand{\N}{{\mathord{\mathbb N}}}
\newcommand{\C}{{\mathord{\mathbb C}}}

\def\chib {\overline{\chi}}

\newcommand{\HH}{\mathcal{H}}

\newcommand{\FF}{\mathcal{F}}

\newcommand{\WW}{\mathcal{W}}
\newcommand{\hh}{\mathfrak{h}}
\newcommand{\UU}{\mathcal{U}}

\newcommand{\umm}{\underline{m}}
\newcommand{\unn}{\underline{n}}
\newcommand{\upp}{\underline{p}}
\newcommand{\uqq}{\underline{q}}
\newcommand{\uzz}{\underline{0}}

\newcommand{\ran}{{\rm Ran}}

\newcommand{\un}[1]{\underline{#1}}

\newcommand{\inn}[1]{\left\langle {#1} \right\rangle }

\DeclareMathOperator*{\esssup}{ess\,sup}

\newcommand{\ben}{\begin{displaymath}}
\newcommand{\een}{\end{displaymath}}
\newcommand{\beqn}{\begin{equation}}
\newcommand{\eeqn}{\end{equation}}
\newcommand{\beqna}{\begin{eqnarray*}}
\newcommand{\eeqna}{\end{eqnarray*}}

\def\inf{{\rm inf}\,}

\def\dist{{\rm dist}\,}
\def\supp{\operatorname{supp}}

\newcommand{\sfrac}[2]{\textrm{\footnotesize $\frac{#1}{#2}$}}

\newtheorem{lemma}{Lemma}
\newtheorem{theorem}[lemma]{Theorem}
\newtheorem{remark}[lemma]{Remark}
\newtheorem{proposition}[lemma]{Proposition}
\newtheorem{corollary}[lemma]{Corollary}
\newtheorem{definition}[lemma]{Definition}

\newtheorem{hypothesis}{Hypothesis}

\newtheorem{hypothesisR}{Hypothesis}

\numberwithin{equation}{section}

\numberwithin{lemma}{section}

\begin{document}
\title{On dilation analyticity and spatial exponential decay of  atomic ground states in non-relativistic qed}
\author{\vspace{5pt} D. Hasler$^1$\footnote{
E-mail: david.hasler@uni-jena.de}  \ and C. Lejsek$^1$\footnote{E-mail: christian.lejsek@uni-jena.de} \\
\vspace{-4pt} \small{$1.$ Department of Mathematics,
Friedrich Schiller University,} \\ \small{Jena, Germany}\\
}
\date{}
\maketitle

\begin{abstract}
We consider the ground state and the ground state energy  of an atom with spinless electrons  in the framework of non-relativistic qed.
We show that the ground state energy as well as the ground state depend analytically
on the parameters   of the group  of  dilations, the parameter of a   group of spatial dependent phase changes,
and on  the minimal coupling constant.
As a corollary we obtain spatial exponential decay of   the ground state as well
as of  its dilation analytic extension.
No infrared regularization is needed for the result.  Our result is based on operator theoretic renormalization.
\end{abstract}

\section{Introduction}

Non-relativistic quantum electrodynamics (qed) is a mathematically rigorous theory describing
low energy phenomena of matter interacting with quantized electromagnetic radiation.
The matter is described  by non-relativistic quantum mechanics and the  quantized field is described by
a transversal field of massless bosons, called photons, with relativistic dispersion relation. The matter couples minimally
to the quantized field. The strength of the coupling is described by the coupling constant, which we denote by $g$.
This theory allows a mathematically rigorous treatment of
various physical aspects. 
Its origins date back to the birth of quantum field theory.  It was used for example
in 1938 by   Pauli and Fierz \cite{PauliFierz.1938}  to study the emission of quantized radiation.

Expansions  of the ground state as well as the ground state energy as a function
of the  coupling constant  $g$  are of interest, since they carry the
physical structure originating from the interactions of  bound electrons with photons.
 Lowest order expansions  were used for example in 1947 by Bethe
 \cite{bet47} to obtain radiative corrections which   contribute to the Lamb shift \cite{lamret47}.

Expansions of the ground  state energy as well as  the    ground state   have been intensively studied with mathematical rigour for non-relativistic qed
as well as for related models \cite{HaiSei02,BarCheVouVug10,BacFroPiz06,BacFroPiz09,HasHer11-3,Ara14,BraHasLan18}.
In fact, even analyticity has been shown to hold in certain situations \cite{GriHas09,HasHer11-1,AbdHas12,HasLan18-1,HasLan22}.
Analyticity results are non-trivial to establish, since for these models the ground state energy is not isolated from the rest of the spectrum.
However, once   analyticity has been mathematically established, the expansion coefficients can then be calculated by the expressions of
formal  Rayleigh-Schr\"odinger
perturbation theory \cite{HasHer11-1,ReeSim4}.

In particular, it has been shown in  \cite{HasHer11-1} that the ground state as well as the ground state energy in non-relativistic
qed are analytic  functions of the  coupling constant $g$.
In this paper we  extend that result to analytic  extensions of dilations 
as well as 
spatial dependent phase changes.
That is, the main result of this paper, Theorem   \ref{thm:main1}, shows that
the ground state as well as the ground state energy of the atom or ion are jointly analytic
functions of the coupling constant, $g$, the  dilation parameter $\theta$,
and the  parameter $\alpha$ given by the unitary phase change $\exp( { i \alpha \sqrt{ 1 +  x^2  }})$, where $x$ denotes
the vector of  the spatial coordinates of the electrons.
As a consequence we obtain spatial exponential decay of the analytically
dilated ground state, Corollary   \ref{cor:expdecay}.

Analytic extensions with respect to the  dilation parameter $\theta$  are referred as analytic dilations
and are used to   study of resonances. 
Initially,  analytic dilations  were introduced to study resonances of  atoms and molecules, cf.  \cite{Simon.1973-resonances,ReeSim4},
without any coupling to a  quantized field. Later  
the notation of analytic dilations was  extended  to  include models of non-relativistic quantum field theories, cf.    \cite{BacFroSig98-1,BacFroSig98-2,BacFroSig99},
to obtain properties of resonances \cite{HasHerHub08,AboFauFroSig09}.
In particular, existence of eigenvectors  of the   dilation analytic extension of the Hamiltonian of non-relativistic qed  was  shown
 in \cite{BacFroSig98-1,BacFroSig98-2,BachBallesterosPizzo.2017} for  a mild infrared regularization  and in \cite{Sig09,BachBallesterosIniestaPizzo.2021} without any such regularization. In \cite{BalDecHan.2019} it was
shown for an infrared regularized spin boson model
  that eigenvectors of
 dilation  analytic extensions of the Hamiltonian exist and  depend analytically on the dilation parameter $\theta$
and on the coupling constant.  We note that for the spin boson model a result of this type  could also be obtained  from a direct  application of  results in \cite{GriHas09}
for the ground state and results in \cite{HasLan22} for the resonance state.
The $(g,\theta)$-analyticity statement
of   Theorem  \ref{thm:main1}  of the  present paper now shows  such a    property  for the ground state of   atoms or ions in the framework of non-relativistic qed
without any infrared regularization.

Analytic extensions of Hamiltonians describing quantum mechanical atoms  or molecules
 with respect to spatially dependent  phase transformations  were studied
to obtain exponential decay of  eigenvectors \cite{OConnor.1973} as well as resonance states \cite{CombesThomas.1973}, i.e.,  eigenvectors of analytically dilated Hamiltonians.
 In this paper we  apply   this  techniques to   non-relativistic quantum field theories  and  obtain   spacial exponential decay
of the analytically dilated ground state,  Corollary   \ref{cor:expdecay}. In the framework of
quantum field theory
the method as well as the result  seem new.
Our  result  extends   exponential decay  results  for  the ground of self-adjoint  Hamiltonians (i.e.,  without any
analytic dilation)  
  \cite{GriLieLos01,Griesemer.2004},  which were obtained  using a variant of   Agmon's method \cite{Agmon.1982}.

The novelty of our result is that it gives analyticity of the ground
state jointly  in $g$,  $\theta$,  as well as   $\alpha$,  and moreover it shows spatial exponential decay for the analytically dilated ground state.
In particular, the  result,
 that the ground state has an analytic extension in $\theta$ which decays  exponentially,
is used  in \cite{Hasler.2023}
as an assumption to prove a formula for one photon scattering.

Let us now address the proof of the main result.
It is well known that the ground state energy is embedded in the continuous
spectrum. In such a situation regular perturbation theory is
typically not applicable and other methods have to be employed.
To prove our result about   existence and  analyticity
 we use a variant of the operator theoretic renormalization analysis as introduced
in \cite{BacFroSig98-2} and further developed in \cite{BCFS}. In particular,
we extend the analysis of    \cite{HasHer11-2} (which in turn is inspired by    \cite{HasHer11-1})  to include  analyticity in  the dilation parameter and spatially dependent phase transformations.
As in  \cite{HasHer11-2} we use rotation invariance  which is given naturally for atoms
to control the marginal terms.
 Therefore,  we do need to assume  any   infrared regularization of the interaction  (as was needed in \cite{GriHas09}).
We note that  a possible  alternative approach
to control the marginal terms
would be to use  a generalized Paul-Fierz transformation \cite{Sig09}.
For simplicity we assume that the electrons of the atom are spin-less,  and we assume that the atomic
Hamiltonian has a unique ground state. This non-degeneracy assumption
is typically made in the context of  operator theoretic renormalization.
We believe that this non-degeneracy assumption could
be relaxed by exploiting  symmetry properties. For  recent papers addressing this issue we refer the
reader to  \cite{HasLan18-1,HasLan22}.

In the next section we introduce the model and state the main results. The following  sections are devoted to the proofs.
  In particular, in  Section \ref{sec:outline} we give an outline of the proof.

\section{Model and Statement of Results}

\label{sec:modalmainres}

First  we introduce the atomic Hilbert space, which describes the configuration of $N$ electrons, by
$$
\HH_{\rm at} := \{ \psi \in L^2([\R^{3}]^N) :  \psi(x_{\pi(1)},...,x_{\pi(N)}) = {{\rm sgn}(\pi)} \psi(x_1,...,x_N) , \pi \in \mathfrak{S}_N \} ,
$$
where $\mathfrak{S}_N$ denotes the group of permutations of $N$ elements, ${\rm sgn}$ denotes the signum of the permutation,
and $x_j \in \R^3$ denotes the coordinate of the $j$-th electron. We will use the notation $x = (x_1,...,x_N)$.
The atomic Hamiltonian is the following operator in $\HH_{\rm at}$ defined by
$$
H_{\rm at} := - \Delta + V ,
$$
where we introduced the Laplacian $\Delta := \sum_{j=1}^N \sum_{l=1}^3 \partial_{x_{j,l}}^2$
and the potential  $V : \R^{3 N} \to \R$  which we assume to be infinitesimally  bounded with respect to $-\Delta$
and to be symmetric with respect to permutations of particle coordinates, cf. Hypothesis \ref{potential}.
In this paper we will adopt   the physicists  convention that for  vector valued expressions $a =(a_1,...,a_d)$
we write $a^2$ as a short hand notation for  $\sum_{j=1}^d a_j^2$. Introducing  $p_j = - i \partial_{x_j}$   we can write $-\Delta = \sum_{j=1}^N p_j^2$.
We shall denote by $\mathcal{D}(-\Delta)$ the natural domain of the Laplacian.
We will use the notation for $p \in \C$ and $r > 0$
$$
D_r(p) := \{ z \in \C : |z-p| < r \} , \quad D_r := D_r(0)  .
$$

Let $(\hh, \langle \cdot , \cdot \rangle_\hh)$ be a
Hilbert space. We introduce the direct sum of the $n$-fold tensor product of $\hh$ and define
$$
\mathcal{F}(\hh) := \bigoplus_{n=0}^\infty \mathcal{F}^{(n)}(\hh) , \qquad \mathcal{F}^{(n)}(\hh) = \hh^{\otimes^n} ,
$$
where we have set $\hh^{\otimes 0} := \C$.
We introduce the vacuum vector $\Omega := (1,0,0,...) \in \FF(\hh)$.
The space $\mathcal{F}(\hh)$ is an inner product space
where the inner product is induced from the inner product in $\hh$. That is, on vectors $\eta_1 \otimes \cdots \eta_n, \varphi_1 \otimes \cdots \varphi_n \in \FF^{(n)}(\hh)$
we have
$$
\langle \eta_1 \otimes \cdots \eta_n , \varphi_1 \otimes \cdots \varphi_n \rangle = \prod_{i=1}^n
\langle \eta_{i} , \varphi_{i} \rangle_\hh .
$$
This  extends uniquely to all of $\FF(\hh)$ by sesquilinearity and continuity. We introduce the bosonic Fock space
$$
\FF_s(\hh) := \bigoplus_{n=0}^\infty \mathcal{F}_s^{(n)}(\hh) , \qquad \mathcal{F}_s^{(n)}(\hh) := S_n \FF^{(n)}(\hh) ,
$$
where $S_n$ denotes the orthogonal projection onto the subspace of  symmetric
tensors in $\FF^{(n)}(\hh)$. For $h \in \hh$ we introduce the so called creation operator $a^*(h)$ in $\FF_s(\hh)$
which is defined on vectors $\eta \in \FF^{(n)}_s(\hh)$
by
\beqn \label{eq:formala}
a^*(h) \eta := \sqrt{n+1} S_{n+1} ( h \otimes \eta ) \; .
\eeqn
The operator $a^*(h)$ extends by linearity to a densely defined linear operator on $\FF(\hh)$.
One can show that $a^*(h)$ is closable, c.f. \cite{ReeSim2}, and we denote its closure by the same symbol.
We introduce the annihilation operator by $a(h) := (a^*(h))^*$.
For a unitary  operator $U$ in  $\hh$  we introduce the operator $\Gamma(U)$  in $\FF(\hh)$ by
$$
\Gamma(U) \eta := U \eta_1 \otimes \cdots \otimes U \eta_n .
$$
It is straight forward to see that $\Gamma(U)$ is unitary and  leaves the subspace $\FF_s(\hh)$ invariant.
For  $A$ a self-adjoint  operator with domain $\mathcal{D}(A)$ we define  $d\Gamma(A)$ in $\FF(\hh)$ defined
on vectors $\eta = \eta_1 \otimes \cdots \otimes \eta_n \in \FF^{(n)}(\hh)$, with $\eta_i \in \mathcal{D}(A)$, by
$$
d \Gamma(A) \eta := \sum_{i=1}^n \eta_1 \otimes \cdots \otimes \eta_{i-1} \otimes A \eta_i \otimes \eta_{i+1} \otimes \cdots \otimes \eta_n
$$
and extended by linearity to a densely defined linear operator on $\FF(\hh)$. One can show
that $d \Gamma(A)$ is  closable, c.f. \cite{ReeSim2}, and we denote their closure by the same symbol.
The operator $d \Gamma(A)$ leaves the subspace $\FF_s(\hh)$ invariant, that is,
its  restriction  to $\FF_s(\hh)$ is  densely defined, closed, and has range contained in $\FF_s(\hh)$.
To define quantum electrodynamics, we fix
$$
\hh := L^2( \R^3 \times \Z_2 )
$$
and set $\FF := \FF_s(\hh)$. We denote the norm of $\hh$ by $\| \cdot \|_\hh$.
We define the operator of the free field energy
by
$$
H_{\rm f} := d \Gamma(M_\omega) ,
$$
where $\omega(k,\lambda) := \omega(k) := |k|$ and $M_\varphi$ denotes the operator of multiplication with the function $\varphi$.
We shall denote by $\mathcal{D}(H_{\rm f})$ the natural domain of $H_{\rm f}$. 
For $f \in \hh$ we write
$$
a^*(f) = \sum_{\lambda=1,2} \int_{\R^3} dk f(k,\lambda) a^*(k,\lambda) , \qquad a(f) = \sum_{\lambda=1,2} \int_{\R^3} dk \overline{f(k,\lambda)} a(k,\lambda) ,
$$
where $a(k,\lambda)$ and $a^*(k,\lambda)$ are operator-valued distributions. For a precise definition of the operator
valued distributions, see Appendix \ref{sec:estfock}.
They satisfy the
following commutation relations, which are  understood in the sense of distributions,
$$
[a(k,\lambda), a^*(k',\lambda') ] = \delta_{\lambda \lambda'} \delta(k - k') , \qquad [a^\#(k,\lambda), a^\#(k',{\lambda'}) ] = 0 \; ,
$$
where $a^{\#}$ stands for $a$ or $a^*$.
 For $\lambda=1,2$ we introduce the so called polarization vectors
$$
\varepsilon(\cdot , \lambda) : S^2 := \{ k \in \R^3 :  |k| = 1 \} \to \R^3
$$
to be measurable maps such that for each $k \in S^2$ the vectors $\varepsilon(k,1), \varepsilon(k,2),k$ form an orthonormal basis of $\R^3$.
We extend $\varepsilon(\cdot , \lambda)$ to $\R^3 \setminus \{ 0 \}$ by setting
$
\varepsilon(k,\lambda) := \varepsilon(k/|k|,\lambda)
$
for all nonzero $k$. For $y \in \R^3$ we define the field operator
\beqn \label{eq:afield}
A(y) := \sum_{\lambda=1,2} \int_{\R^3} \frac{dk}{\sqrt{2 |k|}} \left[  \kappa(k) e^{-i \beta k \cdot y}
\varepsilon(k,\lambda) a^*(k,\lambda) + \overline{\kappa(k)} e^{i \beta k \cdot y} \varepsilon(k, \lambda) a(k, \lambda) \right] \ ,
\eeqn
where the function $\kappa : \R^3 \to \C$ serves as a cutoff  which we assume to be  rotation invariant and to  satisfy
\begin{align}\label{boundonint}
\frac{\kappa}{\sqrt{\omega}}   ,  \, \frac{\kappa}{\omega} \in \hh   ,
\end{align}
and $\beta \in \R$ is model  parameter depending on the specific choice of units.
The Hilbert space describing the full system is
$$\HH := \HH_{\rm at} \otimes \FF. $$
The Hamiltonian describing the interaction is  the following operator
\begin{equation} \label{eq:hamiltoniandefinition}
H_{g} :=  \sum_{j=1}^N    ( p_j + g A( x_j) )^2    + V + H_{\rm f} ,
\end{equation}
where we introduce the coupling constant $g \in \C$, which  is of  interest for the main result,  Theorem~\ref{thm:main1}.
For the definition of  \eqref{eq:hamiltoniandefinition} in terms of forms, see Appendix \ref{sec:estfock}.
In fact, one can show the following result.
\begin{theorem}[\cite{Hiros02,HasHer08-2}]  For $g \in \R$ the operator
$H_{g}$ is a self-adjoint operator with domain $\mathcal{D}(-\Delta + H_{\rm f})$ and that $H_{g}$ is essentially
self adjoint on any operator core for $-\Delta +H_{\rm f}$.
\end{theorem}

Next we define the notation of analytic dilation.

\begin{definition}  The group of unitary operators $u(\theta)$ on $L^2(\R^3)$ given by
\begin{align*}
(u(\theta) \psi )(x) = e^{ 3 \theta/2} \psi(e^\theta x)
\end{align*}
is called the {\bf group of dilation operators on} $\R^3$.
\end{definition}

The factor $e^{3 \theta/2}$ is included to make $u$ unitary.
We extend the definition of  $u$ to a unitary transformation
on $\HH_{\rm at}$ by defining
\begin{align*}
  U_{\rm el}(\theta) = \bigotimes_{j=1}^N u(\theta)  .
\end{align*}
It is straight forward to check that
\begin{align}
U_{\rm el}(\theta) x_j U_{\rm el}(\theta)^{-1} = e^{\theta} x_j , \quad
U_{\rm el}(\theta) p_j U_{\rm el}(\theta)^{-1} = e^{-\theta} p_j , \label{trafoofxunidil}
\end{align}
and so
\begin{align} \label{dilatedhat}
 H_{\rm at}(\theta) := U_{\rm el}(\theta) H_{\rm at}  U_{\rm el}(\theta)^{-1} = e^{-2 \theta} (-\Delta)   + V_\theta
\end{align}
where
\begin{align} \label{dilatedhat2}
 V_\theta(x_1,....,x_N)  :=  V( e^{\theta} x_1,..., e^{\theta} x_N) .
\end{align}

We now state the assumptions on the potential $V$.

\begin{hypothesis} \label{potential} The potential  $V$ has  the following properties:
\begin{enumerate}[(i)]
\item \label{anabound(i)} $V$ is invariant under rotations and permutations, that is
\begin{align*}
V(x_1,...,x_N) & = V(R^{-1} x_1,...,R^{-1} x_N) , \quad \forall R \in SO(3) ,  \\
V(x_1,...,x_N) & = V(x_{\pi(1)},...,x_{\pi(N)})  , \quad \forall \pi  \in \mathfrak{S}_N .
\end{align*}
\item  \label{anabound(ii)} $V$ is infinitesimally operator bounded with respect to $-\Delta$.
\item \label{anabound(iii)}$E_{\rm at} := \inf \sigma(H_{\rm at} )$ is a non-degenerate isolated eigenvalue of $H_{\rm at}$.
\item  \label{anabound(iv)}  There exists a $\theta_{\rm at} > 0$ such that the map $\theta \mapsto V_\theta (-\Delta + 1 )^{-1} $ with values in the bounded operators in $\HH_{\rm at}$
has an analytic extension into a neighborhood of zero containing   $D_{\theta_{\rm at}}$. For any $a > 0$ there
exists a constant $C_a$ such that for this extension
\begin{align} \label{eq:boundonc}
 \left\| V_\theta(x)  \psi \right\| \leq a \| \Delta\psi \| + C_a \| \psi \| .
 \end{align}
for all $\psi \in D(-\Delta)$ and $\theta \in D_{\theta_{\rm at}}$.
\end{enumerate}
\end{hypothesis}

\begin{remark} {\rm We note that  for any $Z \in \N$ and $e  \in \R$ the Coulomb potential
\begin{align} \label{dilatedhat2}
 V_C(x) :=   -  \sum_{j=1}^N \frac{Z e^2}{|x_j|}  + \sum_{i < j } \frac{e^2}{|x_i - x_j| }
\end{align}
satisfies Parts   \eqref{anabound(i)}, \eqref{anabound(ii)},   and \eqref{anabound(iv)} of Hypothesis \ref{potential} for any $\theta_{\rm at} > 0$.   This is shown in Lemma \ref{anatypeAtheta-12} of the appendix. Part  \eqref{anabound(iii)} is well known to hold
for the hydrogen atom ($N=Z=1$) and reasonable to assume  to hold for the noble gases.

}
\end{remark}

If  Hypothesis   \ref{potential} holds, it follows from Part  \eqref{anabound(iv)}
that  the right hand side of   \eqref{dilatedhat} defined  on $\mathcal{D}(-\Delta)$ extends analytically to a
 well defined operator for all $\theta \in D_{\theta_{\rm at}}$,
which we shall denote by the same symbol $H_{\rm at}(\theta)$.
In fact, we show in the appendix  in Lemma \ref{anatypeAtheta} that this yields a closed operator and
that the right hand side  \eqref{dilatedhat} is   an analytic family of type (A). The notations of  analytic families can be
found in the appendix, cf.  Definition    \ref{defanaA}.

The Fock space inherits a  dilation transformation $U_{\rm f}(\theta)$ on  $\FF(\hh)$ by defining it on $\FF_n(\hh)$
by
\begin{align*}
U_{\rm f}(\theta) |_{ \FF_n(\hh)}= \bigotimes_{j=1}^n u(-\theta)  .
\end{align*}
Note that there is a negative sign in front of $\theta$, since the transformation acts in momentum space.
It  is straight forward to verify
\begin{align*}
U_{\rm f}(\theta)  H_{\rm f} U_{\rm f}(\theta)^{-1} = e^{-\theta} H_{\rm f}
\end{align*}
\begin{align*}
U_{\rm f} a^*(f) U_{\rm f}^* = a^*(u(-\theta) f ) , \quad U_{\rm f} a(f) U_{\rm f}^* = a(u(-\theta) f )
\end{align*}
in the sense of operator valued distributions one  finds
\begin{align*}
U_{\rm f} a^*(k) U_{\rm f}^* = e^{3\theta/2}  a^*(e^{\theta} k) , \quad U_{\rm f} a(k) U_{\rm f}^* = e^{3 \theta/2}  a( e^{\theta} k) . 
\end{align*}

On the full Hilbert space $\HH = \HH_{\rm at} \otimes \FF(\hh)$ define the group of dilations by
\begin{align*}
U(\theta) = U_{\rm el}(\theta)  \otimes U_{\rm f}(\theta)
\end{align*}
for all $\theta \in \R$.
For $\theta \in \R$ and $g \in \C$ we define the operator
\begin{align} \label{hgtheta}
H_g(\theta) := U(\theta)   H_g U(\theta)^{-1} =  H_{\rm at}(\theta) + W_g(\theta) +  e^{-\theta} H_{\rm f} ,
\end{align}
on the domain $\mathcal{D}(-\Delta + H_{\rm f})$,
where
\begin{align*}
W_g(\theta) :=  \sum_{j=1}^N  \left(   e^{-\theta}  p_j \cdot  g  A_\theta(x_j) +  g A_\theta(x_j)  \cdot   e^{-\theta}   p_j  + g^2 A_\theta(x_j) \cdot A_\theta(x_j) \right)
\end{align*}
with
\begin{align} \label{trafoA}
& A_{\theta}(x_j) := U_{\rm f}(\theta)  A_{\theta}( e^\theta x_j) U_{\rm f}(\theta) ^* \\
&=  e^{-\theta} \sum_{\lambda=1,2} \int_{\R^3} \frac{dk }{\sqrt{2 |k|}} \left[ \kappa(e^{-\theta }k) e^{-i \beta k \cdot x_j}
\varepsilon(k,\lambda) a^*(k,\lambda) + \overline{ \kappa(e^{-\overline{\theta} }k)} e^{i \beta k \cdot x_j} \varepsilon(k, \lambda) a(k, \lambda) \right] \ . \nonumber
\end{align}

\begin{hypothesis} \label{kappa} For $\theta \in \R$ let
$\theta \mapsto \kappa_\theta(k) := \kappa( e^{-\theta }k)  $.  There exists a positive $\theta_{\rm I}$
 such that the  maps
\begin{itemize}
\item[(i)] $\theta \mapsto \omega^{-1/2} \kappa_\theta $ with values in $\hh$,
\item[(ii)] $\theta \mapsto \omega^{-1} \kappa_\theta$  with values in $\hh$,
\item[(iii)]  $\theta \mapsto \kappa_\theta $  with values in $L^\infty(\R^3 )$
\end{itemize}   have analytic extensions into a neighborhood of zero containing $D_{\theta_{\rm I}}$.
\end{hypothesis}

\begin{remark} {\rm  For example $\kappa(k) = e^{-|k|}$ satisfies Hypothesis  \ref{kappa}. 
}
\end{remark}

\begin{remark} \label{kdappeana}  {\rm  If  Hypothesis \ref{kappa} holds,  it follows  
that  also the maps  $ \theta \mapsto  \omega^{-1/2}  \overline{ \kappa_{\overline{\theta}}}$  
and $ \theta \mapsto  \omega^{-1}  \overline{ \kappa_{\overline{\theta}}}$ with  values in $\hh$,
and  $ \theta \mapsto   \overline{ \kappa_{\overline{\theta}}}$    with values in $L^\infty(\R^3 )$ 
are analytic on   $D_{\theta_{\rm I}}$. 
Furthermore, it follows 
for all $\theta \in D_{\theta_{\rm I}}$  that 
$ e^{-2 \theta} \int_{\R^3}   |k|^{-1} \overline{\kappa_{\overline{\theta}}}(k)  \kappa_\theta(k)  dk  = 
 \int_{\R^3}   |k|^{-1} |\kappa(k)|^2  dk $,
since the expression is analytic in $\theta$ by (ii) and is constant 
for real $\theta$ by the transformation  formula for integrals.  Here and throughout the paper we shall use the standard notation that for a complex valued function $f$  defined on some set $X$ we define 
the function $\overline{f}$ on $X$ by $\overline{f}(x) := \overline{f(x)}$ for all $x \in X$. 
}
\end{remark}

First observe that in view of  Hypotheses  \ref{potential} and  \ref{kappa}  and
the elementary  bounds of creation and annihilation operators, cf. Lemma \ref{kernelopestimate},
the operator  $H_g(\theta)$, defined  in  \eqref{hgtheta},  
extends analytically
  to  $\theta  \in D_{\rm \theta_{\rm at}} \cap D_{\theta_{\rm I}}$.
 We shall denote this extension again by $H_g(\theta)$.
That this yields indeed a closed operator for $(\theta,g)$ in a neighborhood of zero
follows from Lemma  \ref{anatypeAtheta2Field} in the appendix, cf. Remark \ref{closedop}.
Next we state our first main result.

\begin{theorem} \label{thm:main0} Assume Hypotheses  \ref{potential} and  \ref{kappa} hold.  Then there exist
 positive constants $g_{\rm b}$ and  $\theta_{\rm b}$
 such that for all $(g,\theta)  \in D_{g_{\rm b}} \times  D_{\theta_{\rm b}} $
the operator $H_{g}(\theta)$ has an eigenvalue $E_g(\theta)$ with eigenvector $\psi_g(\theta)$
satisfying the following properties.
\begin{enumerate}[(i)]
\item$(g,\theta) \mapsto E_g(\theta)$ and $(g,\theta) \mapsto \psi_g(\theta)$
are analytic on $  D_{g_{\rm b}} \times  D_{\theta_{\rm b}} $..
\item If  $(g,\theta)  \in  \cap ( D_{g_{\rm b}} \cap \R ) \times  (  D_{\theta_{\rm b}}  \cap \R  )  $,
then  $E_g(\theta)  = \inf \sigma ( H_g(\theta)) $ and $E_g(\theta)$  is a simple eigenvalue.  
\end{enumerate}
\end{theorem}

Now we add an additional deformation, which we will use to obtain exponential decay of the analytically dilated ground state.

\begin{definition}  Define for $\alpha \in \R$ and $\langle x \rangle  = (1 + |x|^2)^{1/2}  $ the operator $F(\alpha)$  on $\HH_{\rm at}$ by
\begin{align*}
F(\alpha) \psi(x)  = e^{ i \alpha \langle x \rangle} \psi(x)   .
\end{align*}
\end{definition}
It is straight forward to see, that
\begin{align*}
F(\alpha) p_j F(\alpha)^{-1} =  p_j  -  \alpha   \nabla_{x_j}  \langle x \rangle ,
\end{align*}
where we note that $\nabla_{x_j}  \langle x \rangle  = \frac{x_j}{\langle x \rangle }$. Assuming Part \eqref{anabound(iv)} of  Hypothesis  \ref{potential} and   Hypothesis \ref{kappa}
we define for  $\theta  \in D_{\rm \theta_{\rm at}} \cap D_{\theta_{\rm I}}$  and $\alpha, g  \in \C$  the operator
\begin{align} \label{defofanathetaalpha}
H_g(\theta,\alpha) :=  F(\alpha) H_g(\theta) F(\alpha)^{-1}
\end{align}
on the domain $\mathcal{D}(-\Delta + H_{\rm f})$.
Using  the definitions it is straight forward to see that
\begin{align} \label{defofanathetaalpha_separate}
H_g(\theta,\alpha) =   H_{\rm at}(\theta,\alpha)     + W_g(\theta,\alpha)  + e^{-\theta} H_{\rm f} ,
\end{align}
where we defined
\begin{align} \label{defofanathetaalpha1}
&  H_{\rm at}(\theta,\alpha)   := F(\alpha) H_{\rm at}(\theta) F(\alpha)^{-1} \nonumber  \\
& =   \sum_{j=1}^N e^{-2 \theta} \left( p_j -  \alpha  x_j \langle x \rangle^{-1} \right)^2  + V_\theta(x)  \nonumber \\
& =   H_{\rm at}(\theta)  -   e^{-2 \theta} \alpha  \sum_{j=1}^N \left(  p_j  \cdot    x_j \langle x \rangle^{-1}  +    x_j \langle x \rangle^{-1}  \cdot  p_j
\right)   + e^{-2 \theta}  \alpha^2 x^2 \langle x \rangle^{-2}
\end{align}
and
\begin{align} \label{diffalphaW}
 &  W_{g}(\theta,\alpha)  := F(\alpha) W_{g}(\theta) F(\alpha)^{-1} \nonumber \\
& = \sum_{j=1}^N  \left(  e^{-\theta}   \left[ \left( p_j -  \alpha  x_j \langle x \rangle^{-1}    \right) \cdot  g  A_\theta(x_j) +  g A_\theta(x_j)  \cdot   \left( p_j -  \alpha  x_j \langle x \rangle^{-1}    \right) \right]   + g^2 A_\theta(x_j) \cdot A_\theta(x_j) \right)   \nonumber  \\
& = W_{g}(\theta) -  2 \alpha   \sum_{j=1}^N   e^{-\theta} x_j \langle x \rangle^{-1}  \cdot  g  A_\theta(x_j)   .
\end{align}
 Note that $F(\alpha)$ and $U(\theta)$ do not commute.
 Let us first discuss  the atomic part  \eqref{defofanathetaalpha1}. Since $p_j$ is infinitesimally $-\Delta$ bounded and $x\langle x \rangle^{-1}$ is bounded
it follows from    Hypothesis \ref{potential}   that the right hand side  \eqref{defofanathetaalpha1}  
is a well defined operator  on $\mathcal{D}(-\Delta)$  for all $(\theta, \alpha)  \in D_{\theta_{\rm at}} \times  \C$.
Again, one obtains a closed operator and one can show
that the right hand side   \eqref{defofanathetaalpha1}  is an analytic family of type (A).  
  This  is shown in Lemma \ref{anaAalphathetaat} of  the appendix.
  Let us now consider the  full Hamiltonian  \eqref{defofanathetaalpha_separate}.
First observe that in view of the basic bounds, cf. Lemma \ref{kernelopestimate},  one can define for all
 $\theta \in   D_{\rm \theta_{\rm at}} \cap D_{\theta_{\rm I}}$ and $\alpha, g \in \C$ the operator $H_g(\theta,\alpha)$ on the domain $\mathcal{D}(-\Delta + H_{\rm f})$.
That this yields indeed a closed operator for $\theta $, $\alpha$,  and $g$ in a neighborhood of zero follows from
Lemma \ref{anaAalphathetaField}  in the appendix, cf. Remark \ref{closedop}.

We  now state the second main theorem of this paper, concerning  the  analytic continuation in $\alpha$ by means of  $F(\alpha)$.

\begin{theorem} \label{thm:main1} Assume Hypotheses \ref{potential} and \ref{kappa} hold.
Then there exist
 positive constants $g_{\rm b}$, $\theta_{\rm b}$, and $\alpha_{\rm b}$
 such that for all $(g,\theta,\alpha)  \in D_{g_{\rm b}} \times  D_{\theta_{\rm b}}  \times  D_{\alpha_{\rm b}}  $
the operator $H_{g}(\theta,\alpha)$ has an eigenvalue $E_g(\theta,\alpha)$ with eigenvector $\psi_g(\theta,\alpha)$ 
satisfying the following properties.
\begin{enumerate}[(i)]
\item $(g,\theta,\alpha)  \mapsto E_g(\theta,\alpha)$ and $(g,\theta,\alpha) \mapsto \psi_g(\theta,\alpha)$
are analytic on $  D_{g_{\rm b}} \times  D_{\theta_{\rm b}}  \times  D_{\alpha_{\rm b}} $.
\item If  $(g,\theta,\alpha)  \in  ( D_{g_{\rm b}} \cap \R ) \times  (  D_{\theta_{\rm b}}  \cap \R  )  \times  ( D_{\alpha_{\rm b}} \cap \R ) $,
then  $E_g(\theta,\alpha)  = \inf \sigma ( H_g(\theta,\alpha)) $ and  $E_g(\theta,\alpha)$ is a simple eigenvalue.
\end{enumerate}
\end{theorem}

\begin{remark}{ \rm  We note that statments as  in Parts (ii) of   Theorems  \ref{thm:main0} and  \ref{thm:main1}  that the eigenvalue is simple,  can also be  shown using  more direct methods  such as photon number bounds, cf.   \cite{BacFroSig98-1}.}
\end{remark} 

\begin{remark} \label{gsatconst} {\rm
We want to point out that since $\theta$ and $\alpha$ are  parameters  of  analytic extensions of unitary
groups, it follows from the analyticity result  (i) in Theorems \ref{thm:main0}  and \ref{thm:main1} that
 $E_{g}(\theta)$ and $E_g(\theta,\alpha) $ are constant functions of $\theta$ and $(\theta,\alpha)$,
 respectively.
   This can be seen as follows.
For real $\theta$ we have  $ H_{\rm at}(\theta,0) = U_{\rm el}(\theta)  H_{\rm at}(0,0) U_{\rm el}(\theta)^{-1}$ and $U(\theta)$ is unitary.
So  $ E_{\rm at}(\theta,0) = E_{\rm at}(0,0)$ for real $\theta$ and hence by analytic  continuation they are equal for all $\theta \in D_{\theta_{\rm b}}$.
Now for real $\alpha$ we have  $ H_{\rm at}(\theta,\alpha) = F(\alpha)  H_{\rm at}(\theta,0) F(\alpha)^{-1}$ and $F(\alpha)$ is unitary.
So  $ E_{\rm at}(\theta,\alpha) = E_{\rm at}(\theta,0)$ for real $\alpha$ and hence by analytic  continuation they are equal for all $\alpha
\in D_{\alpha_{\rm b}}$.
 }
\end{remark}

Next we use  the Lemma of O'Connor to obtain  spacial exponential decay of the ground state as well as its  dilation analytic continuation.
We use the following definition of an analytic vector, see \cite{Nel59} or   \cite[Page 201]{ReeSim2} . 

\begin{definition}  Let $A$ be an operator on a Hilbert space $\HH$.
The set $C^\infty := \bigcap_{n=1}^\infty \mathcal{D}(A^n)$ is called the set of {\bf $C^\infty$-vectors for $A$}.
 A vector $\varphi \in C^\infty(A)$
is called {\bf analytic vector for $A$} if
\begin{align} \label{eq:anadef}
\sum_{n=0}^\infty \frac{ \| A^n \varphi \|}{n!} t^n < \infty
\end{align}
for some $t > 0$.
\end{definition}

We will use the following Lemma, whose proof is  from   \cite[Theorem X.39]{ReeSim2}.

\begin{lemma} \label{expdecaylem}  Let $A$ be a self-adjoint operator and $\psi$ analytic for $A$, explicitly
$  \sum_{n=0}^\infty \frac{ \| A^n \varphi \|}{n!} t^n < \infty$ for some $t > 0$. Then   we have
 $\psi \in \mathcal{D}(\exp ( \frac{s}{2} |A|))$ for any $s  \in [0,t]$.
\end{lemma}
\begin{proof}
This follows from the spectral theorem.  The details are as follows.
Let $\mu$ be the spectral measure of $\psi$ for the operator $A$. Let $0 < s < t$. Then by Schwarz inequality
\begin{align*}
\sum_{n=0}^\infty \frac{s^n}{n!} \int_{-\infty}^\infty |x|^n d\mu & \leq \sum_{n=0}^\infty \frac{s^n}{n!} \left( \int_{-\infty}^\infty x^{2n} d\mu \right)^{1/2} \left( \int_{-\infty}^\infty d\mu \right)^{1/2}  \\
& = \| \psi \| \sum_{n=0}^\infty \frac{s^n}{n!} \| A^n \psi \| < \infty .
\end{align*}
Thus, by Fubini's theorem (or monotone convergence),
\begin{align*}
 \int_{-\infty}^\infty e^{ s |x| } d\mu = \int_{-\infty}^\infty \sum_{n=0}^\infty \frac{s^n}{n!}|x|^n d\mu  < \infty .
\end{align*}
Thus by the unbounded functional calculus we find    $\psi \in \mathcal{D}(\exp( \frac{s}{2} |A| ))$.
\end{proof}

Now let us state the following proposition, a  proof of which  can be found   in  \cite{OConnor.1973} or \cite[Page 196]{ReeSim4}.

\begin{proposition}[O'Conner's Lemma] \label{oconnor} Let $W(\alpha) = \exp(i \alpha A)$ be a one-parameter
unitary group and let $D$ be a connected open set  in $\C$ with $0 \in D$. Suppose that a projection-valued
analytic function $P(\alpha)$ is given on $D$ with $P(0)$ of finite rank so that
$$
W(\alpha_0) P(\alpha) W(\alpha_0)^{-1} = P(\alpha + \alpha_0)
$$
for all pairs $(\alpha,\alpha_0)$ with real $\alpha_0$ and with $\alpha, \alpha + \alpha_0 \in D$. Let $\psi \in \ran P(0)$.
Then the function $\psi(\alpha)  = W(\alpha) \psi$ has an analytic continuation from $D \cap \R$ to $D$. In particular, $\psi$ is an analytic vector
for $A$.
\end{proposition}

As a corollary of the main theorem,  Theorem \ref{thm:main1}, we obatain the following result about spacial exponential decay.

\begin{corollary} \label{cor:expdecay} Let the assumptions  be as in Theorem \ref{thm:main1}. 
Then there exist positive $g_{\rm b}$, $\theta_{\rm b}$, and $\alpha_{\rm b}$ such that the assertions of Theorem \ref{thm:main1} hold
and  for  $(g,\theta,a)  \in D_{g_{\rm b}} \times  D_{\theta_{\rm b}} \times  [0, \alpha_{\rm b})$  we have 
$\psi_g(\theta) := \psi_g(\theta,0)   \in \mathcal{D}(\exp(a \langle x \rangle ))$ (i.e.
 $ \exp(a \langle \cdot  \rangle ) \psi_g(\theta)  \in \HH$).
\end{corollary}

\begin{proof}
We need only show  by Lemma \ref{expdecaylem}    that $\psi_g(\theta)$ is an analytic vector for the  operator $\langle x \rangle$.
But this follows from Theorem \ref{thm:main1} and an application of  O'Connors lemma,   i.e. Proposition \ref{oconnor},
for the  projection $$ P_g(\theta,\alpha) := \frac{ \psi_g(\theta,\alpha ) \langle  \psi_{\overline{g}}(\overline{\theta},\overline{\alpha}) , \cdot \rangle  }{\inn{ \psi_{\overline{g}}(\overline{\theta},\overline{\alpha}) , \psi_g(\theta,\alpha )     }    } , $$
where we possibly make $g_{\rm b}, \theta_{\rm b}$, and $\alpha_{\rm b}$ smaller such that the denominator does not vanish. 
Specifically, it follows  for $	\alpha_0, \alpha \in D_{\alpha_{\rm b}} \cap  \R$ and $g \in D_{g_{\rm b}} \cap \R$ and $\theta \in D_{\theta_{\rm b}} \cap \R$ that \begin{align} \label{projana} P_g(\theta,\alpha+\alpha_0)  = 
F(\alpha_0) P_g(\theta,\alpha) F(\alpha_0)^{-1} \end{align}
whenever $\alpha_0 + \alpha \in D_{\alpha_{\rm b}} $ by uniqueness of the ground state (in the self-adjoint case). By analyticity of both sides of   Eq.  \eqref{projana}  in $g, \theta , \alpha$, it follows  that Eq.  \eqref{projana} in  fact  holds for all $(g,\theta,\alpha ) \in D_{g_{\rm b}}
\times  D_{\theta_{\rm b}} \times D_{\alpha_{\rm b}} $ 
as long as $\alpha + \alpha_0 \in D_{\alpha_{\rm b}}$. Thus the assumptions of O'Connors lemma are satisfied for the projection valued analytic function  $\alpha \mapsto  P_g(\theta,\alpha)$.
\end{proof}

\begin{remark} {\rm  Analytic extensions of spatially dependent phase transformations were studied in \cite{OConnor.1973,CombesThomas.1973}
to obtain exponential decay of resonance states for non-relativistic quantum mechanical matter. In these works,  one used the transformation $\exp( i \xi \cdot  x )$ with $\xi \in \R^{3N}$.
Since this is not invariant by rotations, it is not convenient for the symmetry treatment of the marginal terms
in the renormalization analysis.  Therefore  we use  $\exp( i \alpha \langle x \rangle ) $ in this paper. }
\end{remark}

\begin{remark} {\rm
 It is reasonable to assume that the
exponential decay obtained in Corollary  \ref{cor:expdecay} could also be obtained using methods from Agmon \cite{Agmon.1982}.
This has been shown for the  self-adjoint case \cite{Griesemer.2004}.
}
\end{remark}

\section{Outline of the Proof}
\label{sec:outline}

We only prove Theorem \ref{thm:main1}.  Theorem  \ref{thm:main0} will then follows as a trivial corollary.
The  method used in the proof of Theorem \ref{thm:main1} is
operator theoretic renormalization as introduced in  \cite{BacFroSig98-2,BCFS} and the fact that renormalization preserves
analyticity \cite{GriHas09,HasHer11-1}.   As mentioned,   Theorem \ref{thm:main1}  is a generalization
of the main result in \cite{HasHer11-2} to a larger class of analytical extensions of the Hamiltonian.
We follow closely  the proof given in  \cite{HasHer11-2}. This allows us, to use     results from that paper and to focus 
on the  parts   originating from the larger class of  analytic extensions.

The renormalization procedure is an iterated application of
the so called smooth Feshbach map. The smooth Feshbach map is reviewed in Appendix  \ref{sec:smo} and  properties relevant
for this paper are  summarized.
To control
the marginal terms in the renormalization analysis and  to show that the renormalization transformation
is a suitable contraction, we use  rotation invariance symmetry as in \cite{HasHer11-2}.

Let us give  an outline of the remaining part of the paper.
 In Section \ref{sec:symmetries} we define an $SO(3)$ action
on the atomic Hilbert space and the Fock space, which leaves the Hamiltonian invariant.
In Section \ref{sec:ban} we introduce Banach spaces which are used
to control and define  the renormalization transformation.
In Section \ref{sec:ini} we show that after an initial Feshbach transformation the Feshbach map lies 
in a suitable Banach space. This allows us to use results of \cite{HasHer11-2} which are collected
in Section \ref{sec:ren:def}. In section \ref{sec:prov} we put all the pieces together and
prove Theorem \ref{thm:main1}.
In Appendix  \ref{ana:res} we collect a few elementary results from complex analysis and operator theory.
In Appendix \ref{sec:estfock} we collect a few estimates and identities which hold for operators in Fock spaces.
In Appendix \ref{ana:con}  we collect   results about   analyticity properties for  specific  operators which we consider in this paper.

We will use the notation that $\R_+ := [0,\infty)$.
Using an appropriate
scaling, see Remark \ref{rem:scvaling},   we will  assume without loss of generality that the distance between the lowest eigenvalue of $H_{\rm at}$ and the rest of the spectrum is one, i.e.,
\begin{equation} \label{eq:hatscale}
E_{\rm at,1} - E_{\rm at} = 1 ,
\end{equation}
where $E_{\rm at,1} := \inf ( \sigma( H_{\rm at}) \setminus\{ E_{\rm at}\}) $.

\begin{remark}\label{rem:scvaling} {\rm
Any Hamiltonian of the form
\eqref{eq:hamiltoniandefinition}   is up to a positive multiple unitarily equivalent to an operator
of the same  form
\eqref{eq:hamiltoniandefinition}, but with a rescaled potential and with different values for
$\kappa$ and $g$.
More explicitly, let  $\delta := E_{\rm at,1} - E_{\rm at}$. Then  choosing $\tau_{\rm at} \in \R$ such that
  $e^{-2 \tau_{\rm at}} \delta^{-1} = 1$  we find from  \eqref{dilatedhat} that
\begin{align} \label{dilatedhat00}
\delta^{-1}  U_{\rm el}(\tau_{\rm at}) H_{\rm at}  U_{\rm el}(\tau_{\rm at})^{-1} =  -\Delta  + \delta^{-1} V_{\tau_{\rm at}} .
\end{align}
Now choose $\tau_{\rm ph} \in \R$ such that
  $e^{- \tau_{\rm ph}} \delta^{-1} = 1$. Then we find using   \eqref{trafoA}
\begin{align} \label{dilatedhat000}
& \delta^{-1} ( U_{\rm el}(\tau_{\rm at})  \otimes U_{\rm f}(\tau_{\rm ph}) )  (\tau_{\rm at}) H_{g} ( U_{\rm el}(\tau_{\rm at})\otimes U_{\rm f}(\tau_{\rm ph}) )^{-1} \nonumber  \\
& =  \sum_{j=1}^N   ( p_j + \widetilde{g} \tilde{A}( x_j ))^2   + \delta^{-1} V_{\tau_{\rm at}}    + H_{\rm f}
\end{align}
with $\tilde{g} = g e^{-\tau_{\rm ph}} \delta^{-1/2}$ ,
\begin{align*}
\tilde{A}(x) := \sum_{\lambda=1,2} \int_{\R^3} \frac{dk }{\sqrt{2 |k|}} \left[ \tilde{\kappa}(k) e^{-i \tilde{\beta} k \cdot x}
\varepsilon(k,\lambda) a^*(k,\lambda) + \overline{\tilde{\kappa}(k) } e^{i \tilde{\beta} k \cdot x} \varepsilon(k, \lambda) a(k, \lambda) \right] \ ,
\end{align*}
 where $\tilde{\kappa}(k)= \kappa(  e^{-\tau_{\rm ph} } k)  $ and  $\tilde{\beta} = \beta e^{\tau_{\rm at}} $.
From the definition of $\delta$ it follows immediately  that the atomic part of the  right hand side of \eqref{dilatedhat000}
satisfies \eqref{eq:hatscale} in view of \eqref{dilatedhat00}.
}
\end{remark}

\section{Symmetries}

\label{sec:symmetries}

Let us introduce the following canonical representation of $SO(3)$ on $\HH_{\rm at}$ and $\hh$. For $R    \in SO(3)$ and $\psi \in \HH_{\rm at}$ we define
$$
\mathcal{U}_{\rm at}(R) \psi(x_1,...,x_N) = \psi(R^{-1} x_1, ... , R^{-1} x_N) .
$$
Next we  introduce  an $SO(3)$  representation on the one particle space $\hh$.
For $R \in SO(3)$ and $f \in \hh$ we define
\begin{equation} \label{eq:repofh}
(\mathcal{U}_\hh(R) f )(k,\lambda) = \sum_{\widetilde{\lambda}=1,2}
D_{\lambda , \widetilde{\lambda}}(R,k) f(R^{-1} k , \widetilde{\lambda}) ,
\end{equation}
where
$
D_{\lambda , \widetilde{\lambda}}(R,k) := \varepsilon(k,\lambda) \cdot R \varepsilon(R^{-1}k, \widetilde{\lambda})
$.
It is straight forward to verify that this defines a unitary representation.
We lift this  representation canonically  to a representation on Fock space and define $\UU_\FF = \Gamma(\UU_\hh)$.
This representation can  be  characterized by
\begin{align} \label{eq:repofcreation2}
\UU_\FF(R) a^\#(f) \UU_\FF(R)^* = a^\#(\UU_\hh(R) f) \quad , \quad \UU_\FF(R) \Omega = \Omega .
\end{align}
We denote the representation on $\HH_{\rm at} \otimes \FF$ by
 $\mathcal{U} =  \mathcal{U}_{\rm at} \otimes \mathcal{U}_\FF$.
The following transformation properties of the operators $(x_j)_l$ and $(p_j)_l$, with $j=1,...,N$ and $l=1,2,3$, are straight forward
to verify
\begin{align} \label{eq:transx}
\mathcal{U}(R) (x_j)_l \mathcal{U}(R)^* &= \sum_{m=1}^3 R_{m,l} (x_j)_m = (R^{-1} x_j)_l  \ , \\
\mathcal{U}(R) (p_j)_l \mathcal{U}(R)^* &= \sum_{m=1}^3 R_{m,l} (p_j)_m = (R^{-1} p_j)_l  \ ,
\end{align}
where we  used the matrix notation $R=(R_{m,n})_{m,n=1,2,3}$.
Moreover, the transformation property of the $l$-th component of the field operator $A_{l}(x_j)$ is
\begin{equation} \label{eq:transA}
\mathcal{U}(R) A_{\theta,l}(x_j) \mathcal{U}(R)^* = \sum_{m=1}^3 R_{m,l} A_{\theta,m}(x_j) = (R^{-1} A_\theta)_l( x_j) .
\end{equation}
This can be seen as follows. For fixed $y \in \R^3$ and $l=1,2,3$ define the function
\begin{equation} \label{defofintkernela}
f_{(\theta, l,y)}(k,\lambda) := e^{-\theta} \frac{\kappa_{\theta}(k)}{\sqrt{ 2 |k|}} \varepsilon(k,\lambda)_l e^{- i \beta k \cdot y} .
\end{equation}
Eq. \eqref{eq:transA} follows inview of \eqref{eq:repofcreation2} and   $\UU_\hh(R) f_{(\theta, l,y)} = \sum_{m=1}^3 R_{m,l} f_{(\theta, m, R y)} $, which in turn follows from  \eqref{eq:repofh}.
We call a linear  operator $T$ in the Hilbert space  $\HH$ rotation invariant if
$T = \UU(R) T \UU(R)^*$ for all $R \in SO(3)$ and likewise for operators in $\FF$ and $\HH_{\rm at}$.
From \eqref{eq:transx}--\eqref{eq:transA} it is evident to see  that the Hamiltonian
$H_g(\theta,\alpha)$
 defined in
 \eqref{defofanathetaalpha},   is rotation invariant.
We will make use of the following Lemma from \cite{HasHer11-2}, see also \cite{HasHer12-3}.

\begin{lemma} \label{lem:mainidea} Let $f \in \hh$.  If   $a^\#(f)$ is an operator which is invariant under rotations,  then $f=0$.
\end{lemma}

\section{Banach Spaces of Hamiltonians}
\label{sec:ban}

In this section we introduce Banach spaces of integral kernels, which
parameterize certain subspaces of the space of bounded 
operators on Fock space.
These subspaces are suitable to study an iterative application of the Feshbach map
and to formulate the contraction property. We closely  follow the exposition in  \cite{HasHer11-2}
and   \cite{BCFS}.

The renormalization transformation will be defined on operators acting on the reduced Fock space
$\mathcal{H}_{\rm red}:= P_{\rm red} \FF$,
where we introduced the notation $P_{\rm red}:= 1_{H_{\rm f} \leq 1}$.
We will investigate bounded operators in $\mathcal{B}(\mathcal{H}_{\rm red})$ of the form
\beqn \label{eq:sum}
H(w) := \sum_{m+n \geq 0} H_{m,n}(w) ,
\eeqn
with $H_{m,n}(w) := H_{m,n}(w_{m,n})$ and
\begin{align}
& H_{m,n}(w_{m,n}) := P_{\rm red} \int_{\un{B}_1^{m+n}} \frac{ d \mu( {K}^{(m,n)})}{|{K}^{(m,n)}|^{1/2}} a^*({K}^{(m)}) w_{m,n}(H_{\rm f}, {K}^{(m,n)}) a(\widetilde{{K}}^{(n)}) P_{\rm red} , \quad m+n \geq 1 , \label{eq:defhmn11} \\
& H_{0,0}(w_{0,0}) := w_{0,0}(H_{\rm f}) , \nonumber
\end{align}
where $w_{m,n} \in L^\infty([0,1] \times \un{B}_1^m \times \un{B}_1^n)$
is an integral kernel for $m+n \geq 1$, $w_{0,0} \in L^\infty([0,1])$, and $w$ denotes the sequence of
integral kernels $(w_{m,n})_{m,n \in \N_0^2}$.
We have used and will henceforth use the following notation. We set $K = (k, \lambda ) \in \R^3 \times \Z_2$, and write
\begin{align*}
& \un{X} := X \times \Z_2 \quad , \quad B_1 := \{ x \in \R^3 : |x|< 1 \} \\
& K^{(m)} := ({K}_1, ... ,{K}_m ) \in \left( {\R}^{3} \times \Z_2 \right)^m ,
\quad \widetilde{{K}}^{(n)} := (\widetilde{{K}}_1, ... , \widetilde{{K}}_n ) \in \left( {\R}^{3} \times \Z_2 \right)^n , \\
& {K}^{(m,n)} := ({K}^{(m)}, \widetilde{{K}}^{(n)}) \\
& \int_{\un{X}^{m+n}} d {K}^{(m,n)} := \int_{{X}^{m+n}} \sum_{(\lambda_1,...,\lambda_m,\widetilde{\lambda}_1,...,\widetilde{\lambda}_n) \in \Z_2^{m+n} } dk^{(m)} d\widetilde{k}^{(n)} \\
& dk^{(m)} := \prod_{i=1}^m {d^3 k_i} , \quad d\widetilde{k}^{(n)} := \prod_{j=1}^n {d^3 \widetilde{k}_j} , \quad
d K^{(m)} := d K^{(m,0)} , \quad d \widetilde{K}^{(n)} := d K^{(0,n)} , \\
& d \mu (K^{(m,n)}) := (8 \pi )^{-\frac{{m+n}}{2}} d K^{(m,n)} \\
& a^*({K}^{(m)}) := \prod_{i=1}^m a^*({K}_i) , \quad a(\widetilde{{K}}^{(m)}) := \prod_{j=1}^m a(\widetilde{{K}}_j) \\
& | {K}^{(m,n)}| := | {K}^{(m)} | \cdot | \widetilde{{K}}^{(n)}| , \quad | {K}^{(m)} | := |k_1| \cdots |k_m | , \quad | \widetilde{{K}}^{(m)} | := |\widetilde{k}_1| \cdots |\widetilde{k}_m | , \\
& \Sigma[{K}^{(m)}] := \sum_{j=1}^m |k_j | .
\end{align*}
Note that in view of the pull-through formula, Lemma \ref{lem:pullthrough}, Eq.  \eqref{eq:defhmn11} is equal to
\beqn \label{eq:defintegralkernel}
\int_{\underline{B}_1^{m+n}} \frac{ d \mu(K^{(m,n)})}{|K^{(m,n)}|^{1/2}} a^*(K^{(m)}) 1_{H_{\rm f} + \Sigma[K^{(m)}] \leq 1 } w_{m,n}(H_{\rm f} , K^{(m,n)})
1_{H_{\rm f} + \Sigma[\tilde{K}^{(n)}] \leq 1}  a(\tilde{K}^{(n)} ) \; .
\eeqn
Thus we can restrict attention to integral kernels $w_{m,n}$ which are essentially supported on the sets
\begin{align*}
\underline{Q}_{m,n} &:= \{ ( r , K^{(m,n)}) \in [0,1] \times \underline{B}_1^{m+n} \ :  \ r \leq 1 -
\max(\Sigma[K^{(m)}],
\Sigma[\widetilde{K}^{(m)}]) \} , \quad m + n \geq 1 .
\end{align*}
Moreover, note that integral kernels can always be assumed to be symmetric. That is, they lie in the range of the symmetrization operator,
which is defined as follows,
\begin{eqnarray} \label{eq:symmetrization}
w_{M,N}^{({\rm sym})}(r , K^{(M,N)})
 := \frac{1}{M!N!} \sum_{\pi \in \mathfrak{S}_M} \sum_{\widetilde{\pi} \in \mathfrak{S}_N} {w}_{M,N}(r,
K_{\pi(1)},\ldots,K_{\pi(N)}, \widetilde{K}_{\widetilde{\pi}(1)},\ldots,\widetilde{K}_{\widetilde{\pi}(M)}).
\end{eqnarray}
Note that \eqref{eq:defhmn11} is understood in the sense of forms. It defines a densely defined form
which can be seen to be bounded using the expression \eqref {eq:defintegralkernel} and Lemma \ref{kernelopestimate}.
Thus it uniquely determines a bounded operator which we denote by $H_{m,n}(w_{m,n})$. This is explained in more
detail in Appendix 
\ref{sec:estfock}. We have the following lemma.
\begin{lemma}\label{lem:estopwithfac}  For $w_{m,n} \in L^\infty([0,1] \times \un{B}_1^m \times \un{B}_1^n)$\label{lem:operatornormestimates} we have
\beqn \label{eq:operatornormestimate1}
\|H_{m,n}(w_{m,n}) \|_{} \leq \| w_{m,n} \|_{\infty} ( n! m!)^{-1/2} \; .
\eeqn
\end{lemma}
The proof follows using Lemma \ref{kernelopestimate}
and the estimate
\begin{equation} \label{eq:intofwKminus2}
 \int_{\underline{S}_{m,n}} \frac{d K^{(m,n)}}{|K^{(m,n)}|^{2}} \leq \frac{(8 \pi)^{m+n}}{{n! m!}} ,
\end{equation}
where $\underline{S}_{m,n} := \{ (K^{(m)},\widetilde{K}^{(n)}) \in \underline{B}_1^{m+n}
 \ : \Sigma[K^{(m)}] \leq 1 , \Sigma[\widetilde{K}^{(n)}] \leq 1 \}$.
The renormalization procedure will involve kernels which lie in the following Banach spaces.
We shall identify the space $L^\infty(\underline{B}_1^{m+n}; C[0,1])$ with a subspace of $L^\infty([0,1]\times \underline{B}_1^{m+n})$ by
setting $$w_{m,n}(r,K^{(m,n)}) = w_{m,n}(K^{(m,n)})(r)$$ for $w_{m,n} \in L^\infty(\underline{B}_1^{m+n}; C[0,1])$. For example in  (i) and (ii)
of  Definition \ref{def:wgartenhaag}, below,  we use this identification.
The norm in $L^\infty(\underline{B}_1^{m+n}; C[0,1])$
is given by
$$
\| w_{m,n} \|_{\underline{\infty}} := \esssup_{K^{(m,n)} \in \underline{B}_1^{m+n}} \sup_{r \geq 0}| w_{m,n}(K^{(m,n)})(r)| .
$$
We note that for $w \in L^\infty(\underline{B}_1^{m+n}; C[0,1])$ we have $ \| w \|_\infty \leq \| w \|_{\underline{\infty}}$.
Conditions (i) and (ii) of the
following definition are needed for the injectivity property stated in Theorem \ref{thm:injective}, below.

\begin{definition} \label{def:wgartenhaag}
We define $\WW_{m,n}^\#$ to be the Banach space consisting of functions $w_{m,n} \in
L^\infty(\underline{B}_1^{m+n};C^1[0,1])$ satisfying the following properties:
\begin{itemize}
\item[(i)] $ w_{m,n} (1 - 1_{\underline{Q}_{m,n}} ) = 0$, for $m + n \geq 1$,
\item[(ii)] $w_{m,n}(\cdot,K^{(m)}, \widetilde{K}^{(n)})$ is totally
symmetric in the variables $K^{(m)}$ and $\widetilde{K}^{(n)}$
\item[(iii)] the following norm is finite
$$
\| w_{m,n} \|^\# := \| w_{m,n} \|_{\underline{\infty}} + \| \partial_r w_{m,n} \|_{\underline{\infty}} .|
$$
\end{itemize}
Hence for almost all $K^{(m,n)} \in \underline{B}_1^{m+n}$ we have $w_{m,n}(\cdot,K^{(m,n)}) \in C^1[0,1]$, where
the derivative is denoted by $\partial_r w_{m,n}$.
For $0<\xi < 1$, we define the Banach space
$$
 \mathcal{W}^\#_{\xi} := \bigoplus_{(m,n) \in \N_0^2 } \mathcal{W}_{m,n}^\# \
$$
to consist of all sequences $w =( w_{m,n})_{m,n \in \N_0}$ satisfying
$$
\| w \|_\xi^\# := \sum_{(m,n)\in \N_0^2} \xi^{-(m+n)} \| w_{m,n}\|^\# < \infty .
$$
\end{definition}

\begin{remark}
{\em
We shall also use the norm $\| w_{m,n} \|^\#$ for any integral kernel $w_{m,n} \in L^\infty( \underline{B}_1^{m+n}; C^1[0,1])$.
 Note that by  the triangle inequality $\| w_{m,n}^{({\rm sym})} \|^\# \leq \| w_{m,n} \|^\#$.}
\end{remark}

Given $w \in \mathcal{W}_\xi^\#$, we
write $w_{\geq r}$ for the vector in $\mathcal{W}_\xi^\#$ given by
$$
(w_{\geq r})_{m + n} = \left\{ \begin{array}{ll} w_{m,n} & , \quad {\rm if} \ m+n \geq r \\ 0 & , \quad {\rm otherwise} . \end{array} \right.
$$
We will use the following balls to define the renormalization transformation
\begin{align*}
\mathcal{B}^\#(\delta_1,\delta_2,\delta_3) := \left\{ w \in \mathcal{W}_\xi^\# : \| \partial_r w_{0,0} - 1 \|_\infty \leq \delta_1 , \
|w_{0,0}(0) | \leq \delta_2
, \ \| w_{\geq 1} \|_{\xi}^\# \leq \delta_3  \right\} .
\end{align*}
For $w \in \mathcal{W}^\#_{\xi}$, it is easy to see using \eqref{eq:operatornormestimate1} that
$
H(w) := \sum_{m,n} H_{m,n}(w)
$
converges in operator norm with bounds
\begin{align} \label{eq:opestimatgeq12}
& \| H(w) \|_{} \leq \| w\|_\xi^\# , \\
 \label{eq:opestimatgeq123}
& \| H(w_{\geq r} ) \|_{} \leq \xi^r \| w_{\geq r} \|_\xi^\# .
\end{align}
We shall use the notation
$$
W[w] := \sum_{m+n \geq 1} H_{m,n}(w) .
$$
We will use the following theorem, which is a straightforward generalization of a theorem proven in \cite{BCFS}, a proof
can  be found in \cite{HasHer11-1}.

\begin{theorem} \label{thm:injective} The map $H : \WW_\xi^\# \to \mathcal{B}(\HH_{\rm red})$ is injective and bounded.
\end{theorem}

\begin{definition}
Let $\WW_\xi$ denote the Banach space consisting of strongly analytic functions on $D_{1/2}$ with values
in $\WW_\xi^\#$ and norm given by
$$
\| w  \|_\xi := \sup_{z \in D_{1/2}} \| w(z) \|_\xi^\# .
$$
\end{definition}
For $w \in \WW_\xi$ we will use the notation $w_{m,n}(z, \cdot) := (w_{m,n}(z))(\cdot)$.
We extend the definition of $H(\cdot)$ to $\WW_\xi$ in the natural way: for $w \in \WW_\xi$, we set
$$
\left( H(w) \right) (z) := H(w(z))
$$
and likewise for $H_{m,n}(\cdot)$ and $W[\cdot]$.

\begin{definition} \label{defkernel} 
We say that a kernel $w \in \WW_\xi$ is {\bf  symmetric } if $w_{m,n}(\overline{z}) = \overline{w_{n,m}(z)}$  for all $z \in D_{1/2}$. 
\end{definition}

Note that because of Theorem
\ref{thm:injective} we have the following lemma.
\begin{lemma} \label{lem:symmetry} Let $ w \in \WW_\xi$. Then
 $w$ is symmetric if and only if $H(w(\overline{z}))= H(w(z))^*$ for all $z \in D_{1/2}$.
\end{lemma}
The renormalization transformation will be defined on the following balls in $\mathcal{W}_\xi$
\begin{eqnarray*}
\lefteqn{ \mathcal{B}(\delta_1,\delta_2,\delta_3) } \\
&& := \left\{ w \in \mathcal{W}_\xi : \sup_{z \in D_{1/2}} \| \partial_r w_{0,0}(z) - 1 \|_\infty \leq \delta_1 , \,
\sup_{z \in D_{1/2}} | w_{0,0}(z,0) + z | \leq \delta_2
, \, \| w_{\geq 1} \|_{\xi} \leq \delta_3  \right\} .
\end{eqnarray*}
We define on the space of kernels $\WW_{m,n}^\#$ a
natural representation of $SO(3)$, $\mathcal{U}_\WW$, which by Theorem \ref{thm:injective}
is uniquely determined by
\begin{equation} \label{eq:rinvkerop}
H (\mathcal{U}_\WW(R) w_{m,n}) = \mathcal{U}_\FF(R) H(w_{m,n}) \mathcal{U}_\FF^*(R) , \quad \forall R \in SO(3) .
\end{equation}
Using the definition \eqref{eq:repofh} of $\UU_\hh$ one can show that
 $\mathcal{U}_\WW(R) w_{0,0}(r) = w_{0,0}(r)$ and
\begin{eqnarray}
\lefteqn{
\left( \mathcal{U}_\WW(R)w_{m,n} \right)(r , k_1,\lambda_1,\ldots,\widetilde{k}_n,\widetilde{\lambda}_n) } \label{eq:defofronkernels} \\
&&= \sum_{(\lambda_1',..., \widetilde{\lambda}_n') \in \Z_2^{m+n}}
     D_{\lambda_1 , \lambda_1'}(R,k_1) \cdots D_{\widetilde{\lambda}_n ,  \widetilde{\lambda}_n'}(R,\widetilde{k}_n)
     w_{m,n}(r , R^{-1} k_1,\lambda_1',\ldots ,R^{-1} \widetilde{k}_n, \widetilde{\lambda}_n')  \nonumber
\end{eqnarray}
 for $m+n \geq 1$..
 The representation on $\WW_{m,n}^\#$ yields a natural representation
on $\WW_\xi^\#$, which is given  by $(\mathcal{U}_\WW(R) w )_{m,n} = \mathcal{U}_\WW(R) w_{m,n}$ for all $R \in SO(3)$.
We say that a kernel $w_{m,n} \in \WW_{m,n}^\# $ is rotation invariant if $\mathcal{U}_\WW(R) w_{m,n} = w_{m,n}$ for all $R \in SO(3)$
and we say
a kernel $w \in \WW_{\xi}^\# $ is rotation invariant if each component is rotation invariant.
For a proof of the next Lemma we refer the reader to \cite{HasHer11-2}.

\begin{lemma} \label{lem:wHinvequiv} \quad
\begin{itemize}
\item[ (i)]
Let $w_{m,n} \in \WW_{m,n}^\#$. Then $H(w_{m,n})$ is rotation invariant if and only if $w_{m,n}$ is rotation invariant.
\item[(ii)]   Let $w \in \WW_{\xi}^\#$. Then $H(w)$ is rotation invariant if and only if $w$ is rotation invariant.
\item[(iii)] If
$w_{m,n} \in \WW_{m,n}^\#$ with $m+n=1$  is rotation invariant, then $w_{m,n} = 0$.
\end{itemize}
\end{lemma}

The contraction property of the renormalization transformation will be obtained by restricting to  balls of integral kernels
which are invariant under rotations
$$
\mathcal{B}_0(\delta_1,\delta_2,\delta_3) := \{ w \in \mathcal{B}(\delta_1,\delta_2,\delta_3)  :  \
w_{m,n}(z) \ {\rm is \ rotation \ invariant \ for \ all \ }  z \in D_{1/2} \ \} .
$$

\section{Initial Feshbach Transformations}

\label{sec:ini}

Throughout  this section we shall assume that   Hypotheses  \ref{potential}  and \ref{kappa}  hold. Without loss of generality, see Section \ref{sec:outline}, we
assume that the distance between the lowest eigenvalue of $H_{\rm at}$ and the rest of the spectrum is one, that is
\begin{equation} \label{eq:hatscale1}
 \inf \left( \sigma( H_{\rm at}) \setminus \{ E_{\rm at} \} \right) - E_{\rm at} = 1 .
\end{equation}
Let $\chi_1$ and $\chib_1$ be two functions in $C^\infty(\R_+;[0,1])$ with
\begin{align} \label{propofchi}
\begin{array}{l}
 \chi_1^2 + \chib_1^2 = 1 , \\
 \chi_1 = 1  \text{ on  }  [0,3/4), \text{ and  } \supp \chi_1 \subset [0,1].
\end{array}
\end{align}

For an explicit choice of $\chi_1$ and $\chib_1$ see for example
\cite{BCFS}.
We use the abbreviation $\chi_1 = \chi_1(H_{\rm f})$ and $\chib_1 = \chib_1(H_{\rm f})$.
It should be clear from the context whether $\chi_1$ or $\chib_1$ denotes a function or an operator.

We recall that since  $E_{\rm at}$ is an isolated eigenvalue by  Hypothesis \ref{potential}
it follows from Lemma \ref{anaAalphathetaat}  and  the Kato-Rellich theorem cf. \cite[Theorem XII.8]{ReeSim4}
that there exist  $\theta_{{\rm at},0} \in (0, \theta_{\rm at} ]$ (with $\theta_{\rm at}$   as in Hypothesis   \ref{potential})   and $\alpha_{{\rm at},0} >0$, such that for all $(\theta,\alpha) \in D_{\theta_{{\rm at},0}} \times D_{\alpha_{{\rm at},0}}$ the number 
$E_{\rm at}(\theta,\alpha)$  is the only point of $\sigma(H_{\rm at}(\theta,\alpha))$ near $E_{\rm at}$ and this point is isolated and a simple eigenvalue.
$E_{\rm at}(\theta,\alpha)$ is an analytic function of $(\theta,\alpha)$
on  the polydisc $D_{\theta_{{\rm at},0}} \times D_{\alpha_{{\rm at},0}}$, 
and
there is an analytic eigenvector $\varphi_{\rm at}(\theta,\alpha)$
for  $(\theta,\alpha) \in  D_{\theta_{{\rm at},0}} \times D_{\alpha_{{\rm at},0}}$.
In fact, one can show  that $E_{\rm at}(\theta,\alpha)$ is independent of $\theta$ and $\alpha$, cf.
 Remark \ref{gsatconst}.  Nevertheless, for consistency of notation  we keep these variables in the notation of $E_{\rm at}(\theta,\alpha)$.
If $\theta $ and $\alpha$ are real, then  $\varphi_{\rm at}(\theta,\alpha)$ can be chosen to be 
 normalized. For every $(\theta,\alpha) \in   D_{\theta_{{\rm at},0}} \times D_{\alpha_{{\rm at},0}}$ the Riesz-projection for $\epsilon > 0$ sufficiently small
$$
P_{\rm at}(\theta , \alpha) = \frac{1}{2\pi i }  \ointctrclockwise\limits_{|z-E_{\rm at}(\theta,\alpha)|=\epsilon} \frac{1}{z-H_{\rm at}(\theta,\alpha) } dz
$$
projects onto the eigenvector $\varphi_{\rm at}(\theta,\alpha)$. 
From the definition we see that the Riesz projection depends analytically on $(\theta,\alpha)$.
Below, we shall assume that $(\theta,\alpha) \in D_{\theta_{{\rm at},0}} \times D_{\alpha_{{\rm at},0}}$.

We define \begin{align} \label{relchiis}  \chi^{(I)}_{\theta,\alpha}(r) := P_{\rm at}(\theta,\alpha) \otimes \chi_1(r) ,  \quad
\chib^{(I)}_{\theta,\alpha}(r) := \overline{P}_{\rm at}(\theta,\alpha) \otimes 1 + P_{\rm at}(\theta,\alpha) \otimes \chib_1(r) ,
\end{align}
with $\overline{P}_{\rm at}(\theta,\alpha) := 1 - P_{\rm at}(\theta,\alpha)$. We set
$\chi^{(I)}_{\theta,\alpha}:= \chi^{(I)}_{\theta,\alpha}(H_{\rm f})$ and $\chib^{(I)}_{\theta,\alpha} := \chib^{(I)}_{\theta,\alpha}(H_{\rm f})$. It is
evident to see that $[{\chi^{(I)}_{\theta,\alpha}}]^2 + [{\chib^{(I)} _{\theta,\alpha}}]^2 = 1$.

For any linear operator $L$ in $\HH_{\rm at} \otimes \FF$ that is defined and bounded on $\ran \, P_{\rm at}(\theta,\alpha) \otimes \FF$,
there exists  a   unique bounded linear
transformation $\langle L \rangle_{{\rm at},\theta,\alpha}$ on $\FF$,  cf.  Lemma \ref{atprojexprem},
such that
\begin{align}
 \label{eq:eff2}
( P_{\rm at}(\theta,\alpha) \otimes  1 ) L ( P_{\rm at}(\theta,\alpha) \otimes 1  ) =
P_{\rm at}(\theta,\alpha) \otimes \langle L \rangle_{{\rm at},\theta,\alpha}.
\end{align}

\begin{lemma}\label{atprojexprem} Suppose $\ran P_{\rm at}(\theta,\alpha) $ is one dimensional,   let
$V_{\theta,\alpha} :  \ran \, P_{\rm at}(\theta,\alpha)  \to \C $, $
w \, \varphi_{\rm at}(\theta,\alpha) \mapsto w $,
and let $\phi$  be   the canonical isomorphism $ \C \otimes \FF \to \FF $, $z \otimes \xi \mapsto z \xi$.
Then for any linear operator $L$ in  $\HH_{\rm at} \otimes \FF$ that is defined and bounded on $\ran \, P_{\rm at}(\theta,\alpha) \otimes \FF$ the operator 
\begin{align}
 \label{eq:eff23}
 \langle L \rangle_{{\rm at},\theta,\alpha} =  \phi  \, ( V_{\theta,\alpha} \otimes 1 )
( P_{\rm at}(\theta,\alpha) \otimes  1 ) L ( P_{\rm at}(\theta,\alpha) \otimes 1  )   ( V_{\theta,\alpha}^{-1} \otimes 1 )  \, \phi^{-1}  
\end{align}
is the unique bounded linear transformation on $\FF$ satisfying  \eqref{eq:eff2}.
\end{lemma}
\begin{proof} First observe that  uniqueness follows by applying \eqref{eq:eff2} to elements in its  domain.
To see  \eqref{eq:eff23} we
 apply the vector $\varphi_{\rm at}(\theta,\alpha) \otimes f$  for  any $f \in \FF$ such that the vector
is in the  domain of $L$ to  both sides  of \eqref{eq:eff2}.
We find
\begin{align}
& [ P_{\rm at}(\theta,\alpha)  \otimes  
  \langle L \rangle_{{\rm at},\theta,\alpha}    ] ( \varphi_{\rm at}(\theta,\alpha) \otimes f )  \nonumber \\
 & =  \varphi_{\rm at}(\theta,\alpha)  \otimes  \left[  \langle L \rangle_{{\rm at},\theta,\alpha} f  \right]  \nonumber \\
& =  \varphi_{\rm at}(\theta,\alpha)  \otimes  \left[   \phi  \, ( V_{\theta,\alpha} \otimes 1 )
( P_{\rm at}(\theta,\alpha) \otimes  1 )    L   ( \varphi_{\rm at}(\theta,\alpha)  \otimes f  ) \right]  \nonumber
 \\
& =
( P_{\rm at}(\theta,\alpha) \otimes  1 )    L   ( \varphi_{\rm at}(\theta,\alpha)  \otimes f  ) ,  \label{eq:atexp}
\end{align}
where we used 
 $ \phi ( V_{\theta,\alpha} \otimes 1 ) (\varphi_{\rm at}(\theta,\alpha) \otimes \xi) 
=    \phi (  1  \otimes \xi)    =  \xi$ 
for all $\xi \in \FF$.
On the other hand  applying $\varphi_{\rm at}(\theta,\alpha) \otimes f$
to the left hand side \eqref{eq:eff2} it is straight forward to see that we obtain  \eqref{eq:atexp} as well.
Thus by uniqueness   \eqref{eq:eff23} now follows.

\end{proof}

For the renormalization analysis  to be applicable in its standard form we  will work with   the operators
\begin{align} \label{eq:defoftrafoham}
\widehat{H}_{\rm at}(\theta,\alpha)  :=  e^\theta H_{\rm at}(\theta,\alpha)  , \quad \widehat{H}_g(\theta,\alpha) :=  e^\theta \left( H_g(\theta,\alpha) -   c_{N,g,\kappa}  \right)   .
\end{align}
where we introduced an auxiliary  energy shift   (finite by \eqref{boundonint})
\begin{align}
c_{N,g,\kappa} :=  N g^2 \int_{\R^3} dk \frac{|\kappa(k)|^2}{|k|} ,
\end{align}
which will simplify the notation, since   it   allows us to work with normal ordered expressions.
To this end,   assuming   Hypothesis \ref{kappa}, we observe  using    analytic continuation, that  with
\begin{align*} 
& A^+_{\theta,l}(y) :=   \sum_{\lambda=1,2} \int_{\R^3} f_{(\theta,l,y)}(k,\lambda)  a^*(k,\lambda)  , \quad   A^-_{\theta,l}(y) :=   \sum_{\lambda=1,2} \int_{\R^3} \overline{ f_{(\overline{\theta},l,y)}}(k,\lambda)  a(k,\lambda)
\end{align*}
we    have,    $A_{\theta}(x_j) =  A^+_{\theta}(x_j)+ A^-_{\theta}(x_j)$, cf. Remark \ref{kdappeana} and \eqref{defofintkernela}.
Now   using the canonical commutation relations
we find that for all $\theta \in D_{\theta_{\rm I}} $ 
\begin{align} \label{normalorder}
c_{N,g,\kappa} =  g^2 \sum_{j=1}^N  ( A_\theta^-(x_j) \cdot A_\theta^+(x_j)  -  A_\theta^+(x_j) \cdot A_\theta^-(x_j)  )  .
\end{align}
Clearly,  the expression on the left hand  side of  \eqref{eq:defoftrafoham}  is an  analytic of type (A) if and only if 
 the expressions on the right is.
Furthermore, for any  linear operator $A$ on a vector space $V$, vector $v \in V$, and complex numbers $\lambda $ and $ a $ the
following trivial  relation holds for all $\tau \in \C$
\begin{align} \label{trivialtrafo}
A v = \lambda v  \quad \Leftrightarrow \quad (e^{-\tau} A  +  a  ) v =  (e^{-\tau}  \lambda +  a ) v .
\end{align}
Thus we conclude that
\begin{align}
\widehat{E}_{\rm at}(\theta,\alpha)  := e^\theta  {E}_{\rm at}(\theta,\alpha)
\end{align}
is an eigenvalue of $\widehat{H}_{\rm at}(\theta,\alpha)$ with eigenvector $\varphi_{\rm at}(\theta,\alpha)$ and eigenprojection $P_{\rm at}(\theta,\alpha)$.

\begin{theorem} \label{thm:inimain1} Assume Hypotheses \ref{potential} and \ref{kappa}. For any $0 < \xi < 1$ and any positive numbers $\delta_1,\delta_2,\delta_3$ there exist  positive numbers
$g_{\rm f}$, $\theta_{\rm f}$,   and   $\alpha_{\rm f}$  such that following is satisfied.

\vspace{0.2cm}

\noindent {\rm (I)}
For all
$(g, \theta,\alpha,z) \in  D_{g_{\rm f}} \times D_{\theta_{\rm f}}  \times    D_{\alpha_{\rm f}}  \times D_{1/2}$
the pair of operators
$(\widehat{H}_{g}(\theta,\alpha) - z - \widehat{E}_{\rm at}(\theta,\alpha), \widehat{H}_0(\theta,\alpha) - z - \widehat{E}_{\rm at}(\theta,\alpha) )$ is a Feshbach pair for $\chi^{(I)}_{\theta,\alpha}$.
The operator valued function
\begin{equation} \label{eq:qdef111}
 Q_{\chi^{(I)}}(g,\theta,\alpha,z) := Q_{\chi^{(I)}}( \widehat{H}_{g}(\theta,\alpha) - z - \widehat{E}_{\rm at}(\theta,\alpha), \widehat{H}_0(\theta,\alpha) - z -\widehat{E}_{\rm at}(\theta,\alpha) )
\end{equation}
defined  on $ D_{g_{\rm f}} \times D_{\theta_{\rm f}}  \times    D_{\alpha_{\rm f}}  \times D_{1/2}$ is bounded and analytic.

\vspace{0.2cm}

\noindent  {\rm (II)}  For $(g, \theta,\alpha,z) \in  D_{g_{\rm f}} \times D_{\theta_{\rm f}}  \times    D_{\alpha_{\rm f}}  \times D_{1/2}$
there exists a unique kernel
$\widehat{w}^{(0)}(g,\theta,\alpha,z) \in \WW_\xi^\#$ such that
\begin{align} \label{eq:inimainA}
& H^{(0)}_{g,\theta,\alpha}(z) := H(\widehat{w}^{(0)}(g,\theta,\alpha,z))\\
&  = \langle  F_{\chi^{(I)}}( \widehat{H}_{g}(\theta,\alpha) - z - \widehat{E}_{\rm at}(\theta,\alpha) , \widehat{H}_0(\theta,\alpha) - z - \widehat{E}_{\rm at}(\theta,\alpha ))|_{ \ran ( P_{\rm at}(\theta,\alpha)  \otimes 1_{H_{\rm f} \leq 1}  )}\rangle_{{\rm at},\theta,\alpha} . \nonumber
\end{align}
Moreover, $\widehat{w}^{(0)}$ satisfies the following properties.
\begin{itemize}
\item[(a)] We have
$\widehat{w}^{(0)}(g,\theta,\alpha) := \widehat{w}^{(0)}(g,\theta,\alpha, \cdot ) \in \mathcal{B}_0(\delta_1,\delta_2,\delta_3)$
for all $(g,\theta,\alpha) \in D_{g_{\rm f}}\times  D_{\theta_{\rm f}} \times  D_{\alpha_{\rm f}}  $.
\item[(b)]
$\widehat{w}^{(0)}(g,\theta,\alpha)$ is a symmetric kernel for all $(g,\theta,\alpha) \in ( D_{g_{\rm f}} \cap \R) \times  ( D_{\theta_{\rm f}} \cap  \R) \times (D_{\alpha_{\rm f}} \cap  \R )  $.
\item[(c)]
The function $(g,\theta,\alpha,z) \mapsto \widehat{w}^{(0)}(g,\theta,\alpha,z)$ is a $\WW_\xi^\#$-valued analytic function on
 $D_{g_{\rm f}}\times  D_{\theta_{\rm f}}  \times  D_{\alpha_f}  \times D_{1/2}$.
\end{itemize}
\end{theorem}

 The remaining part of this section is devoted to the proof of  Theorem   \ref{thm:inimain1}.
First we show (I) and then (II).
To prove Theorem \ref{thm:inimain1}, we write the interaction part of the Hamiltonian using  \eqref{normalorder}   in terms
of integral kernels
$$
\widehat{H}_{g}(\theta,\alpha) = 
\widehat{H}_{\rm at}(\theta,\alpha)  + H_{\rm f} +  \widehat{W}_{g}(\theta,\alpha)    ,
$$
\begin{equation} \label{eq:sumofws}
\widehat{W}_{g,\theta,\alpha} := \sum_{m+n=1,2} \widehat{W}_{g,m,n}(\theta,\alpha) ,
\end{equation}
where $\widehat{W}_{g,m,n}(\theta,\alpha) := \underline{H}_{m,n}(\widehat{w}_{g,m,n}^{(I)}(\theta,\alpha))$
with
\begin{align}
 \label{eq:defhlinemn}
& \underline{H}_{m,n}(w_{m,n}) := \int_{{(\underline{\R}^3)}^{m+n}} \frac{ dK^{(m,n)}}{|K^{(m,n)}|^{1/2}}
a^*(K^{(m)}) w_{m,n}(K^{(m,n)}) a(\widetilde{K}^{(n)}) ,
\end{align}
and 
\begin{align}
&\widehat{w}^{(I)}_{1,0}(g,\theta,\alpha)( K) := 2 e^{-\theta}  g \sum_{j=1}^N  \frac{\kappa_{\theta}(k)e^{ -i  \beta k \cdot x_j }}{\sqrt{2}}
\varepsilon(k,\lambda) \cdot ( p_j - \alpha x_j \langle x \rangle^{-1} )
 , \label{defofwI} \\
&\widehat{w}^{(I)}_{0,1}(g,\theta,\alpha)( K) := 2 e^{-\theta}  g \sum_{j=1}^N ( p_j - \alpha x_j \langle x \rangle^{-1} ) \cdot \varepsilon(k,\lambda) \frac{\tilde{\kappa}_{\theta}(k)e^{ i \beta  k \cdot x_j }}{\sqrt{2}} , \nonumber \\
&\widehat{w}^{(I)}_{1,1}(g,\theta,\alpha)(K,\widetilde{K}) := 2 e^{-\theta} g^2  \sum_{j=1}^N \varepsilon(k,\lambda) \cdot \varepsilon(\widetilde{k},\widetilde{\lambda}) \frac{\kappa_{\theta}(k)e^{- i \beta  k \cdot x_j }}{\sqrt{2}} \frac{ \tilde{\kappa}_{\theta}(\widetilde{k})e^{ i  \beta \widetilde{k}\cdot x_j }}{\sqrt{2}} ,
\nonumber \\
&\widehat{w}^{(I)}_{2,0}(g,\theta,\alpha)( K_1, K_2 ) := e^{-\theta} g^2 \sum_{j=1}^N \varepsilon(k_1,\lambda_1) \cdot \varepsilon({k}_2,{\lambda}_2) \frac{\kappa_{\theta}(k_1)e^{ - i \beta  k_1 \cdot x_j }}{\sqrt{2}} \frac{\kappa_{\theta}(k_2)e^{ - i \beta k_2 \cdot x_j }}{\sqrt{2}} ,
\nonumber
\\
&\widehat{w}^{(I)}_{0,2}(g,\theta,\alpha)( K_1, K_2 ) := e^{-\theta} g^2 \sum_{j=1}^N \varepsilon(k_1,\lambda_1) \cdot \varepsilon({k}_2,{\lambda}_2) \frac{\tilde{ \kappa}_{\theta}(k_1)e^{  i  \beta  k_1 \cdot x_j }}{\sqrt{2}} \frac{ \tilde{ \kappa}_{\theta}(k_2)e^{ i \beta  k_2 \cdot x_j }}{\sqrt{2}} ,
\nonumber
\end{align}
with $\tilde{\kappa}_\theta(k) := \overline{\kappa_{\overline{\theta}}}(k)$. Note that in view of Remark  \ref{kdappeana} 
to Hypothesis  \ref{kappa}
we know that $\tilde{\kappa}_\theta$ depends analytically on $\theta$.
We note that \eqref{eq:defhlinemn} is understood in the sense of forms, c.f. Appendix
\ref{sec:estfock}. Furthermore,  we used that the  divergence of $A(x)$ vanishes, i.e.,  $\nabla_x \cdot A(x) = 0$.
We set
\begin{align} \label{defofw00}
\widehat{w}^{(I)}_{0,0}(\theta,\alpha,z)(r) = \widehat{H}_{\rm at}(\theta,\alpha) - z + r .
\end{align}
By $\hat{w}^{(I)}$ we denote the vector consisting of the components $\hat{w}^{(I)}_{m,n}$ with $m+n=0,1,2$.

Next  we show the Feshbach  pair property of Theorem \ref{thm:inimain1}. It   will  follow from Lemma \ref{ddd2}, below, and  Theorem \ref{fesh:thm2}.

\begin{lemma} \label{ddd}
 Suppose  that Hypothesis \ref{kappa} holds. Then for all compact subsets $K$ of $D_{\theta_{\rm I}} \times \C $
there exists a constant $C_K$ such that
\begin{align}
&\sup_{(\theta,\alpha) \in K } \| \widehat{W}_{g}(\theta,\alpha)(-\Delta + H_{\rm f} + 1 )^{-1}  \| \leq C_K |g|   \label{uniformanabound1}  \\
&
 \sup_{(\theta,\alpha) \in K } \| (-\Delta + H_{\rm f} + 1 )^{-1} \widehat{W}_{g}(\theta,\alpha) \| \leq C_K |g|
\label{uniformanabound2}
\end{align}
The maps
\begin{align} \label{analytelem1}
& (g,\theta,\alpha) \mapsto\widehat{W}_{g}(\theta,\alpha)(-\Delta + H_{\rm f} + 1 )^{-1}   \\
& (g,\theta,\alpha) \mapsto (-\Delta + H_{\rm f} + 1 )^{-1} \widehat{W}_{g}(\theta,\alpha)  \label{analytelem2}
\end{align}
are analytic on $\C \times D_{\theta_{\rm I}} \times \C $.
\end{lemma}

\begin{proof}
We first show \eqref{uniformanabound2}. To this end,  we find by the triangle inequality
\begin{align} \label{eq:estonhfw}
\| (-\Delta + H_{\rm f} + 1 )^{-1} \widehat{W}_{g}(\theta,\alpha) \|  \leq \sum_{m+n =  2} \|  (-\Delta + H_{\rm f} + 1 )^{-1} \widehat{W}_{g,m,n}({\theta,\alpha}) \|  .
\end{align}
 For $(m,n)=(0,2)$ we estimate as follows,
\begin{eqnarray} \lefteqn{
 \left\| ( H_{\rm f} + 1 )^{-1} \widehat{W}_{g,0,2}(\theta,\alpha) \right\|} \nonumber \\
 && \leq \frac{|g|^2 |e^{-\theta}|N}{2} \left[ \int_{({\underline{\R}^3})^2 } \frac{ d
\widetilde{K}^{(2)}}{|\widetilde{K}^{(2)}|^2} \left| \tilde{\kappa}_{ {\theta}      }(\widetilde{k}_1) \right|^2
 \left| \tilde{\kappa}_{ {\theta}     }(\widetilde{k}_2) \right|^2 \sup_{r \geq 0} \frac{ ( r + |\widetilde{k}_1| + |\widetilde{k}_2|)^2}{(r + 1)^2} \right]^{1/2}
\nonumber \\
 && \leq \frac{|g|^2  |e^{-\theta}| N }{2} \left[ 3 \| \tilde{\kappa}_{{\theta}}/ \omega \|_\hh^4 + 6 \| \tilde{\kappa}_{{\theta}}/ \omega \|_\hh^2
 \| \tilde{\kappa}_{{\theta}} \|_\hh^2 \right]^{1/2} , \label{eq:someestimate}
\end{eqnarray}
where in the first inequality we used Lemma \ref{kernelopestimate} and
 in the last inequality we used the following estimate for $r \geq 0$,
$$\frac{ ( r + |\widetilde{k}_1| + |\widetilde{k}_2|)^2}{(r + 1)^{2}} \leq 3 ( 1 + |\widetilde{k}_1|^2 + |\widetilde{k}_2|^2) .
$$
To estimate the
right hand side of \eqref{eq:estonhfw} for $(m,n)=(2,0)$ we
use the  fact that the norm of an operator is equal to the norm of its adjoint, the pull-through formula, and a similar estimate
as used in \eqref{eq:someestimate},
$$
 \left\| ( H_{\rm f} + 1 )^{-1} \widehat{W}_{g,2,0}( \overline{\theta},\overline{\alpha}) \right\| =
\left\| \widehat{W}_{g,0,2}(  \theta,\alpha ) ( H_{\rm f} + 1 )^{-1} \right\| \leq {\rm r.h.s.} \ \eqref{eq:someestimate} .
$$
 To estimate the
term for $(m,n)=(1,1)$ we first use the pull-through formula and then
 Lemma \ref{kernelopestimate} to obtain
\begin{eqnarray}
\lefteqn{ \left\| ( H_{\rm f} + 1 )^{-1} \widehat{W}_{g,1,1}(\theta,\alpha) \right\|} \nonumber \\
&& \leq |g|^2 |e^{-\theta}| N \left[ \int_{{(\underline{\R}^3)}^2 } \frac{ d K^{(1,1)}}{|K^{(1,1)}|^2} \left| \kappa_{\theta}({k}_1) \right|^2
 \left| \tilde{\kappa}_{{\theta}}(\widetilde{k}_1) \right|^2 \sup_{r \geq 0} \frac{ ( r + |{k}_1|)( r + |\widetilde{k}_1|)}{(r + 1)^2}\right]^{1/2} \nonumber \\
 && \leq |g|^2|e^{-\theta}|   N  \left[ 2 \| \kappa_{\theta}/ \omega \|_\hh^2 \| \tilde{\kappa}_{{\theta}}/ \omega \|_\hh^2+  \| \tilde{\kappa}_{{\theta}}/ \omega \|_\hh^2 \| \kappa_{\theta} \|_\hh^2  +  \| \kappa_{\theta}/ \omega \|_\hh^2 \| \tilde{\kappa}_{{\theta}} \|_\hh^2\right]^{1/2} , \label{eq:usedforqcont1}
\end{eqnarray}
where in the last inequality we used the following estimate for $r \geq 0$,
$$
\frac{ ( r + |{k}_1|)( r + |\widetilde{k}_1|)}{(r + 1)^2} \leq 2 + |k_1|^2 + |\widetilde{k}_1|^2 .
$$
To estimate the summands with $m+n=1$ on the right hand side of   \eqref{eq:estonhfw}
make use of $\|  (H_{\rm f}  - \Delta + 1 )^{-1} (H_{\rm f} + 1)^{1/2} (-\Delta + 1)^{ 1/2}\| \leq 1 $ (use spectral theorem).
\begin{align}
& \lefteqn{ \left\| (-\Delta + 1 )^{-1/2} (H_{\rm f} + 1 )^{-1/2} \widehat{W}_{g,m,n}({\theta,\alpha}) \right\| } \nonumber \\
 &\leq 2 |g| |e^{-\theta}| \sum_{j=1}^N \sum_{l=1}^3 \left(
\left\| \frac{ (p_j)_l }{ (-\Delta + 1 )^{1/2}} \right\|  + | \alpha|  \right) \nonumber  \\
& \times \left\| (H_{\rm f} + 1 )^{-1/2} \left[ \delta_{m,0} \underline{H}_{1,0}( \omega^{1/2}f_{(\theta,l, x_j)}) +
  \delta_{n,0} \underline{H}_{0,1}( \omega^{1/2}\overline{f_{(\overline{\theta}, l, x_j)}}) \right] \right\| \nonumber \\
& \leq 6 N |g| |e^{-\theta}| (1 + |\alpha| ) \left( \delta_{m,0}  \| \kappa_{\theta} /\omega \|_\hh^2 + \delta_{n,0} ( \| \tilde{\kappa}_{{\theta}} / \sqrt{\omega}\|_\hh^2  +  \| \tilde{\kappa}_{{\theta}} /\omega \|_\hh^2 )  \right)^{1/2} , 
 \label{eq:usedforqcont2}
\end{align}
where in the first inequality we used the triangle inequality and  the notation introduced in  \eqref{defofintkernela},
and in the second inequality we used the pull-through formula,  Lemma \ref{kernelopestimate}. 
This shows \eqref{uniformanabound2}. Now  \eqref{uniformanabound1}  from \eqref{uniformanabound2}
by taking the adjoint (or alternatively using an analogous estimate).

 To show the  statements about the analyticity we use  the criterion of Lemma \ref{critanaop}.
Explicitly we apply  the  criterion   to $X = D(-\Delta +H_{\rm f})$ equipped with the graph norm and $Y = \HH$.
The bounds in  \eqref{uniformanabound1}  and \eqref{uniformanabound2} and  the analyticity on a fundamental set of
vectors, by   Lemma  \ref{anaint}, show that $(g,\theta,\alpha) \mapsto \widehat{W}_g(\theta,\alpha)$ is a bounded analytic function from $X$
to $Y$, in view of Lemma \ref{critanaop}. This implies the analyticity of \eqref{analytelem1} and \eqref{analytelem2}.
\end{proof}

The next lemma will be used to control the reduced resolvent.
For notational compactness we set
\begin{align} \label{shortnotP}  \overline{P}(\theta,\alpha) = \overline{P}_{\rm at}(\theta,\alpha) \otimes 1 .
\end{align}

\begin{lemma}\label{lemredresbound0} \label{lemredresbound}  Suppose that Hypothesis \ref{potential} holds.
  Then there  exist positive $\theta_{{\rm at},1} \in (0, \theta_{{\rm at},0}]$ and $\alpha_{{\rm at},1} \in (0, \alpha_{{\rm at},0}]$ such that for $\mathcal{U} = D_{\theta_{{\rm at},1}} \times D_{\alpha_{{\rm at},1}} \times  D_{1/2}$
the following holds.
\begin{enumerate}[(a)]
\item  For all
 $(\theta,\alpha , z  ) \in \mathcal{U}$ and $ r \geq 0$  the number
 $\widehat{E}_{\rm at}(\theta,\alpha)  + z - r  $  is in the resolvent set of
${H}_{\rm at}(\theta,\alpha) \overline{P}_{\rm at}(\theta,\alpha)$  and
$$
 \sup_{ (\theta,\alpha , z  ) \in \mathcal{U} } \sup_{r\geq 0 } \left\|  ( \widehat{H}_{\rm at}(\theta,\alpha) + r  - \widehat{E}_{\rm at}(\theta,\alpha) - z )^{-1} \overline{P}_{\rm at}(\theta,\alpha) \right\|   < \infty  .
$$
\item For all
 $(\theta,\alpha , z  ) \in \mathcal{U} $   the number
 $\widehat{E}_{\rm at}(\theta,\alpha)  + z $  is in the resolvent set of
$\widehat{H}_{0}(\theta,\alpha) \overline{P}(\theta,\alpha)$  and
$$
 ( \widehat{H}_0(\theta,\alpha) - \widehat{E}_{\rm at}(\theta,\alpha) - z )^{-1} \overline{P}(\theta,\alpha)
$$
is an analytic function of $(\theta,\alpha , z  ) \in
  \mathcal{U}$ and uniformly bounded on that set.
\end{enumerate}
\end{lemma}
\begin{proof}
(a) This follows from  Theorem \ref{thm:veryIII}, whose assumptions are satisfied by Lemma  \ref{anaAalphathetaat}.
This yields the statement. \\
 (b) That $\widehat{E}_{\rm at}(\theta,\alpha)  + z $  is in the resolvent set of
$\widehat{H}_{0}(\theta,\alpha) \overline{P}(\theta,\alpha)$ and the boundedness of resolvent   is derived from a spectral representation of $H_{\rm f}$ and (a). The  analyticity   follows from  Proposition \ref{Prop27inGriesemerHasler} in the appendix.
\end{proof}

\begin{lemma}
\label{ddd2First}
Suppose that Hypothesis \ref{potential} holds.
  Then there  exist positive $\theta_{{\rm at},2}, \alpha_{{\rm at},2}$ and a constant $C$
such that for $\mathcal{U} = D_{\theta_{{\rm at},2}} \times D_{\alpha_{{\rm at},2}} \times  D_{1/2}$  and $r \geq 0$
the following holds.
\begin{align}
& \sup_{(\theta,\alpha,z) \in \mathcal{U}  } \| (-\Delta + H_{\rm f} + 1)  ( \widehat{H}_0(\theta,\alpha) + r  -  \widehat{E}_{\rm at}(\theta,\alpha)  - z )^{-1} \chib^{(I)}_{\theta,\alpha}(H_{\rm f} + r)   \| < \infty   \label{firstfreeintbound}  \\
& \sup_{(\theta,\alpha,z) \in \mathcal{U}  }\|   ( \widehat{H}_0(\theta,\alpha) + r -  \widehat{E}_{\rm at}(\theta,\alpha)  - z )^{-1} \chib^{(I)}_{\theta,\alpha}(H_{\rm f} + r)   (-\Delta + H_{\rm f} + 1) \|  < \infty  .\label{secondfreeintbound}
\end{align}
The maps
\begin{align}
& (\theta,\alpha,z) \mapsto (-\Delta + H_{\rm f} + 1)  ( \widehat{H}_0(\theta,\alpha) + r -  \widehat{E}_{\rm at}(\theta,\alpha)  - z )^{-1} \chib^{(I)}_{\theta,\alpha}(H_{\rm f} + r )   \label{firstfreeintboundana}  \\
&  (\theta,\alpha,z) \mapsto   ( \widehat{H}_0(\theta,\alpha) + r -  \widehat{E}_{\rm at}(\theta,\alpha)  - z )^{-1} \chib^{(I)}_{\theta,\alpha}(H_{\rm f} + r ) (-\Delta + H_{\rm f} + 1)  \label{secondfreeintboundana}
\end{align}
are analytic on $\mathcal{U}$.
\end{lemma}

\begin{proof} For \eqref{firstfreeintbound} and  \eqref{secondfreeintbound} it suffices to consider the case $r=0$. The case  $ r > 0$ then follows  from the spectral theorem  and by 
replacing   $H_{\rm f}$ by $H_{\rm f}+r$ .
Let  $\theta_{{\rm at},1}$ and $\alpha_{{\rm at},1}$   be as in Lemma  \ref{lemredresbound}.
Let $ \theta_{{\rm at},2} \in (0,  \theta_{{\rm at},1})$ and  $\alpha_{{\rm at},2} \in (0, \alpha_{{\rm at},1})$
and $\mathcal{U} = D_{\theta_{{\rm at},2}} \times D_{\alpha_{{\rm at},2}} \times  D_{1/2}$.
 We estimate first \eqref{firstfreeintbound}.
Pick $z_0 \leq   \inf \sigma (\widehat{H}_0(0,0))-1$. Then  $z_0 \in \rho(\widehat{H}_0(0,0))$
and it follows by analyticity   of  $(\theta,\alpha) \mapsto \widehat{H}_0(\theta,\alpha)$, cf.   Lemma  \ref{anaAalphathetaField},
that $z_0 \in \rho(\widehat{H}_0(\theta,\alpha) )$ for $(\theta, \alpha) \in D_{\theta_{{\rm at},2}} \times D_{\alpha_{{\rm at},2}}$
if we choose $\theta_{{\rm at},2} $ and $ \alpha_{{\rm at},2} $ sufficiently small but positive.
 Thus for $(\theta, \alpha,z) \in \mathcal{U}$
we  can write by Lemma \ref{lemredresbound}
\begin{align}
&
(-\Delta + H_{\rm f} + 1 )   ( \widehat{H}_0(\theta,\alpha) -  \widehat{E}_{\rm at}(\theta,\alpha)  - z )^{-1}\chib^{(I)}_{\theta,\alpha} \nonumber  \\
& =  (-\Delta +  {H}_f + 1 )  ( \widehat{H}_0(0,0)  - z_0  )^{-1} \label{eq:relativeenergybound2} \\
& \times  ( \widehat{H}_0(0,0)  - z_0  )   (  \widehat{H}_0(\theta,\alpha) -  \widehat{E}_{\rm at}(\theta,\alpha)  - z_0  )^{-1} \label{eq:relativeenergybound3}\\
& \times  ( \widehat{H}_0(\theta,\alpha) -  \widehat{E}_{\rm at}(\theta,\alpha)  - z_0   )   (  \widehat{H}_0(\theta,\alpha) -  \widehat{E}_{\rm at}(\theta,\alpha)  - z  )^{-1} \chib^{(I)}_{\theta,\alpha}  \label{eq:relativeenergybound4} .
\end{align}
To estimate \eqref{eq:relativeenergybound2} we use the bound \eqref{eq:boundonc} of Hypothesis \ref{potential}.
To estimate \eqref{eq:relativeenergybound3}
we use that $\widehat{H}_0(\theta,\alpha)$, by Lemma   \ref{anaAalphathetaField}, is an analytic family of type (A) and Lemma \ref{lemtypaAest2}.
To show that  \eqref{eq:relativeenergybound4} is bounded we use
\begin{align}
 & ( \widehat{H}_0(\theta,\alpha)  -  \widehat{E}_{\rm at}(\theta,\alpha)   - z_0   )   (  \widehat{H}_0(\theta,\alpha) -  \widehat{E}_{\rm at}(\theta,\alpha)  - z  )^{-1} \chib^{(I)}_{\theta,\alpha}  \nonumber \\
&  =   \left( 1 -   (z - z_0 )    (  \widehat{H}_0(\theta,\alpha) -  \widehat{E}_{\rm at}(\theta,\alpha)  - z  )^{-1} \right) \chib^{(I)}_{\theta,\alpha} .
\label{anaprodfac1}
\end{align}
Now we write the resolvent occurring on the right hand side using \eqref{relchiis}  (recalling the notation \eqref{shortnotP}) as
\begin{align}
 &( \widehat{H}_0(\theta,\alpha) -  \widehat{E}_{\rm at}(\theta,\alpha)  - z  )^{-1}  \chib^{(I)}_{\theta,\alpha}  \label{anaprodfac1v2}\\
& =   ( \widehat{H}_0(\theta,\alpha) -  \widehat{E}_{\rm at}(\theta,\alpha)  - z  )^{-1}  \overline{P}(\theta,\alpha) +   ( \widehat{H}_0(\theta,\alpha) -  \widehat{E}_{\rm at}(\theta,\alpha)  - z  )^{-1}  P_{\rm at}(\theta,\alpha) \otimes \chib_1(H_{\rm f}) . \nonumber
\end{align}
Now the first term on the right hand side is estimated using Lemma
\ref{lemredresbound} (b). The second term is estimates as follows  observing  that 
\begin{align}\label{anaprodfac1v3}
 ( \widehat{H}_0(\theta,\alpha) -  \widehat{E}_{\rm at}(\theta,\alpha)  - z  )^{-1}  P_{\rm at}(\theta,\alpha) \otimes \chib_1(H_{\rm f}) &=  (  H_{\rm f}   - z  )^{-1}  P_{\rm at}(\theta,\alpha) \otimes \chib_1(H_{\rm f}) . 
\end{align}
Now  $\|  (  H_{\rm f}   - z  )^{-1}  P_{\rm at}(\theta,\alpha) \otimes \chib_1(H_{\rm f}) \| \leq  
\|  (  H_{\rm f}   - z  )^{-1}  \chib_1(H_{\rm f}) \| \| P_{\rm at}(\theta,\alpha) \| $  and  by the spectral theorem
\begin{align*}
&  \left\|  (   H_{\rm f}   - z  )^{-1} \chib_1(H_{\rm f}) \right\| \leq  \sup_{r \geq 0} \frac{| \chib_1(r) |}{| H_{\rm f}   - z |}  \leq \frac{1}{\frac{3}{4} - | z |}  \leq 4 .
\end{align*}
Collecting estimates we obtain \eqref{firstfreeintbound}.
Now \eqref{secondfreeintbound} follows by taking adjoints.
\\
Now the analyticity of    \eqref{firstfreeintboundana}  can be seen by noting
that the expressions are a product of bounded analytic functions. Explicitly,
 \eqref{eq:relativeenergybound3} depends analytically on $(\theta, \alpha)$  on $D_{\theta_{{\rm at},2}} \times D_{\alpha_{{\rm at},2}} $, since it is the resolvent
of an analytic family. 
To show the analyticity of   \eqref{eq:relativeenergybound4}  we use the identity \eqref{anaprodfac1}, respectively
\eqref{anaprodfac1v2}. Clearly, $  \chib^{(I)}_{\theta,\alpha} $
is analytic.
The first term on the right hand side of \eqref{anaprodfac1v2} is analytic by Lemma  \ref{lemredresbound}.
 The second  term on the right hand side of \eqref{anaprodfac1v2} is analytic in view of \eqref{anaprodfac1v3}.
This shows analyticity  of  \eqref{firstfreeintboundana}. Analyticity of   \eqref{secondfreeintboundana}  follows similarly
(or by taking adjoints and complex conjugates).
\end{proof}

\begin{lemma}
\label{ddd2} Suppose that Hypotheses \ref{potential} and
 Hypothesis \ref{kappa} hold for $\theta_{\rm I}$. Then there exist positive $\theta_{{\rm at},3} \in (0,\theta_{\rm I})$ and $\alpha_{{\rm at},3}$ and a constant $C$
such that for $\mathcal{U} = D_{\theta_{{\rm at},3}} \times D_{\alpha_{{\rm at},3}} \times  D_{1/2}$
the following holds.
\begin{align}
& \sup_{(\theta,\alpha,z) \in \mathcal{U}} \| \widehat{W}_{g}(\theta,\alpha)  ( \widehat{H}_0(\theta,\alpha) -  \widehat{E}_{\rm at}(\theta,\alpha)  - z )^{-1} \chib^{(I)}_{\theta,\alpha}   \| \leq C |g| \\ 
& \sup_{(\theta,\alpha,z) \in \mathcal{U} } \|   ( \widehat{H}_0(\theta,\alpha) -  \widehat{E}_{\rm at}(\theta,\alpha)  - z )^{-1} \chib^{(I)}_{\theta,\alpha}  \widehat{W}_{g}(\theta,\alpha) \| \leq C |g| 
\end{align}
The  following maps  are analytic on  $\C \times \mathcal{U}$
\begin{align}
&(g, \theta,\alpha, z) \mapsto \widehat{W}_{g}(\theta,\alpha)  ( \widehat{H}_0(\theta,\alpha) -  \widehat{E}_{\rm at}(\theta,\alpha)  - z )^{-1} \chib^{(I)}_{\theta,\alpha}   \label{firstfreeintboundana1}  \\
& (g, \theta,\alpha, z) \mapsto  ( \widehat{H}_0(\theta,\alpha) -  \widehat{E}_{\rm at}(\theta,\alpha)  - z )^{-1} \chib^{(I)}_{\theta,\alpha}  \widehat{W}_{g}(\theta,\alpha) \label{secondfreeintboundana2} .
\end{align}
\end{lemma}

\begin{proof}   Pick $\theta_{{\rm at},3}   \in (0, \theta_{\rm I})$ with    $\theta_{{\rm at},3} \leq  \theta_{{\rm at},2}  $   and     $  \alpha_{{\rm at},3} = \alpha_{{\rm at},2}         $,
where  $\theta_{{\rm at},2} $ and $\alpha_{{\rm at},2} $   are  as in Lemma  \ref{ddd2First}.
 The lemma now follows  as  an immediate consequence of  Lemma  \ref{ddd},  Lemma  \ref{ddd2First}, and the
following two identities
\begin{align*}
&
\widehat{W}_{g}(\theta,\alpha)   ( \widehat{H}_0(\theta,\alpha) -  \widehat{E}_{\rm at}(\theta,\alpha)  - z )^{-1}\chib^{(I)}_{\theta,\alpha} \nonumber  \\
& =\widehat{W}_{g}(\theta,\alpha) (- \Delta + {H}_f + 1 )^{-1} (- \Delta + {H}_f + 1 )
 ( \widehat{H}_0(\theta,\alpha) -  \widehat{E}_{\rm at}(\theta,\alpha)  - z )^{-1}\chib^{(I)}_{\theta,\alpha}
\end{align*}
and
\begin{align*}
&
\chib^{(I)}_{\theta,\alpha}  ( \widehat{H}_0(\theta,\alpha) -  \widehat{E}_{\rm at}(\theta,\alpha)  - z )^{-1} \widehat{W}_{g}(\theta,\alpha)   \nonumber  \\
& =  \chib^{(I)}_{\theta,\alpha}  ( \widehat{H}_0(\theta,\alpha) -  \widehat{E}_{\rm at}(\theta,\alpha)  - z )^{-1} (- \Delta + {H}_f + 1 ) (- \Delta + {H}_f + 1 )^{-1} \widehat{W}_{g}(\theta,\alpha) .
\end{align*}
\end{proof}

\begin{proof}[Proof of Theorem  \ref{thm:inimain1} Part {\rm  (I)}]
To show the Feshbach property we verify the assumption of
Lemma  \ref{fesh:thm2}.  Now Assumptions (a') and (b') of that Lemma  follow directly from the definition. Furthermore, we
see from   Lemma \ref{ddd2}  that Assumption (c')  of Lemma  \ref{fesh:thm2} holds for $|g|$ sufficiently small.
The statement about analyticity of   \eqref{eq:qdef111}  will follow from   Lemma \ref{ddd2}  and   the definition  given in \eqref{eq:defofQ}   of the auxiliary operator. Thus
inserting into the definition  \eqref{eq:defofQ}  and using the abbreviation $$T(\theta,\alpha;z) =  \widehat{H}_0(\theta,\alpha)-  z - \widehat{E}_{\rm at}(\theta,\alpha) $$
we find using a Neumann expansion, which is justified by   Lemma \ref{ddd2}, provided  $|g|$ is sufficiently small, that
\begin{align*}
  & Q_{\chi^{(I)}}(g,\theta,\alpha,z)  \\
 &  = Q_{\chi^{(I)}}( \widehat{H}_{g}(\theta,\alpha) - z - \widehat{E}_{\rm at}(\theta,\alpha), T(\theta,\alpha;z)  ) \\
      & = \chi^{(I)}_{\theta,\alpha}  - \chib^{(I)}_{\theta,\alpha} \left( T(\theta,\alpha;z)   +   \chib^{(I)}_{\theta,\alpha}
       \widehat{W}_g(\theta,\alpha)  \chib^{(I)}_{\theta,\alpha}  \right)^{-1}  \chib^{(I)}_{\theta,\alpha} \widehat{W}_g(\theta,\alpha)  \chi^{(I)}_{\theta,\alpha} \\
     & =  \chi^{(I)}_{\theta,\alpha}  -   \chib^{(I)}_{\theta,\alpha} \sum_{n=0}^\infty \left( - { T(\theta,\alpha;z)  }^{-1} \chib^{(I)}_{\theta,\alpha}
     \widehat{W}_g(\theta,\alpha)
      \chib^{(I)}_{\theta,\alpha} \right)^n
{ T(\theta,\alpha;z)}^{-1}   \chib^{(I)}_{\theta,\alpha}  \widehat{W}_g(\theta,\alpha) \chi^{(I)}_{\theta,\alpha} .
\end{align*}
Now observe that the right hand side  is  an absolutely  convergent power series
 of  expressions, which are analytic by   Lemma \ref{ddd2}, and so the analyticity statement regarding   \eqref{eq:qdef111} follows.
\end{proof}

The remaining part of this section is devoted to the proof of  Part (II) of Theorem  \ref{thm:inimain1}. Throughout  the remaining part of this  section we assume that Hypotheses \ref{potential} and
 Hypothesis \ref{kappa} hold.  In the following let \begin{align*} \theta_{\rm f} = {\rm min} \{ \theta_{\rm at}, \theta_{{\rm at},0} , \theta_{{\rm at},1},\theta_{{\rm at},2},\theta_{{\rm at},3}\} < \theta_{\rm I} \quad  \text{ and }  \quad \alpha_{\rm f} = {\rm min} \{ \alpha_{{\rm at},0} , \alpha_{{\rm at},1},\alpha_{{\rm at},2},\alpha_{{\rm at},3}\} \end{align*}
with the constants as in   Lemmas  \ref{lemredresbound0}, \ref{ddd2First},   \ref{ddd2},
and  
 Hypotheses   \ref{potential} and \ref{kappa}, respectively.
Henceforth,  let  also $(\theta,\alpha) \in D_{\theta_{\rm f}} \times D_{\alpha_{\rm f}}$.
 Furthermore,  if not stated otherwise we assume that 
\begin{align}\label{eq:zzetarelation}
 \zeta = z +  \widehat{E}_{\rm at}(\theta,\alpha) \quad \text{with } z \in D_{1/2} .
\end{align}

Next we want to show that there exists a $\widehat{w}^{(0)}(g,\theta,\alpha,z) \in \WW_\xi^\#$ such that \eqref{eq:inimainA} holds.
Uniqueness will follow from Theorem \ref{thm:injective}.
In view of Lemmas  \ref{lemredresbound} and \ref{ddd2}  we can define for $z = \zeta - \widehat{E}_{\rm at}(\theta,\alpha) \in D_{1/2}$ and $g$ sufficiently small
the Feshbach map and express it in terms of a Neumann series
\begin{align}\label{feshexpansion} 
& F_{\chi^{(I)}}( \widehat{H}_{g}(\theta,\alpha)  - \zeta, \widehat{H}_0(\theta,\alpha) - \zeta) \nonumber  \\
& = \left( T + \chi^{} W \chi^{} - \chi^{} W \chib^{} ( T + \chib^{} W_{}\chib^{} )^{-1}
\chib^{} W_{} \chi^{} \right) \nonumber \\ 
& = \left( T^{} + \chi W^{} \chi - \chi W^{} \chib \sum_{n=0}^\infty \left( - {T^{}}^{-1} \chib W^{} \chib \right)^n
{T^{}}^{-1} \chib W^{} \chi \right) \; ,   
\end{align}
where here we used the abbreviations
$T^{} = \widehat{H}_0(\theta,\alpha)  - \zeta$, $W^{} = \widehat{W}_{g}(\theta,\alpha) $, $\chi = \chi^{(I)}_{\theta,\alpha}$, $\chib = \chib^{(I)}_{\theta,\alpha}$.
We normal order above expression, using the pull-through formula. To this end we use the identity of
Theorem \ref{thm:wicktheorem}, see Appendix B. This will show  \eqref{eq:inimainA}, i.e.,
\begin{align} \label{eq:inimainA2}
& H(\widehat{w}^{(0)}(g,\theta,\alpha,z))\\
&  = \langle  F_{\chi^{(I)}}( \widehat{H}_{g}(\theta,\alpha) - \zeta , \widehat{H}_0(\theta,\alpha) - \zeta)|_{ \ran ( P_{\rm at}(\theta,\alpha)  \otimes 1_{H_{\rm f} \leq 1} )} \rangle_{{\rm at},\theta,\alpha} ,\nonumber
\end{align}
where $\widehat{w}^{(0)}$ is  given  as follows.
First,  we introduce   the definition
\begin{align} \label{eq:defofWW}
& \underline{W}_{p,q}^{m,n}[w](K^{(m,n)}) \\
&:= \int_{{(\underline{\R}^3)}^{p+q}} \frac{d X^{(p,q)}}{|X^{(p,q)}|^{1/2}} a^*(X^{(p)}) w_{m+p,n+q}(K^{(m)}, X^{(p)}, \widetilde{K}^{(n)},
 \widetilde{X}^{(q)}) a(\widetilde{X}^{(q)}) .  \nonumber
\end{align}
Thus normal ordering \eqref{feshexpansion} by means of  Theorem \ref{thm:wicktheorem} and calculating  \eqref{eq:eff23},
we obtain a sequence of integral kernels
$\widetilde{w}^{(0)}$, which are given as follows.
For $M+N \geq 1$,
\begin{eqnarray} \label{eq:defofwmnschlange}
\lefteqn{ \widetilde{\widehat{w}}^{(0)}_{M,N}(g,\theta,\alpha,z)(r , K^{(M,N)})} \label{initial:eq7} \\ &&=
( 8 \pi )^{\frac{M+N}{2}} \sum_{L=1}^\infty (-1)^{L+1}
\sum_{\substack{ (\umm,\upp,\unn,\uqq) \in \N_0^{4L}: \\ |\umm|=M, |\unn|=N, \\ 1 \leq m_l+p_l+q_l+n_l \leq 2 } } \prod_{l=1}^L \left\{
\binom{ m_l + p_l}{ p_l} \binom{ n_l + q_l}{ q_l }
\right\} \nonumber \\ && \times
V_{(\umm,\upp,\unn,\uqq)}[\widehat{w}^{I}(g,\theta,\alpha,\zeta)](r,K^{(M,N)}). \nonumber
\end{eqnarray}
 Furthermore,
\begin{align*}
\widetilde{\widehat{w}}^{(0)}_{0,0}(g,\theta,\alpha,z)(r) = - z + r + \sum_{L=2}^\infty (-1)^{L+1}
\sum_{(\upp,\uqq)\in \N_0^{2L}: p_l+q_l = 1, 2}
V_{(\uzz,\upp,\uzz,\uqq),\theta,\alpha}[\widehat{w}^{(I)}(g,\theta,\alpha,\zeta)](r) \; .
\end{align*}
Above we have introduced  the definition  for $(\umm,\upp,\unn,\uqq) \in \N_0^{4L}$
\begin{eqnarray} \label{eq:defofV}
\lefteqn{ V_{\umm,\upp,\unn,\uqq,\theta,\alpha}[w](r, K^{(|\umm|,|\unn|)}) := } \\
&&
\left\langle  \Omega  , \left\langle  F_0[w](H_{\rm f} + r) \prod_{l=1}^L \left\{
\underline{W}_{p_l,q_l}^{m_l,n_l}[w](K^{(m_l,n_l)})
 F_l[w](H_{\rm f} + r + \widetilde{r}_l ) \right\} \right\rangle_{{\rm at},\theta,\alpha} \Omega \right\rangle
\nonumber ,
\end{eqnarray}
where for $l=0,L$ we set $F_l[w](r) := \chi_1(r )$ , and for $l=1,...,L-1$ we set
\begin{eqnarray*}
F_l[w](r) := F[w](r) := \frac{ { \chib}^{(I)}_{\theta,\alpha}( r  )^2}{ w_{0,0}(r )} .
\end{eqnarray*}
Moreover, see \eqref{eq:rltildedef} for the definition of $\widetilde{r}_l$.
We define
$\widehat{w}^{(0)}(g,\theta,\alpha,z) := \big(\widetilde{ \widehat{w}}^{(0)} \big)^{({\rm sym})}(g,\theta,\alpha,z) $.
So far we have determined $\widehat{w}^{(0)}$ on a formal level as a sequence of integral kernels. We have not yet shown that the involved series lies in the Banach space $ \WW^\#_\xi$.
Our next goal is to show estimates \eqref{eq:feb2:1}, \eqref{eq:feb2:2}, and \eqref{eq:feb2:3},
below. These estimates will then imply that
 $\widehat{w}^{(0)}(g,\theta,\alpha,z) \in \WW^\#_\xi$ and they will be used to show part (II) (a) of Theorem \ref{thm:inimain1}.
To this end, we need an estimate on $V_{\umm,\upp,\unn,\uqq}[\widehat{w}^{(I)}]$, which is given in the next lemma.

\begin{lemma} \label{initial:thmE22N} There exists finite constants $C_W$ and $C_F$
 such that with $C_W(g) := C_{W}( |g| +|g|^2) $  for all $(g,\theta, \alpha,\zeta) \in \C \times D_{\theta_{\rm f}}  \times  D_{\alpha_{\rm f}} \times \C $ such that   $|\zeta- \widehat{E}_{\rm at}(\theta,\alpha) | < {1/2} $,  the following inequality holds
\begin{align} \label{initial:thmmain:eq2}
 \| V_{\umm,\upp,\unn,\uqq,\theta,\alpha}[\widehat{w}^{(I)}(g,\theta, \alpha,\zeta)] \|^\# \leq (L + 1) C_F^{L+1} C_W(g)^L 
\end{align}
  for all $(\umm,\upp,\unn,\uqq) \in \N_0^{4L}$, $L \in \N$.
\end{lemma}

For the proof of Lemma  \ref{initial:thmE22N}  we
will make use of the estimates collected in  Lemma \ref{ini:lemelemestimatesN}, below, 
and  introduce an  auxiliary operator
$$
G_0 := - \Delta + H_{\rm f} + 1 .
$$

\begin{lemma} \label{ini:lemelemestimatesN} There exist finite constants $C_W$ and $C_F$ such that the following holds
for all $(g,\theta,\alpha,\zeta) \in \C \times D_{\theta_{\rm f}} \times D_{\alpha_{\rm f}} \times \C$.   If   $m+n+p+q \geq 1$ and
$K^{(m,n)} \in   \underline{B}_1^{m+n}$, then
\begin{align}
& \| G_0^{-1/2}  \underline{W}_{p,q}^{m,n}[w^{(I)}(g,\theta,\alpha,\zeta)](K^{(m,n)}) G_0^{-1/2} \| \leq C_{W} (|g| + |g|^2).  \label{eq1:ini:lemelemestimatesN}
\end{align}
If  $ |\zeta- \widehat{E}_{\rm at}(\theta,\alpha) | < {1/2}$ and $r \in[0, \infty) $,
then
\begin{align}
& \| G_0^{1/2} F[\widehat{w}^{(I)}( g,\theta,\alpha,\zeta ) ](r + H_{\rm f} ) G_0^{1/2}  \| \leq C_F \label{eq2:ini:lemelemestimatesN} , \\
& \|   G_0^{1/2}  \partial_r F[\widehat{w}^{(I)}( g,\theta,\alpha,\zeta)](r + H_{\rm f} )  G_0^{1/2} \| \leq C_F \label{eq3:ini:lemelemestimatesN} .
\end{align}
\end{lemma}

\begin{proof}
First we show \eqref{eq1:ini:lemelemestimatesN}.
For simplicity we drop the $(g, \theta, \alpha, \zeta)$--dependence in the notation.
If $p=q=0$ it follows directly from the definitions  given  in  \eqref{eq:defofWW}  and   \eqref{defofwI}    that   
$$
{\rm l . h. s. \ of \ } \eqref{eq1:ini:lemelemestimatesN} \ \leq 2 |e^{-\theta}| |g|^{m+n} N  (1+|\alpha|)  \| \kappa_\theta \|_\infty^{m}  \| \tilde{\kappa}_{{\theta}} \|_\infty^{n} ,
$$
where the right hand side can be bounded by means of     Hypothesis \ref{kappa}.
To see the corresponding estimate for $ p + q \geq 1$ we first introduce the notation
\begin{equation} \label{eq:defofbor}
B_0(r) := (- \Delta + r + 1 )^{-1}.
\end{equation}
Hence
by definition $B_0(H_{\rm f}) = G_0^{-1}$.
Using the pull-through formula and Lemma \ref{kernelopestimate} we see that
\begin{eqnarray}
\lefteqn{
I^{m,n}_{p,q} := \left\|  G_0^{-1/2} \underline{W}_{p,q}^{m,n}[\widehat{w}^{(I)}(g,\theta,\alpha)](K^{(m,n)}) G_0^{-1/2} \right\| } \nonumber \\
 && \leq \Bigg\{ \int_{{(\underline{\R}^3)}^{p+q}} \frac{d X^{(p,q)}}{|X^{(p,q)}|^2}
\sup_{r \geq 0} \Bigg[ \Big\| B_0( r + \Sigma[X^{(p)}])^{1/2} \widehat{w}_{m+p,n+q}^{(I)}(g,\theta,\alpha)( K^{(m)}, X^{(p)} , \widetilde{K}^{(n)},
 \widetilde{X}^{(q)})
 \nonumber \\
 &&\quad  \times   B_0( r + \Sigma[\widetilde{X}^{(q)}])^{1/2} \Big\|^2 \left( r + \Sigma[{X}^{(p)}]\right)^p \left( r + \Sigma[\widetilde{X}^{(q)}]\right)^q \Bigg] \Bigg\}^{1/2} ,
 \label{eq:contsigmainitial1}
\end{eqnarray}
where we used the trivial estimate for $r \geq 0$,
\begin{equation} \label{eq:trivestimateforkernel}
\prod_{l=1}^p \left( r + \Sigma[{X}^{(l)}]\right) \leq \left( r + \Sigma[{X}^{(p)}]\right)^p .
\end{equation}
Now we insert   \eqref{defofwI}   into  \eqref{eq:contsigmainitial1} to estimate the remaining cases for $m , n,p,q$ 
and a straightforward  estimate         gives
$$
I^{m,n}_{p,q} \leq
\left\{ \begin{array}{ll} |g| |e^{-\theta}| (1+|\alpha|) 2^{1/2} N \|  {{\kappa}_{\theta}} / \omega \|_\hh^p \|  \tilde{\kappa}_\theta / \omega \|_\hh^q  ,  & {\rm if } \ \ S =1,\ p+q =1 , \\
 |g|^2 |e^{-\theta}|  N \|  {{\kappa}_{\theta}}/\omega \|_\hh^{p}  \|  {\kappa}_\theta \|_\infty^{m} 
 \|  \tilde{\kappa}_{\theta}/\omega \|_\hh^{q}  \|  \tilde{\kappa}_{\theta} \|_\infty^{n}
,  &   {\rm if } \ \ S = 2 ,\ \max(p,q) = 1 , \\
 |g|^2|e^{-\theta}|  N \left( 2^{-1}\|  {{\kappa}_{\theta}}/\omega \|_\hh^2 +  \|  {\kappa}_{\theta} / \omega \|_\hh
 \| { \kappa}_{\theta} / \omega^{1/2} \|_\hh \right)
 , & {\rm if } \ \ S = 2 , \ p = 2 ,
\\
 |g|^2|e^{-\theta}|  N \left( 2^{-1}\|  \tilde{\kappa}_{\theta}/\omega \|_\hh^2 +  \|  \tilde{\kappa}_{\theta} / \omega \|_\hh
 \| \tilde{ \kappa}_{\theta} / \omega^{1/2} \|_\hh \right)
 , & {\rm if } \ \ S = 2 , \ q = 2 ,
\end{array} \right.
$$
with $S := m + n + p + q$.  
Collecting estimates Eq. \eqref{eq1:ini:lemelemestimatesN} follows,   since $\theta_{\rm f} < \theta_{\rm I}$ (for 
    $\theta_{\rm I}$  such that   Hypothesis \ref{kappa} holds). 

Next we show \eqref{eq2:ini:lemelemestimatesN}.
Using the boundedness of $\chib_{\theta,\alpha}^{(I)}(H_{\rm f} + r )$ it follows from the bounds \eqref{firstfreeintbound}  and \eqref{secondfreeintbound} that there exists a constant $C$ such that 
\begin{align*}
& \| G_0  F[\widehat{w}^{(I)}( g,\theta,\alpha,\zeta ) ](r + H_{\rm f} )   \| \leq C  \\ 
& \|    F[\widehat{w}^{(I)}( g,\theta,\alpha,\zeta)](r + H_{\rm f} )  G_0 \| \leq C ,   
\end{align*}
 for all $r \geq 0$,  $(\theta,\alpha) \in  D_{\theta_{\rm f}} \times D_{\alpha_{\rm f}} $,
 and $\zeta \in \C$ with    $ |\zeta- \widehat{E}_{\rm at}(\theta,\alpha) | < {1/2}$.
This implies   \eqref{eq2:ini:lemelemestimatesN} using interpolation of operators, cf.  Lemma \ref{intpol} in the appendix.
Likewise, we show  \eqref{eq3:ini:lemelemestimatesN}.  Calculating a derivative we find
$$
\partial_r F[\widehat{w}^{(I)}(g,\theta,\alpha,\zeta)](r) = \frac{- \left[ \chib_{\theta,\alpha}^{(I)}(r) \right]^2}{\left(\widehat{w}_{0,0}^{(I)}(g,\theta,\alpha,\zeta)(r)\right)^2} +
\frac{2 \chib_{\theta,\alpha}^{(I)}(r) \partial_r \chib^{(I)}(r)}{\widehat{w}_{0,0}^{(I)}(g,\theta,\alpha,\zeta)(r)}  .
$$
Using again the boundedness of  $\chib_{\theta,\alpha}^{(I)}(H_{\rm f} + r )$ and  $\partial_r \chib_{\theta,\alpha}^{(I)}(H_{\rm f} + r )$ it follows from the bounds \eqref{firstfreeintbound}  and \eqref{secondfreeintbound} that
there exists a constant $C$ such that  
\begin{align*}
& \| G_0  \partial_r F[\widehat{w}^{(I)}( g,\theta,\alpha,\zeta ) ](r + H_{\rm f} )   \| \leq C  \\ 
& \|  \partial_r   F[\widehat{w}^{(I)}( g,\theta,\alpha,\zeta)](r + H_{\rm f} )  G_0 \| \leq C  , 
\end{align*}
for all $r \geq 0$,  $(\theta,\alpha) \in  D_{\theta_{\rm f}} \times D_{\alpha_{\rm f}} $,
 and $\zeta \in \C$ with    $ |\zeta- \widehat{E}_{\rm at}(\theta,\alpha) | < {1/2}$.
This implies   \eqref{eq3:ini:lemelemestimatesN} using interpolation.
\end{proof}

\begin{proof}[Proof of Lemma \ref{initial:thmE22N}.]

We estimate $\| V_{\umm,\upp,\unn,\uqq}[w^{(I)}(g,\theta,\alpha,\zeta)] \|_{ \underline{\infty}}$ inserting several of the identities  $1=G_0^{1/2} G_0^{-1/2}$ and $1=G_0^{-1/2} G_0^{1/2}$,  recalling   \eqref{eq:eff23},
\begin{equation} \label{eq:babyestimate1N}
| \langle  \Omega , A_1 A_2 \cdots A_n \Omega \rangle | \leq \| A_1 \|_{\rm op} \| A_2 \|_{\rm op} \cdots \| A_n \|_{\rm op},
\end{equation}
where $\| \cdot \|_{\rm op}$ denotes the operator norm,
and using Inequalities \eqref{eq1:ini:lemelemestimatesN} and \eqref{eq2:ini:lemelemestimatesN}.
To estimate $\| \partial_r V_{\umm,\upp,\unn,\uqq}[w^{(I)}(g,\theta,\alpha,\zeta)] \|_{ \underline{\infty}}$ we first calculate
the derivative using the Leibniz rule. The resulting expression is estimated using again \eqref{eq:babyestimate1N}
and Inequalities \eqref{eq1:ini:lemelemestimatesN}--\eqref{eq3:ini:lemelemestimatesN}.
\end{proof}

Now we are ready to give the proof of Part (II) of the main theorem of this section.

\begin{proof}[Proof of Theorem  \ref{thm:inimain1} Part  {\rm (II)}]
Recall that we assume \eqref{eq:zzetarelation}.
Let $S^L_{M,N}$ denote the set of tuples $(\umm,\upp,\unn,\uqq) \in \N_0^{4L}$ with
$|\umm|=M$, $|\unn|=N$, and $1 \leq m_l+p_l+q_l+n_l \leq 2$.
We estimate the norm of \eqref{eq:defofwmnschlange} using \eqref{initial:thmmain:eq2} and
find, with $\widetilde{\xi} := (8 \pi)^{-1/2} \xi$,
\begin{align}
\| \widehat{w}^{(0)}_{\geq 1}(g,\theta,\alpha,z) \|_\xi^\# & =   \sum_{M+N \geq 1} {\xi}^{-(M+N)}
 \| \widehat{\widetilde{w}}_{M,N}(g,\theta,\alpha,z) \|^\# \nonumber \\
&\leq \sum_{M+N\geq 1} \sum_{L=1}^\infty \sum_{(\umm,\upp,\unn,\uqq) \in S^L_{M,N}}
\widetilde{\xi}^{-(M+N)} 4^L \| V_{\umm,\upp,\unn,\uqq,\theta,\alpha}[\widehat{w}^{(I)}(g,\theta,\alpha,\zeta)] \|^\# \nonumber \\
&\leq \sum_{L=1}^\infty \sum_{M+N\geq 1} \sum_{(\umm,\upp,\unn,\uqq) \in S^L_{M,N} }
\widetilde{\xi}^{-|\umm|-|\unn|} (L+1) C_F \left( 4 C_W(g) C_F \right)^L \nonumber \\
&\leq\sum_{L=1}^\infty (L+1) (14)^L \widetilde{\xi}^{-2L} C_F \left( 4 C_W(g) C_F \right)^L \; , \label{eq:feb2:1}
\end{align}
for all $(g,\theta,\alpha) \in \C \times D_{\theta_{\rm f}}$  $\times D_{\alpha_{\rm f}} $, 
where in the second line we used $\binom{ m + p }{ p} \leq 2^{m+p}$
and in the last line we used  $|\umm|+|\unn| \leq 2L$ and that the number of elements  $(\umm,\upp,\unn,\uqq) \in \N_0^L$ with
$1 \leq m_l+n_l+p_l+q_l \leq 2$ is bounded by $(14)^{L}$.
A similar but simpler estimate yields
\begin{align}
\sup_{r \in [0,1]} | \partial_r \widehat{w}^{(0)}_{0,0}(g,\theta,\alpha,z)(r) - 1 |
&\leq \sum_{L=2}^\infty \sum_{(\upp,\uqq) \in \N_0^{2L}: p_l+q_l= 1, 2}
 \| V_{\uzz,\upp,\uzz,\uqq,\theta,\alpha}[\widehat{w}^{(I)}(g,\theta,\alpha,\zeta)] \|^\# \nonumber \\
&\leq \sum_{L=2}^\infty 3^L (L+1) C_F \left( C_W(g) C_F \right)^L \; , \label{eq:feb2:2}
\end{align}
for all  $(g,\theta,\alpha) \in \C \times D_{\theta_{\rm f}} \times D_{\alpha_{\rm f}}.$ 
Analogously we have for all  $(g,\theta,\alpha) \in \C \times D_{\theta_{\rm f}}  \times D_{\alpha_{\rm f}} $ 
\begin{align}
| \widehat{w}^{(0)}_{0,0}(g,\theta,\alpha,z)(0) + z |
& \leq \sum_{L=2}^\infty \sum_{(\upp,\uqq) \in \N_0^{2L}: p_l+q_l= 1, 2}
\| V_{\uzz,\upp,\uzz,\uqq,\theta,\alpha}[\widehat{w}^{(I)}(g,\beta,\sigma,\zeta)] \|^\# \nonumber
\\ \label{eq:feb2:3}
& \leq \sum_{L=2}^\infty 3^L (L+1) C_F \left( C_W(g) C_F \right)^L .
\end{align}
In view of the definition of $C_W(g)$ the right hand sides in \eqref{eq:feb2:1}--\eqref{eq:feb2:3} can be made arbitrarily small for sufficiently small $|g|$.
This implies that there exists a $g_{\rm f} > 0$
such that for $g \in D_{g_{\rm f}}$ the kernel $w^{(0)}(g,\theta,\alpha,z)$ is in $\WW_\xi^\#$ and that the inequalities in the definition of $\mathcal{B}_0(\delta_1,\delta_2,\delta_3)$
are satisfied. Rotation invariance of $\widehat{w}^{(0)}$ follows since the right hand side of \eqref{eq:inimainA} is invariant under rotations and Lemma \ref{lem:wHinvequiv}.
(b) follows from the properties of the right hand side of \eqref{eq:inimainA} and Lemma \ref{lem:symmetry}.
It remains to show (c). This part will from the convergence established in \eqref{eq:feb2:1}--\eqref{eq:feb2:3}, which is
uniform in $(g,\theta,\alpha,z) \in D_{g_{\rm f}} \times  D_{\theta_{\rm f}} \times D_{\alpha_{\rm f}} \times D_{1/2}$, and Lemma \ref{lem:analytwI}, shown below.
\end{proof}

\begin{lemma} \label{lem:analytwI} The mapping $(g,\theta,\alpha, z) \mapsto V_{\umm,\upp,\unn,\uqq,\theta,\alpha}[\widehat{w}^{(I)}(g,\theta,\alpha, \widehat{E}_{\rm at}(\theta,\alpha)  + z )]$ is a $\WW_{|\umm|,|\unn|}^\#$-valued
analytic function on $\C \times  D_{\theta_{\rm f}} \times D_{\alpha_{\rm f}} \times D_{1/2}$.
\end{lemma}

\begin{proof}
The analyticity in $g$ follows since $ V_{\umm,\upp,\unn,\uqq,\theta,\alpha}[w^{(I)}(g,\theta,\alpha, \widehat{E}_{\rm at}(\theta,\alpha) + z )]$ is a polynomial in $g$ and
the coefficients of this polynomial are elements in $\WW_{|\umm|,|\unn|}^\#$ because of \eqref{initial:thmmain:eq2}.
To show the analyticity in the other variables $\theta$, $\alpha$, and $z$ first observe that $V_{\umm,\upp,\unn,\uqq,\theta,\alpha}$
as well as  its derivative $\partial_r V_{\umm,\upp,\unn,\uqq,\theta,\alpha}$ (by the product rule of differentiation)  is multilinear expression of integral kernels.
As before we will insert   $1=G_0^{1/2} G_0^{-1/2}$ and $1=G_0^{-1/2} G_0^{1/2}$  and we will use the following algebraic identity
\begin{eqnarray} \label{eq:telescoping2}
\lefteqn{\frac{ A_1(s) \cdots A_n(s) - A_1(s_0) \cdots A_n(s_0) }{s-s_0} } \\
&& - \sum_{i=1}^n A_1(s_0) \cdots A_{i-1}(s_0) A'_i(s_0) A_{i+1}(s_0) \cdots A_n(s_0) \nonumber \\
&&= \sum_{i=1}^n A_1(s) \cdots A_{i-1}(s) \left[ \frac{A_i(s) - A_i(s_0)}{s-s_0} - A'_i(s_0) \right] A_{i+1}(s_0) \cdots A_n(s_0) \nonumber \\
&& + \sum_{i=1}^n \left[ A_1(s) \cdots A_{i-1}(s) - A_1(s_0) \cdots A_{i-1}(s_0) \right] A'_i(s_0) A_{i+1}(s_0) \cdots A_n(s_0) \nonumber .
\end{eqnarray}
to  show differentiability  with respect to $\theta$, $\alpha$ and $z$.
Thus using \eqref{eq:babyestimate1N} analyticity will follow from complex  differentiability  of each of the terms $A_i(s)$ in $s$.
There are the following  type of factors  in $ V_{\umm,\upp,\unn,\uqq,\theta,\alpha}[w^{(I)}(g,\theta,\alpha, \widehat{E}_{\rm at}(\theta,\alpha) + z )]$ as
well as its derivative $ \partial_r V_{\umm,\upp,\unn,\uqq,\theta,\alpha}[w^{(I)}(g,\theta,\alpha,  \widehat{E}_{\rm at}(\theta,\alpha) + z  )]$  which we have
to estimate, namely
\begin{align}
  \mathfrak{F}_{\theta,\alpha}^{(I)}(r) (z) &  :=
G_0^{1/2}
\frac{\left[\chib^{(I)}(r)\right]^2}{\widehat{H}_{\rm at}(\theta,\alpha) + H_{\rm f} + r - \widehat{E}_{\rm at}(\theta,\alpha) - z  } G_0^{1/2}
\label{eq:defofF} ,  \\
  \partial_r \mathfrak{F}_{\theta,\alpha}^{(I)}(r) (z)  & =  G_0^{1/2}
\frac{2 \chib^{(I)}(r) \partial_r \chib^{(I)}(r)}{  \widehat{H}_{\rm at}(\theta,\alpha) + H_{\rm f} + r - \widehat{E}_{\rm at}(\theta,\alpha) - z   }
G_0^{1/2}
\nonumber  \\
& \quad  - G_0^{1/2}
\frac{ \left[ \chib^{(I)}(r) \right]^2}{( \widehat{H}_{\rm at}(\theta,\alpha) + H_{\rm f} + r - \widehat{E}_{\rm at}(\theta,\alpha) - z )^2    }
G_0^{1/2}
  \label{eq:defofdrF}
\end{align}
as well as
\begin{align}
& G_0^{-1/2} \underline{W}^{m,n}_{p,q}[  \widehat{w}^{(I)}(g,\theta,\alpha)](K^{(m,n})G_0^{-1/2} \label{eq:defofW} .
\end{align}
The analyticity of \eqref{eq:defofF} and \eqref{eq:defofdrF}
 with respect to $\theta$, $\alpha$, and $z$ in the $\sup_{r \in [0,1] }| \cdot |$-norm can be shown to  follow from Lemma \ref{lemredresbound}. This follows by means of the  mean value theorem using  that the second derivative with respect to
each  of the variables $\theta$, $\alpha$, and $z$ is uniformly bounded on compact subsets and $r \in [0,1]$.
But that follows from the analyticity statement of  Lemma \ref{lemredresbound} and compactness.

Next we show the analyticity of \eqref{eq:defofW} with respect to $\theta$, $\alpha$, and $z$.
The analyticity in $z$ is trivial as it does not depend on that variable.
The analyticity in $\alpha$ follows in view of \eqref{eq:telescoping2} since $w^{(I)}$ is polynomial function of $\alpha$ (in fact an affine function).
To show  the analyticity in $\theta$ we first factor out the trivial factor $e^\theta$ and observe
 that by Hypothesis \ref{kappa} and  Remark   \ref{kdappeana} there exist   functions  $\tau_\theta , \tilde{\tau}_\theta \in L^\infty(\R^3) \cap \omega^{1/2} \hh \cap \omega \hh $ which we denote by $\partial_\theta \kappa_\theta$ and $\partial_\theta \tilde{\kappa}_{\theta}$, respectively, 
 such that    for $\theta \in D_{\theta_{\rm I}}$  and $h \to 0$ 
  \begin{align} \label{convkappa1} 
 &   \| h^{-1} ( \kappa_{\theta + h} - \kappa_\theta )  - \partial_\theta \kappa_\theta \|_\infty \to 0 , \quad   \| h^{-1} ( \tilde{\kappa}_{\theta + h} - \tilde{\kappa}_\theta )  -  \partial_\theta \tilde{\kappa}_\theta  \|_\infty \to 0 \\
  &   \| \omega^{-1/2} ( h^{-1} ( \kappa_{\theta + h} - \kappa_\theta )  - \partial_\theta \kappa_\theta )  \|_\hh \to 0 , \quad   
  \|   \omega^{-1/2} (  
  h^{-1} ( \tilde{\kappa}_{\theta + h} - \tilde{\kappa}_\theta )  -  \partial_\theta \tilde{\kappa}_\theta  ) \|_\hh \to 0 \\
    &   \| \omega^{-1} ( h^{-1} ( \kappa_{\theta + h} - \kappa_\theta )  - \partial_\theta \kappa_\theta )  \|_\hh \to 0 , \quad   
  \|   \omega^{-1} (  
  h^{-1} ( \tilde{\kappa}_{\theta + h} - \tilde{\kappa}_\theta )  -  \partial_\theta \tilde{\kappa}_\theta  ) \|_\hh \to 0 .  \label{convkappa3}
  \end{align} 
  Note that the  uniqueness of the derivatives follows in view of  \cite[Corollary 2.32,Theorem 6.8 (c)]{Folland}.  
  Although not needed for the proof we note that $\tilde{\tau}_\theta =\overline{ \tau_{\overline{\theta}}}$.
Define
\begin{align}
&D_{1,0}^{(I)}(g,\theta,\alpha)( K) :=  2 g \sum_{j=1}^N ( p_j - \alpha x_j \langle x \rangle^{-1} ) \cdot \varepsilon(k,\lambda) \frac{\partial_\theta \kappa_{\theta}(k)e^{ -i  \beta k \cdot x_j }}{\sqrt{2}} , \nonumber \\
&D_{0,1}^{(I)}(g,\theta,\alpha)( K) :=  2 g \sum_{j=1}^N ( p_j - \alpha x_j \langle x \rangle^{-1} ) \cdot \varepsilon(k,\lambda)
\frac{ \partial_\theta  \tilde{\kappa}_{   \theta}(k)e^{ i \beta  k \cdot x_j }}{\sqrt{2}} , \nonumber\\
&D_{1,1}^{(I)}(g,\theta,\alpha)(K,\widetilde{K}) \nonumber \\
&
:=  2 g^2  \sum_{j=1}^N \varepsilon(k,\lambda) \cdot \varepsilon(\widetilde{k},\widetilde{\lambda})
\left(\partial_\theta \kappa_{\theta}(k) \tilde{\kappa}_{\theta}(\widetilde{k}) +  \kappa_{\theta}(k) \partial_\theta \tilde{\kappa}_{\theta}(\widetilde{k}) \right)
\frac{e^{- i \beta  k \cdot x_j }}{\sqrt{2}} \frac{ e^{ i  \beta \widetilde{k}\cdot x_j }}{\sqrt{2}} ,
\nonumber \\
&D_{2,0}^{(I)}(g,\theta,\alpha)( K_1, K_2 ) \nonumber \\
& :=  g^2 \sum_{j=1}^N \varepsilon(k_1,\lambda_1) \cdot \varepsilon({k}_2,{\lambda}_2)
\left(\partial_\theta \kappa_{\theta}(k_1) \kappa_{\theta}(k_2) +  \kappa_{\theta}(k_1) \partial_\theta \kappa_{\theta}(k_2) \right)
 \frac{e^{ - i \beta  k_1 \cdot x_j }}{\sqrt{2}} \frac{ e^{ - i \beta k_2 \cdot x_j }}{\sqrt{2}} ,
\nonumber
\\
&D_{0,2}^{(I)}(g,\theta,\alpha)( \widetilde{K}_1, \widetilde{K}_2 ) \nonumber \\
& :=  g^2 \sum_{j=1}^N \varepsilon(\tilde{k}_1,\tilde{\lambda}_1) \cdot \varepsilon(\widetilde{k}_2, \widetilde{\lambda}_2)
\left(\partial_\theta \tilde{\kappa}_{\theta}(\widetilde{k}_1) \tilde{\kappa}_{\theta}(\widetilde{k}_2) +  \tilde{\kappa}_{\theta}(\widetilde{k}_1) \partial_\theta \tilde{\kappa}_{\theta}(\widetilde{k}_2) \right)
 \frac{e^{  i \beta  \widetilde{k}_1 \cdot x_j }}{\sqrt{2}} \frac{ e^{  i \beta \tilde{k}_2 \cdot x_j }}{\sqrt{2}}
\nonumber
\end{align}
Using the pull-through formula and Lemma \ref{kernelopestimate} we see that
\begin{align}
 & J^{m,n}_{p,q}(h,K^{(m,n)}) \nonumber \\
 &  :=  \bigg\| \frac{1}{h} G_0^{-1/2}  \left(  \underline{W}^{m,n}_{p,q}[e^\theta \widehat{w}^{(I)}(g,\theta+h,\alpha)](K^{(m,n)}) -
  \underline{W}^{m,n}_{p,q}[e^\theta \widehat{w}^{(I)}(g,\theta,\alpha)](K^{(m,n)}) ] \right) G_0^{-1/2} \nonumber \\
 &   \qquad   -  G_0^{-1/2}   \underline{W}^{m,n}_{p,q}[D^{(I)}(g,\theta,\alpha)](K^{(m,n)})  G_0^{-1/2}  \bigg\|_{\rm op} \nonumber  \\
 & =
  \bigg\|   \underline{W}^{m,n}_{p,q}[h^{-1} G_0^{-1/2}   \left( e^\theta \widehat{w}^{(I)}(g,\theta+h,\alpha)    - e^\theta \widehat{w}^I(g,\theta,\alpha) \right)   -  D^{(I)}(g,\theta,\alpha)]  (K^{(m,n)}) G_0^{-1/2} 
   \bigg\|_{\rm op} \nonumber \\
 & \leq \Bigg\{ \int_{{(\underline{\R}^3)}^{p+q}} \frac{d X^{(p,q)}}{|X^{(p,q)}|^2} \nonumber
  \sup_{r \geq 0} \Bigg[ \Big\|B_0( r + \Sigma[X^{(p)}])^{1/2} \nonumber  \\
& \times  ( h^{-1} ( e^\theta \widehat{w}_{m+p,n+q}^{(I)}    (g,\theta+h,\alpha) -
e^\theta \widehat{w}_{m+p,n+q}^{(I)}(g,\theta,\alpha))  - D_{m+p,n+q}^{(I)}(g,\theta,\alpha) )( K^{(m)}, X^{(p)} , \widetilde{K}^{(n)},
 \widetilde{X}^{(q)}) \nonumber   \\
 & \times B_0( r + \Sigma[\widetilde{X}^{(q)}])^{1/2} \Big\|^2 \times \left( r + \Sigma[{X}^{(p)}]\right)^p \left( r + \Sigma[\widetilde{X}^{(q)}]\right)^q \Bigg] \Bigg\}^{1/2} ,
 \label{eq:contsigmainitial12}
\end{align}
where we used the trivial estimate \eqref{eq:trivestimateforkernel} for $r \geq 0$.
Now we use \eqref{eq:contsigmainitial12} to estimate the remaining cases for $m , n,p,q$ separately. For this we define
$$
\mathfrak{D}_h  \kappa_\theta  := h^{-1} ({{\kappa}_{\theta+h}} -  {{\kappa}_{\theta}}) - \partial_\theta \kappa_\theta , 
\quad \mathfrak{D}_h  \tilde{\kappa}_{\theta}
 := h^{-1} (\tilde{\kappa}_{\theta +h} -  \tilde{\kappa}_{\theta}) - \partial_\theta \tilde{\kappa}_{\theta} . 
$$

Let $S = m+n+p+q$.  In view of the definition of $\widehat{w}^{(I)}$ and $\underline{W}_{p,q}^{m,n}$ we only need to
consider $S=1$ and $S=2$.
First, if  $S =1$ and $q =1$ we estimate
\begin{align*}
 & J^{m,n}_{p,q}(h,K^{m,n}) = J^{0,0}_{0,1}(h)  \\
 & \leq \Bigg\{ \int_{\underline{\R}^3} \frac{d \tilde{Y}}{|\tilde{Y}|^2}
\sup_{r \geq 0} \Bigg[ \Big\|( - \Delta + 1 + r )^{-1/2} \\
&\times \big( h^{-1} ( e^\theta \widehat{w}_{0,1 }^{(I)}    (g,\theta+h,\alpha) -
e^\theta \widehat{w}_{0,1 }^{(I)}(g,\theta,\alpha))  - D_{0,1}^{(I)}(g,\theta,\alpha)\big)(\tilde{Y})\nonumber   \\
 & \times ( - \Delta + 1 + r + |\tilde{Y}|)^{-1/2} \Big\|^2
 \left( r + |\tilde{Y}| \right) \Bigg]  \Bigg\}^{1/2} \\
  & \leq \Bigg\{  \sum_{\tilde{\lambda}=1,2}  \int_{\R^3} \frac{d \tilde{k}}{|\tilde{k}|^2}
\sup_{r \geq 0} \Bigg[ \Big\|( - \Delta + 1 + r )^{-1/2} (
 2 g \sum_{j=1}^N ( p_j - \alpha x_j \langle x \rangle^{-1} ) \cdot \varepsilon(\tilde{k},\tilde{\lambda}) \frac{\mathfrak{D}_h  \tilde{\kappa}_\theta(\tilde{k})e^{ -i  \beta
 \tilde{ k} \cdot x_j }}{\sqrt{2}}\nonumber   \\
 & \times ( - \Delta + 1 + r + |\tilde{k}| )^{-1/2} \Big\|^2
 \left( r + |\tilde{k}| \right) \Bigg] \Bigg\}^{1/2} \\
   & \leq \Bigg\{  \sum_{\tilde{\lambda}=1,2}  \int_{\R^3} \frac{d \tilde{k}}{|\tilde{k}|^2}
 \Bigg[ \Big\|( - \Delta + 1  )^{-1/2} (
 2 g \sum_{j=1}^N ( p_j - \alpha x_j \langle x \rangle^{-1} ) \cdot \varepsilon(\tilde{k},\tilde{\lambda}) \Big\|^2 \Big| \frac{\mathfrak{D}_h  \tilde{\kappa}_\theta(\tilde{k})e^{ -i  \beta
 \tilde{ k} \cdot x_j }}{\sqrt{2}}  \Big|^2
 \Bigg] \Bigg\}^{1/2} \\
 &
 \leq |g| (1+|\alpha|) \sqrt{2}  N  \|  \mathfrak{D}_h  \tilde{\kappa}_{\theta}  /\omega \|_\hh.
\end{align*}
Likewise, if   $S =1$ and $p =1$ we obtain the estimate
\begin{align*}
 & J^{m,n}_{p,q}(h,K^{(m,n)}) = J^{0,0}_{1,0}(h,K^{(m,n)})  \leq
 |g| (1+|\alpha|) \sqrt{2}  N \|  \mathfrak{D}_h  \kappa_\theta  /\omega \|_\hh.
\end{align*}
The case $S=1$ and $p=q=0$ is a simpler version of the above estimation.
Then there is no  integration. If  $(m,n)=(0,1)$, then similarly as before
\begin{align*}
& J^{0,1}_{0,0}(h,K^{(0,1)})  \\
 & \leq \sup_{r \geq 0}  \Big\|( - \Delta + 1 + r )^{-1/2} ( h^{-1} ( e^\theta \widehat{w}_{0,1 }^{(I)}    (g, \theta+h,\alpha) -
e^\theta \widehat{w}_{0,1 }^{(I)}(g, \theta,\alpha))  - D_{  0,1  }^{(I)}(g,\theta,\alpha))(\tilde{K})\nonumber   \\
 & \times ( - \Delta + 1 + r )^{-1/2} \Big\| \\
& \le |g|(1+|\alpha|) \sqrt{2} N \| \mathfrak{D}_h \tilde{\kappa}_{\theta} \|_\infty .
\end{align*}
For the case $(m,n)=(1,0)$  we obtain the  same estimate  by taking the adjoint. \\
Now let  $S=2$. If $S=2$  and $p=q=1$,  then $m=n=0$ and we estimate
\begin{align*}
 & J^{m,n}_{p,q}(h,K^{(m,n)}) = J^{0,0}_{1,1}(h)  \\
 & \leq \Bigg\{ \int_{[\underline{\R}^3]^2} \frac{d Y d \tilde{Y}}{|Y|^2 |\tilde{Y}|^2}
\sup_{r \geq 0} \Bigg[ \Big\|( - \Delta + 1 + r + |Y| )^{-1/2} \nonumber \\
& \times
 \big\{ ( h^{-1} ( e^\theta \widehat{w}_{1,1}^{(I)}    (g,\theta+h,\alpha) -
e^\theta \widehat{w}_{1,1}^{(I)}(g,\theta,\alpha))   - D_{1,1}^{(I)}(g,\theta,\alpha) \big\}(Y, \tilde{Y}) \nonumber \\
& \times  ( - \Delta + 1 + r + |\tilde{Y}|)^{-1/2} \Big\|^2
 \left( r + |\tilde{Y}| \right) \left( r + |Y| \right) \Bigg]  \Bigg\}^{1/2} \\
  & =  \Bigg\{  \sum_{\lambda, \tilde{\lambda}=1,2}  \int_{[\R^3]^2} \frac{dk d \tilde{k}}{|k|^2 |\tilde{k}|^2}
\sup_{r \geq 0} \Bigg[ \Big\|( - \Delta + 1 + r + |k| )^{-1/2}
  2 g^2  \sum_{j=1}^N \varepsilon(k,\lambda) \cdot \varepsilon(\widetilde{k},\widetilde{\lambda}) \\
 & \times
\Big( \frac{1}{h} \left( \kappa_{\theta+h}(k)\tilde{\kappa}_{\theta+h}(\tilde{k}) - \kappa_{\theta}(k)\tilde{\kappa}_{\theta}(\tilde{k}) \right) - \partial_\theta\kappa_\theta(k) 
\tilde{\kappa}_{\theta}(\tilde{k}) -  \kappa_\theta(k) \partial_\theta \tilde{\kappa}_{\theta}(\tilde{k}) \Big) \\
 & \times
\frac{e^{- i \beta  k \cdot x_j }}{\sqrt{2}} \frac{ e^{ i  \beta \widetilde{k}\cdot x_j }}{\sqrt{2}}
( - \Delta + 1 + r + |\tilde{k}| )^{-1/2} \Big\|^2
  \left( r + |k| \right) \left( r + |\tilde{k}| \right) \Bigg] \Bigg\}^{1/2} \\
  & \leq   |g|^2  N   \Bigg\{  \sum_{\lambda, \tilde{\lambda}=1,2}  \int_{[\R^3]^2} \frac{dk d \tilde{k}}{|k|^2 |\tilde{k}|^2}
  \nonumber \\
   & \times
\Big| \frac{1}{h} \left( \kappa_{\theta+h}(k) \tilde{\kappa}_{\theta+h}(\tilde{k}) - \kappa_{\theta}(k)\tilde{\kappa}_{\theta}(\tilde{k}) \right) - \partial_\theta\kappa_\theta(k) \tilde{\kappa}_{\theta}(\tilde{k}) -  \kappa_\theta(k) \partial_\theta \tilde{\kappa}_{\theta}(\tilde{k}) \Big|^2  \Bigg\}^{1/2} \\
&  \leq |g|^2 N  \left(  \|(  \kappa_{\theta + h} - \kappa_\theta )/\omega \|_\hh  \| \partial_\theta  \tilde{\kappa}_{\theta} /\omega \|_\hh+
   \|  \kappa_{\theta + h} /\omega \|_\hh  \| \mathfrak{D}_h \tilde{\kappa}_{\theta} /\omega \|_\hh +\|  \tilde{\kappa}_{\theta} /\omega \|_\hh       \| \mathfrak{D}_h \kappa_\theta /\omega \|_\hh \right)
\end{align*}
where we made use of the algebraic identity
\begin{align}
& \frac{1}{h} \left( \tau_{\theta+h}(X)\sigma_{\theta+h}(Y) - \tau_{\theta}(X)\sigma_{\theta}(Y) \right) - \partial_\theta\tau_\theta(X) \sigma_\theta(Y) -  \tau_\theta(X) \partial_\theta\sigma_\theta(Y)  \nonumber \\
&= \left(\tau_{\theta + h}(X) - \tau_\theta(X) \right) \partial_\theta \sigma_\theta(Y) + \tau_{\theta+h}(X) \left[ \frac{1}{h}  \left( \sigma_{\theta+h}(Y) - \sigma_{\theta}(Y)\right) - \partial_\theta\sigma_\theta(Y) \right]  \nonumber \\
& +  \left[\frac{1}{h} \left( \tau_{\theta+h}(X) - \tau_{\theta}(X)\right) - \partial_\theta\tau_\theta(X)  \right] \sigma_{\theta}(Y)
\label{eq:Kappdiffrel}
\end{align}
which holds for $\tau_\theta $ and $\sigma_\theta$ any of the two functions  $\kappa_\theta$ or $\tilde{\kappa}_{\theta}$.  

Let us now  consider is $S=2$ with $\max(p,q)=2$. If  $(p,q)=(0,2)$, then
\begin{align}
& J^{0,0}_{0,2}(h) \nonumber  \\
& \le \Bigg\{ \int_{[\underline{\R}^3]^2} \frac{d\tilde{Y_1} d\tilde{Y_2}}{|\tilde{Y_1}|^2 |\tilde{Y_2}|^2}  \sup_{r \geq 0} \Bigg[ \Big\|( - \Delta + 1 + r )^{-1/2} \nonumber \\
&  \times \big( h^{-1} ( e^\theta \widehat{w}_{0,2}^{(I)} (g,\theta+h,\alpha) -  e^\theta \widehat{w}_{0,2}^{(I)}(g,\theta,\alpha))  - D_{0,2}^{(I)}(g,\theta,\alpha) \big)(\tilde{Y_1}, \tilde{Y_2}) \nonumber \\
& \times ( - \Delta + 1 + r + |\tilde{Y_1}| + |\tilde{Y_2}|)^{-1/2} \Big\|^2 \left( r + |\tilde{Y_1}| + |\tilde{Y_2}| \right)^2  \Bigg]  \Bigg\}^{1/2} \nonumber  \\
&  \leq |g|^2 N 2^{-1} \left[  \|( \tilde{ \kappa}_{ \theta + h } - \tilde{\kappa}_{\theta} )/\omega \|_\hh  \| \partial_\theta  \tilde{\kappa}_{\theta }  /\omega \|_\hh+
 (  \| \tilde{ \kappa}_{\theta  + h} /\omega \|_\hh  +\| \tilde{ \kappa}_{\theta } /\omega \|_\hh      )  \| \mathfrak{D}_h \tilde{\kappa}_{\theta} /\omega \|_\hh \right. \nonumber \\
 & \left. + 2 \left( \left\|\frac{ \tilde{\kappa}_{\theta + h} - \tilde{\kappa}_{  \theta } }{\sqrt{  \omega }} \right\|_\hh  \| \partial_\theta  \tilde{ \kappa}_{  \theta  }  /\omega \|_\hh+
  \left\| \frac{ \tilde{ \kappa}_{\theta + h} }{ \sqrt{\omega}} \right\|_\hh \| \mathfrak{D}_h \tilde{\kappa}_{ 
\theta }  /\omega \|_\hh  +\|  \tilde{ \kappa}_{ \theta   } /\omega \|_\hh   \left\| \frac{ \mathfrak{D}_h   \tilde{ \kappa}_{  \theta}}{ \sqrt{\omega}} \right\|_\hh \right)  \right] ,   \label{J0002bound}
\end{align}
where we again used  \eqref{eq:Kappdiffrel}, and in addition
we used the estimation for $r\ge 0$
\begin{align*}
\frac{\left( r + |\tilde{Y_1}| + |\tilde{Y_2}| \right)^2}{(1 + r + |\tilde{Y_1}| + |\tilde{Y_2}|) (1 + r)} \le \frac{ r + |\tilde{Y_1}| + |\tilde{Y_2}| }{1 + r} \le 1 + |\tilde{Y_1}| + |\tilde{Y_2}| .
\end{align*}
The case $(p,q)=(2,0)$  is obtained analogously giving the same estimate $ J^{0,0}_{2,0}(h)  \leq \text{R.H.S. of } \eqref{J0002bound}$.

Now let $S=2$ and  $p+q=1$. Then  $m+n=1$ and  $(p,q),  (m,n) \in\{(0,1),(1,0)\}$. If
 $(p,q)=(0,1)$ and $(m,n)=(1,0)$, then we estimate
\begin{align}
& J^{1,0}_{0,1}(h, K^{(1,0)}) \nonumber   \\
& \le \Bigg\{ \int_{\underline{\R}^3} \frac{d \tilde{Y}}{ |\tilde{Y}|^2}
\sup_{r \geq 0} \Bigg[ \Big\|( - \Delta + 1 + r )^{-1/2} \nonumber  \\
& \times \big( h^{-1} ( e^\theta \widehat{w}_{1,1}^{(I)}    (g,\theta+h,\alpha) -
e^\theta \widehat{w}_{1,1}^{(I)}(g,\theta,\alpha))  - D_{1,1}^{(I)}(g,\theta,\alpha)\big)(K, \tilde{Y})  \nonumber   \\
 & \times ( - \Delta + 1 + r + |\tilde{Y}|)^{-1/2} \Big\|^2  \left( r + |\tilde{Y}| \right) \Bigg]  \Bigg\}^{1/2} \nonumber   \\
&  \leq |g|^2 N  \left(  \|( \tilde{ \kappa}_{\theta + h} - \tilde{\kappa}_\theta )/\omega \|_\hh  \| \partial_\theta  \kappa_\theta \|_\infty +
  \|  \tilde{\kappa}_{\theta + h}  /\omega \|_\hh \| \mathfrak{D}_h \kappa_\theta  \|_\infty +\|  \kappa_{\theta }  \|_\infty      \| \mathfrak{D}_h \tilde{\kappa}_\theta /\omega \|_\hh  \right) , \label{J1001bound}
\end{align}
where we again used  \eqref{eq:Kappdiffrel}. The remaining cases for   $S=2$ and $p+q=1$   are  estimated analogously
giving a bound  as \eqref{J1001bound}
\begin{align*}
& J^{1,0}_{1,0}(h, K^{(1,0)}) ,  \ J^{0,1}_{0,1}(h, K^{(0,1)})  ,   J^{0,1}_{1,0}(h, K^{(0,1)})\\
&    \leq |g|^2 N \sup_{\sigma,\tau \in \{ \kappa, \tilde{\kappa} \} } \left(  \|( \tau_{\theta + h} - \tau_\theta )/\omega \|_\hh  \| \partial_\theta  \sigma_\theta \|_\infty +
  \|  \tau_{\theta + h}  /\omega \|_\hh \| \mathfrak{D}_h \sigma_\theta  \|_\infty +\|  \sigma_{\theta }  \|_\infty      \| \mathfrak{D}_h \tau_\theta /\omega \|_\hh  \right) ,
\end{align*}
where the supremum takes care of the different cases, which may occur.

It remains to consider the  case $S=2$ with $p=q=0$.  Then  there is no integral and we have  $(m,n)=(0,2)$, $(m,n)=(1,1)$, or  $(m,n)=(2,0) $.
Now  we estimate the case where  $(m,n)=(1,1)$. Then
\begin{align}
& J^{1,1}_{0,0}(h, K^{(1,1)})  \nonumber  \\
 & \leq \sup_{r \geq 0} \Bigg[ \Big\|( - \Delta + 1 + r )^{-1/2} ( h^{-1} ( e^\theta \widehat{w}_{1,1}^{(I)}    (g, \theta+h,\alpha) -
e^\theta \widehat{w}_{1,1}^{(I)}(g,\theta,\alpha))  - D_{1,1}^{(I)}(g, \theta,\alpha))({K}, \tilde{K})\nonumber   \\
 & \times ( - \Delta + 1 + r )^{-1/2} \Big\| \nonumber  \\
& \le  |g|^2 N  \left[  \|  \kappa_{\theta + h} - \kappa_\theta  \|_\infty  \| \partial_\theta  \tilde{\kappa}_\theta \|_\infty+
  \|  \kappa_{\theta + h}  \|_\infty  \| \mathfrak{D}_h \tilde{\kappa}_\theta  \|_\infty +  \| \mathfrak{D}_h \kappa_\theta  \|_\infty \|  \tilde{\kappa}_{\theta }  \|_\infty  \right]   ,  \label{J0200bound}
\end{align}
where we again used  \eqref{eq:Kappdiffrel}.
The remaining cases for   $S=2$ and $(p,q)=(0,0)$, i.e., $(m,n)=(0,2)$ and $(m,n)=(2,0)$  are  estimated analogously
to  \eqref{J0200bound}
giving the   bound
\begin{align*}
 & J^{0,2}_{0,0}(h, K^{(0,2)}) ,  \ J^{2,0}_{0,0}(h, K^{(2,0)})   \\
&  \leq  |g|^2 N  \sup_{\tau \in \{ \kappa, \tilde{\kappa} \} } \left[  \|  \tau_{\theta + h} - \tau_\theta  \|_\infty  \| \partial_\theta  \tau_\theta \|_\infty+
  \|  \tau_{\theta + h}  \|_\infty  \| \mathfrak{D}_h \tau_\theta  \|_\infty +  \| \mathfrak{D}_h \tau_\theta  \|_\infty \|  \tau_{\theta }  \|_\infty  \right]    .
\end{align*}
The above estimates  and Hypothesis  \ref{kappa}, specifically \eqref{convkappa3}--\eqref{convkappa3}, imply the convergence of $J^{m,n}_{p,q}(h,K^{m,n})$ to zero as  $h \to 0$
uniformly in $K^{m,n} \in B_1^{m+n}$. The analyticity of   \eqref{eq:defofW} in $\theta$  now follows in view of \eqref{eq:contsigmainitial12}.
\end{proof}

\section{Renormalization Transformation}
\label{sec:ren:def}

In this section we define the renormalization transformation as in \cite{BCFS}.
We follow very closely the exposition of Section 8 in \cite{HasHer11-2}, which  uses  results from \cite{HasHer11-1}.
Let $0<\xi<1$ and $0 < \rho < 1$.
For $w \in \mathcal{W}_\xi$ we define the analytic function
$$E_\rho[w](z) := \rho^{-1} E[w](z) := - \rho^{-1} w_{0,0}(z,0) = - \rho^{-1} \langle \Omega , H(w(z)) \Omega \rangle$$
 and the set
$$
U[w] := \{ z \in D_{1/2} : | E[w](z) | < \rho / 2 \} .
$$

\begin{lemma} \label{renorm:thm3} Let $0< \rho\leq 1/2$. Then for all $w \in \mathcal{B}(\rho/8, \rho/8, \rho/8 )$, the
function $E_{\rho}[w]: U[w] \to D_{1/2}$ is an analytic bijection.
\end{lemma}
For a proof of the lemma see \cite{BCFS} or \cite{HasHer11-1} (Lemma 21).
For the smooth functions $\chi_1, \chib_1 \in C^\infty(\R_+;[0,1])$  satisfying   \eqref{propofchi}  we define
$$
\chi_\rho(\cdot) = \chi_1(\cdot /\rho) \quad , \quad \chib_\rho(\cdot) = \chib_1(\cdot /\rho) \; ,
$$
and use the abbreviation $\chi_\rho = \chi_\rho(H_{\rm f})$ and $\chib_\rho = \chib_\rho(H_{\rm f})$. As before,
it should be clear from the context whether $\chi_\rho$ or $\chib_\rho$ denotes a function or an operator.

\begin{lemma} \label{renorm:thm1} Let $0 < \rho \leq 1/2$. Then for all $w \in \mathcal{B}(\rho/8,\rho/8,\rho/8)$, and
all $z \in D_{1/2}$ the pair of operators $(H(w(E_\rho[w]^{-1}(z)),H_{0,0}(E_\rho[w]^{-1}(z)))$ is a Feshbach pair for $\chi_\rho$.
\end{lemma}

A proof of Lemma \ref{renorm:thm1} can be found in   \cite{BCFS}  or  \cite{HasHer11-1}  (Lemma 23 and Remark 24).
The definition of the renormalization transformation involves a scaling transformation $S_\rho$ which scales the energy value $\rho$ to the value 1.
It is defined as follows.
For operators $A \in \mathcal{B}(\FF)$ set
$$
S_\rho(A) = \rho^{-1} \Gamma_\rho A \Gamma_\rho^* ,
$$
where $\Gamma_\rho$ is the unitary dilation on $\FF$ which is uniquely determined by
\begin{align*}
& \Gamma_\rho a^\#(k) \Gamma_\rho^* = \rho^{-3/2} a^\#(\rho^{-1} k)  , \quad \Gamma_\rho \Omega = \Omega .
\end{align*}
It is easy to verify that $\Gamma_\rho H_{\rm f} \Gamma_\rho^* = \rho H_{\rm f}$. This implies  $\Gamma_\rho \chi_\rho \Gamma_\rho^* = \chi_1$.
We are now ready to define the renormalization transformation, which is well defined by  Lemmas \ref{renorm:thm3} and \ref{renorm:thm1}.

\begin{definition} Let $0 < \rho \leq 1/2$. For $w\in \mathcal{B}(\rho/8,\rho/8,\rho/8)$ we define the renormalization transformation
\begin{equation} \label{eq:defofrenorm}
\left( {R}_\rho H(w) \right)(z) := S_\rho F_{\chi_{\rho}}(H(w(E_\rho[w]^{-1}(z)),H_{0,0}(E_\rho[w]^{-1}(z))) \upharpoonright \HH_{\rm red}
\end{equation}
where $z \in D_{1/2}$.
\end{definition}

Let us cite the following theorem   \cite[Theorem 8.4]{HasHer11-2}, whose proof is based on ideas from  \cite{BCFS}  and  \cite{HasHer11-1}. It states that the renormalization transformation on the operators  induces a renormalization 
transformation on the integral kernels.  

\begin{theorem} \label{thm:maingenerala} Let $0<\rho \leq 1/2$ and $0< \xi \leq 1/2$.
For
$w\in \mathcal{B}(\rho/8,\rho/8,\rho/8)$ there exists a unique integral kernel $\mathcal{R}_\rho(w) \in \WW_{\xi}$ such that
\begin{equation} \label{eq:defofrenormkern}
({R}_\rho H(w))(z) = H(\mathcal{R}_\rho(w)(z)) .
\end{equation}
If $w$ is symmetric then also $\mathcal{R}_\rho(w)$ is symmetric. If $w(z)$ is invariant under rotations for all $z \in D_{1/2}$
than also $\mathcal{R}_\rho(w)(z)$ is invariant under rotations for all $z \in D_{1/2}$.
\end{theorem}

Next we  cite the following theorem   \cite[Theorem 8.5]{HasHer11-2}, whose proof can be found in    \cite[Theorem 9.1]{HasHer11-1}.

\begin{theorem} \label{codim:thm1} For any positive numbers $\rho_0 \leq 1/2$ and $\xi_0 \leq 1/2$ there exist numbers $\rho, \xi, \epsilon_0$ satisfying
$\rho \in (0, \rho_0]$, $\xi \in (0, \xi_0]$, and $0 < \epsilon_0 \leq \rho/8$
such that the following property holds,
\begin{equation} \label{codim:thm1:eq}
\mathcal{R}_\rho : \mathcal{B}_0(\epsilon, \delta_1, \delta_2 ) \to
\mathcal{B}_0( \epsilon + \delta_2/2 , \delta_2/2 , \delta_2/2 ) \quad , \quad \forall \ \epsilon, \delta_1, \delta_2 \in [0, \epsilon_0) .
\end{equation}
\end{theorem}

Using the contraction property of Theorem  \ref{codim:thm1}
we can iterate the renormalization transformation.  To this end we
introduce the following Hypothesis.

\setcounter{hypothesisR}{17}

\begin{hypothesisR} \label{hypR}
Let $\rho, \xi, \epsilon_0$ are positive numbers such that the contraction property \eqref{codim:thm1:eq}
holds and $\rho \leq 1/4$, $\xi \leq 1/4$ and $\epsilon_0 \leq \rho/8$.
\end{hypothesisR}

Hypothesis  \eqref{hypR}
allows us to iterate the renormalization transformation as follows,
\begin{equation} \nonumber
\mathcal{B}_0(\sfrac{1}{2}\epsilon_0 , \sfrac{1}{2} \epsilon_0 , \sfrac{1}{2} \epsilon_0 ) \stackrel{\mathcal{R}_\rho}{\longrightarrow}
\mathcal{B}_0( [ \sfrac{1}{2}+ \sfrac{1}{4} ] \epsilon_0 , \sfrac{1}{4} \epsilon_0 , \sfrac{1}{4} \epsilon_0 ) \stackrel{\mathcal{R}_\rho}{\longrightarrow} \cdots
\mathcal{B}_0( \Sigma_{l=1}^n \sfrac{1}{2^l} \epsilon_0 , \sfrac{1}{2^n} \epsilon_0 , \sfrac{1}{2^n} \epsilon_0 )  \stackrel{\mathcal{R}_\rho}{\longrightarrow} \cdots   .
\end{equation}

Finally, let us cite the following theorem  \cite[Theorem 8.6]{HasHer11-2}, whose proof is based on ideas from  whose proof is  again based on ideas from  \cite{BCFS}  and  \cite{HasHer11-1}. 

\begin{theorem} \label{thm:bcfsmain} Assume Hypothesis \eqref{hypR}.
There exist functions
\begin{align*}
&e_{(0)}[ \cdot ] : \mathcal{B}_0(\epsilon_0/2,\epsilon_0/2,\epsilon_0/2) \to D_{1/2} \\
&\psi_{(0)}[ \cdot ] : \mathcal{B}_0(\epsilon_0/2,\epsilon_0/2,\epsilon_0/2) \to \FF
\end{align*}
such that the following holds.
\begin{enumerate}[(a)]
\item For all $w \in \mathcal{B}_0(\epsilon_0/2,\epsilon_0/2,\epsilon_0/2)$,
\begin{align*} 
{\rm dim} \ker \{ H(w(e_{(0)}[w]) \}  \geq   1 ,
\end{align*} 
and $\psi_{(0)}[w]$ is a nonzero element in the kernel of $H(w(e_{(0)}[w])$.
\item If $w$ is symmetric and $-1/2 < z < e_{(0)}[w]$, then $H(w(z))$ is bounded invertible.
\item \label{anaeigrg} Let $S$ be an open subset of $\C^\nu$. Suppose
\begin{eqnarray*}
w(\cdot, \cdot) : &&S \times D_{1/2} \to \WW_\xi^\# \\
&&(s,z) \mapsto w(s,z)
\end{eqnarray*}
is an analytic  function such that $w(s)(\cdot) := w(s, \cdot)$ is in $\mathcal{B}_0(\epsilon_0/2,\epsilon_0/2,\epsilon_0/2)$.
Then $s \mapsto e_{(0)}[w(s)]$ and $\psi_{(0)}[w(s)]$ are  analytic   functions.
\end{enumerate}
\end{theorem}

\section{Main Theorem}

\label{sec:prov}

In this section, we prove Theorem \ref{thm:main1}, the main result of this paper.
Its proof is based on Theorems \ref{thm:inimain1}, \ref{codim:thm1}and \ref{thm:bcfsmain}.

\begin{proof}[ Proof of Theorem \ref{thm:main1}.]  By Theorem \ref{codim:thm1} we can   choose $\rho, \xi, \epsilon_0$ such that Hypothesis  \eqref{hypR}  holds. Let $g_{\rm b}$,
$\theta_{\rm b}$, and $\alpha_{\rm b}$ be positive numbers  such that    the conclusion
of Theorem \ref{thm:inimain1} holds for $\delta_1=\delta_2=\delta_3 = \epsilon_0/2$.
Assume   $( g,\theta,\alpha) $ is an element of  $ D_{g_{\rm b}}
\times D_{\theta_{\rm b}} \times D_{\alpha_{\rm b}}$

(i)
 It  follows from
Theorem \ref{thm:bcfsmain} (a)
 that
$\psi_{(0)}[\widehat{w}^{(0)}(g,\theta,\alpha)]$ is a   nonzero element in the kernel of $H_{g,\theta,\alpha}^{(0)}(e_{(0)}[\widehat{w}^{(0)}(g,\theta,\alpha)]) = H(\widehat{w}^{(0)}(g,\theta,\alpha,e_{(0)}[\widehat{w}^{(0)}(g,\theta,\alpha)]))$.
From Theorem \ref{thm:inimain1}
the Feshbach property,  cf. Theorem  \ref{thm:fesh},  and the transformation  \eqref{eq:eff23} it follows that
\begin{equation} \label{eq:psiis}
\psi_{g}(\theta,\alpha) := Q_{\chi^{(I)}}(g,\theta,\alpha, e_{(0)}[\widehat{w}^{(0)}(g,\theta,\alpha)])  ( \varphi_{{\rm at},\theta,\alpha} \otimes \psi_{(0)}[\widehat{w}^{(0)}(g,\theta,\alpha)] ) ,
\end{equation}
is nonzero and an eigenvector of $\widehat{H}_{g}(\theta,\alpha)$ with eigenvalue
$$
\widehat{E}_g(\theta,\alpha) := \widehat{E}_{\rm at}(\theta,\alpha) +
e_{(0)}[\widehat{w}^{(0)}(g,\theta,\alpha)] .
$$
It follows from \eqref{eq:defoftrafoham}  and  \eqref{trivialtrafo}  that $\psi_g(\theta,\alpha)$ is an
eigenvector of $H_g(\theta,\alpha)$ with eigenvalue
\begin{align} \label{trafoeiginproof}
E_g(\theta,\alpha) = e^{-\theta} \widehat{E}_g(\theta,\alpha) + c_{N,g,\kappa}  .
\end{align}
By Theorem \ref{thm:inimain1}, we know that $(g,\theta,\alpha,z) \mapsto \widehat{w}^{(0)}(g,\theta,\alpha,z)$ is analytic and hence by Theorem \ref{thm:bcfsmain} \eqref{anaeigrg} it follows that
 $g \mapsto \widehat{E}_g(\theta,\alpha)$  and $g \mapsto \psi_{(0)}[\widehat{w}^{(0)}(g,\theta,\alpha)] $
are analytic. This implies  by   \eqref{trafoeiginproof} that
$g \mapsto E_g(\theta,\alpha)$ is analytic and that
 $(g,\theta,\alpha)  \mapsto \psi_g(\theta,\alpha) $
 is analytic because of the analyticity
of $(g,\theta,\alpha, z) \mapsto Q_{\chi^{(I)}}(g,\theta,\alpha,z)$ and  \eqref{eq:psiis}.
This shows (i). \\

(ii)
If $(g,\theta,\alpha)\in D_{g_0} \times D_{\theta_{\rm b}} \times D_{\alpha_{\rm b}}$ is real
the kernel
$\widehat{w}^{(0)}(g,\theta,\alpha)$ is symmetric by  Theorem \ref{thm:inimain1}.
It now follows from Theorem \ref{thm:bcfsmain} (b) that $\widehat{H}^{(0)}_{g}(\theta,\alpha,z)$ is bounded invertible if $z \in (-\frac{1}{2}, e_{(0,\infty)}[\widehat{w}^{(0)}(g,\theta,\alpha)] )$. Applying the Feshbach property, Theorem  \ref{thm:fesh},
it follows that $\widehat{H}_{g}(\theta,\alpha) - \zeta$ is bounded invertible for $\zeta \in ( \widehat{E}_{\rm at}(\theta,\alpha) -\frac{1}{2}, \widehat{E}_{\rm at}(\theta,\alpha) + e_{(0,\infty)}[\widehat{w}^{(0)}(g,\theta,\alpha)] )$.
We we will show below  that there exists a constant $C$ such that  for all  $\zeta \in (-\infty,  \widehat{E}_{\rm at}(\theta,\alpha) - 1/2)$
the following estimate holds.
\begin{align} \label{estforinfbound}
& \| ( \widehat{H}_0(\theta,\alpha)  - \zeta )^{-1} \widehat{W}_{g}(\theta,\alpha) \| \leq C |g|  .
\end{align}
This estimate implies for $g$ sufficiently small and $\zeta \in (-\infty,  \widehat{E}_{\rm at}(\theta,\alpha) - 1/2)$ the bounded
 invertibility of $\widehat{H}_{g}(\theta,\alpha) - \zeta$. We conclude that for  $|g|$ sufficiently small
 $E_g(\theta,\alpha) = \inf \sigma(H_{g}(\theta, \alpha)$ for real  $(g,\theta,\alpha)\in D_{g_0} \times D_{\theta_{\rm b}} \times D_{\alpha_{\rm b}}$, and hence the claim (ii) follows.
It remains to show the bound \eqref{estforinfbound} (whose proof is analogous that of  \eqref{secondfreeintbound}).
Let
 $\zeta \leq \widehat{E}_{\rm at}(\theta,\alpha) - 1/2$.
\begin{align} \label{estforinfbound1}  \| ( \widehat{H}_0(\theta,\alpha)  - \zeta )^{-1} \widehat{W}_{g}(\theta,\alpha) \|
& \leq  \| ( \widehat{H}_0(\theta,\alpha)  - \zeta )^{-1}   ( \widehat{H}_0(\theta,\alpha)  -  \widehat{E}_{\rm at}(\theta,\alpha) + e^\theta /2) \|
\nonumber \\
& \times \| ( \widehat{H}_0(\theta,\alpha)  - \widehat{E}_{\rm at}(\theta,\alpha) + e^\theta /2 )^{-1} \widehat{W}_{g}(\theta,\alpha) \|  .
\end{align}
To estimate the first factor on the right hand side we use
\begin{align} \label{estforinfbound2}
& \| ( \widehat{H}_0(\theta,\alpha)  - \zeta )^{-1}   ( \widehat{H}_0(\theta,\alpha)  -  \widehat{E}_{\rm at}(\theta,\alpha) + e^\theta /2) \|  \nonumber \\
& \leq  \sup_{\lambda \geq  0}  \left| \frac{ \lambda +  e^\theta  /2 }{\lambda + \widehat{E}_{\rm at}(\theta,\alpha) - \zeta} \right| \leq  \sup_{\lambda \geq  0}  \left|  \frac{ \lambda   +  e^\theta  /2 }{\lambda  + 1/2} \right|  \leq 1 +   | e^\theta| .
\end{align}
Now we estimate the second factor on the right hand  side of  \eqref{estforinfbound1}   using unitarity and  find
\begin{align} \label{estforinfbound3}
 & \| ( \widehat{H}_0(\theta,\alpha)  - \widehat{E}_{\rm at}(\theta,\alpha) + e^\theta /2 )^{-1} \widehat{W}_{g}(\theta,\alpha) \|
 =  \| ( {H}_0(0,0)  - {E}_{\rm at}(0,0) +  1/2 )^{-1} {W}_{g}(0,0) \|  \nonumber \\
& \leq \| ( {H}_0(0,0)  - {E}_{\rm at}(0,0) +  1/2 )^{-1} (-\Delta + H_{\rm f} +  1)\|
\| (-\Delta + H_{\rm f} +  1)^{-1} {W}_{g}(0,0) \| .
\end{align}
Now to bound the first factor on the  right hand side of \eqref{estforinfbound3}
we use that $V_\theta$ is infinitesimally bounded with respect to the Laplacian
and to bound the second factor we use \eqref{uniformanabound2}.
Thus inserting \eqref{estforinfbound2}  and \eqref{estforinfbound3}  into  \eqref{estforinfbound1}, we arrive at  \eqref{estforinfbound}.
Finally,  the fact that   the  eigenvalue $E_g(\theta,\alpha)$ is simple  follows   from \cite{HasHer12-2}.
\end{proof}

\section*{Acknowledgements}

 D. H. wants to thank I. Herbst and M. Griesemer  for interesting conversations on the subject.

\appendix

\section{Results from   complex Analysis}  \label{ana:res}

We use the following definition  of analyticity in several variables from  \cite{Hoe90}.

\begin{definition} Let $R \subset \C^\nu$ be an open set. A function $f : R \to \C$ is called {\bf analytic} if
for every $j=1,...,\nu$  and $z \in R$  the limit
$$
\lim_{h \to 0} \frac{f(z+ h e_j) - f(z)}{h}
$$
exists, where $e_j$ denotes the $j$-th unit vector in $\C^\nu$ and $h \in \C$.
If $X$ is a Banach space we say that a  function $f : R \to X$ {\bf analytic} if
for every $j=1,...,\nu$  and $z \in R$  the limit
$$
\lim_{  h \to 0} \frac{f(z+ h e_j) - f(z)}{h}  .
$$
exists in $X$.
\end{definition}

To show that Banach space valued functions are analytic the  criterion in the following Lemma  is useful.
To formulate it we introduce the following definition.

\begin{definition} A  {\bf fundamental subset} of a Banach space $X$ is  a dense set of  the unit ball.
\end{definition}

\begin{lemma}\label{critanafu}  Let $X$ be a Banach space and $R \subset \C^\nu$ open. The $X$ valued function $u(\kappa)$ is analytic in $\kappa$ if and only if each $\kappa \in R$ has a neighborhood in which $\| u(\kappa) \|$ is bounded and $f(u(\kappa))$ is analytic for all $f$ in a fundamental subset of the dual space $X^*$.
\end{lemma}

\begin{proof} For a proof see \cite{Kat95}  Page 365.
\end{proof}

For operators one obtains similarly the  criterion in the following.

\begin{lemma} \label{critanaop}
Let $X$ and $Y$ be Banach spaces and $R \subset \C^\nu$ open. $T(\kappa) \in \mathcal{B}(X,Y)$ is analytic on $R$  if and only if each $\kappa \in R$ has a neighborhood in which $T(\kappa)$
is bounded and $g( T(\kappa) u )$ is analytic  for every $u$ in a fundamental subset of $X$ and every $g$ in a fundamental subset of the dual  $Y^*$.
\end{lemma}

\begin{proof} For a proof see \cite{Kat95}  Page 365.
\end{proof}

We will work with the following definition of an analytic family of operators, cf.   \cite[Page 14]{ReeSim4}.

\begin{definition} A operator-valued function $T(\beta)$ on an open  set  $R \subset \C^\nu$ is called an {\bf analytic family} or an {\bf analytic family in the sense of Kato}
if and only if:
\begin{itemize}
\item[(i)] For each $\beta \in R$, $T(\beta)$ is closed and has a nonempty resolvent set.
\item[(ii)]  For every $\beta_0 \in R$, there is a  $\lambda_0 \in \rho(T(\beta_0))$ so that $\lambda_0 \in \rho(T(\beta))$ for $\beta$ near $\beta_0$ and $(T(\beta) - \lambda_0)^{-1}$ is an analytic operator-valued function of $\beta$ near $\beta_0$.
\end{itemize}
\end{definition}

We will work with the following definition of an analytic family of type (A) operators, cf.    \cite[Page 16]{ReeSim4}.

\begin{definition} \label{defanaA}
Let $R \subset \C^\nu$ be open and let
$T(\beta)$,  a closed operator with nonempty resolvent set,
be given for each $\beta \in R$. We say that $T(\beta)$ is
an {\bf analytic family of type (A)} if and only if
\begin{itemize}
\item[(i)] The operator domain of $T(\beta)$ is some set $\mathcal{D}$ independent of $\beta$.
\item[(ii)] For each $\psi \in \mathcal{D}$, $T(\beta) \psi$ is a vector-valued analytic function of $\beta$.
\end{itemize}
\end{definition}

One can show that every family of type (A) is an   analytic family  in the sense of Kato, see for example
 \cite[Theorem XII.9]{ReeSim4}   or \cite{Kat95} pp. 375--381.

\begin{lemma}\label{lemtypaAest2}  Let $T(\kappa)$ be an analytic family of type (A) on an open  connected subset  $R \subset \C^\nu$.
Let  $K$ be a compact and subset of $R$ and $\kappa_0 \in K$ . Then  there exist  $a$ and $b$ such that for all $\kappa \in K$  and $\psi \in \mathcal{D}(T(\kappa_0))$
\begin{align}  \label{typaAelemineq}
\| T(\kappa) \psi \| \leq a \|  T(\kappa_0) \psi \|  + b \| \psi \| .
\end{align}
For  $ z \in \rho(T(\kappa_0))$  there exists a constant $C$ such that for all  $\kappa \in K$
\begin{align}  \label{typaAelemineq2}
\| T(\kappa) ( T(\kappa_0) -  z )^{-1} \|  \leq C  .
 \end{align}
\end{lemma}
\begin{proof} Inequality  \eqref{typaAelemineq}  is   shown in Kato page 376, cf.  (2.2).  Inequality  \eqref{typaAelemineq2}  follows from \eqref{typaAelemineq}
by observing that there exists a constant $C$ such that  for any $\psi \in \HH$
\begin{align*}  
\| T(\kappa) ( T(\kappa_0) -  z )^{-1}  \psi \| & \leq a \|  T(\kappa_0)  ( T(\kappa_0) -  z )^{-1} \psi \|  + b \|  ( T(\kappa_0) -  z )^{-1} \psi \| \\
 & \leq a \|  \psi   + z  ( T(\kappa_0) -  z )^{-1} \psi \|  + b \|  ( T(\kappa_0) -  z )^{-1} \psi \| \\
 & \leq C  \|  \psi \| .
\end{align*}
\end{proof}

We will use the following lemma to control the so called reduced resolvent.
It can be viewed as a generalization of Theorem XII.7   in \cite{ReeSim4}.

\begin{lemma} \label{Prop27inGriesemerHasler} Let $R \ni s \mapsto T(s)$ be an analytic family. Suppose
there exists an isolated non-degenerate eigenvalue $E(s)$ with eigenprojection
$P(s)$ depending analytically on $s$. Let $\overline{P}(s) = 1 - P(s)$ and let
\begin{align*}
\Gamma   := \{(s,z) \in R \times \C : T(s) - z   \text{ is a bijection from } \mathcal{D}(T(s)) \cap \ran \overline{P}(s) \text{ to } \ran \overline{P}(s)  \}
\end{align*}
Then $\Gamma$ is open and $(s,z) \mapsto (T(s) - z)^{-1} \overline{P}(s)$ is analytic on $\Gamma$.
\end{lemma}
\begin{proof}
A proof can be found in  \cite[Prop. 27]{GriHas09}.
\end{proof}

\begin{theorem}(Kato-Rellich) \label{katorellich} Let $R \subset \C^\nu$ be open and let $T(\beta)$ be an analytic family in the sense of Kato for $\beta \in R$.
Let $E_0$ be a nondegenerate discrete eigenvalue of $T(\beta_0)$.
\begin{itemize}
\item[(a)]
Then, for $\beta$ near $\beta_0$, there is
exactly on point $E(\beta)$ of $\sigma(T(\beta))$ near $E_0$ and this point is isolated and nondegenerate.
$E(\beta)$ is analytic function of $\beta$ for $\beta$ near $\beta_0$, and there is an analytic eigenvector $\Omega(\beta)$
for $\beta$ near $\beta_0$. If $T(\beta)$ is self-adjoint for $\beta-\beta_0$ real, then $\Omega(\beta)$ can be chosen to be normalized
for $\beta-\beta_0$ real.
\item[(b)]
There exist a neighborhood $N$ of $\beta_0$ and an   $\epsilon > 0$ such that
$$
P(\beta) = \frac{1}{ 2 \pi i }    \ointctrclockwise_{|z-E_0|} (z - T(\beta)  )^{-1} dz
$$
exits for all $\beta \in N$, is a  projection with one dimensional range, depends analytically on $\beta$  and projects onto the eigenvector $\Omega(\beta)$.
\end{itemize}
\end{theorem}
\begin{proof} (a) Is proven in  Theorem XII.8 \cite{ReeSim4}. (b) Follows from the proof given in  \cite{ReeSim4} (see also Theorem XII.5)
\end{proof}

We shall use the following theorem, whose prove can be found in \cite{GriHas09}.
To formulate the theorem we follow closely the exposition given there.
Let $\HH$ be a separable Hilbert space.
We consider families of (unbounded) closed operators $T(s) : \mathcal{D}(T(s)) \subset \HH \to \HH$, $s \in R$, where $R \subset \C^\nu$ is open,
symmetric with respect to complex conjugation and $R \cap \R^\nu \neq \emptyset$.

If there exists a point $s_0 \in R $
such that $e(s_0)$ is an isolated, non-degenerate eigenvalue of $T(s_0)$,
then by the  Kato-Rellich theorem of analytic perturbation theory, cf.  \cite[Theorem XII.8]{ReeSim4}, there is exactly one point $e(s)$ of $\sigma(T(s))$ near $e(s_0)$,
for $s$ near $s_0$, and this point is a non-degenerate eigenvalue of $T(s)$. Moreover,
for $s$ near $s_0$ there is an analytic projection $p(s)$  onto the eigenvector $e(s)$, which is
given by the Riesz projection, cf. \cite[Theorem XII.5]{ReeSim4},
\begin{align} \label{rieszproj}
p(s) = \frac{1}{2\pi i}  \ointctrclockwise_{|e(s)-z|=\epsilon} (z-T(s))^{-1} dz ,
\end{align}
for $\epsilon > 0$ sufficiently small. We set $\overline{p}(s) = 1 - p(s)$.

\begin{theorem} \label{thm:veryIII} Let $R \subset \C^\nu$ be open and suppose $s \mapsto T(s)$ is  on  $V$ an analytic family of type (A) with with $T(s)^* = T(\overline{s})$
for all $s \in R$. Assume that for $s_0 \in R \cap \R^\nu$
the number $e(s_0) = \inf \sigma(T(s_0))$ is a non-degenerate isolated eigenvalue of $T(s_0)$.
Let $\delta = \dist (  e(s_0), \sigma(T(s_0)) \setminus \{ e(s_0) \} )$.
Then there exists a neighborhood $\mathcal{U} \subset R \times \C$ of $(s_0,e(s_0))$ such that for all
$(s,z) \in \mathcal{U}$, one has  $|e(s)-z|< \delta/2$,   $z - r  \subset  \rho(T(s)|_{\ran \overline{p}(s)})$ for all $r \geq 0$,  and
\begin{align*}
\sup_{(s,z) \in \mathcal{U}} \sup_{r \geq 0} \left\| \frac{r+1}{T(s) - z + r } \overline{p}(s) \right\| < \infty .
\end{align*}

\end{theorem}

\begin{proof} A proof can be found in Corollary 2 and Corollary 3 of  \cite{GriHas09}.
\end{proof}

\section{ Elementary Estimates and  Identities in Fock spaces}

\label{sec:estfock}

To give a precise meaning to expressions which occur in \eqref{eq:defhmn11} and \eqref{eq:defhlinemn}, we introduce the following.
For $\psi \in \FF$ having finitely many particles we have
\beqn \label{eq:defofa}
\left[ a(K_1) \cdots a(K_m) \psi \right]_n(K_{m+1},...,K_{m+n}) = \sqrt{\frac{(m+n)!}{n!}} \psi_{m+n}(K_{1},...,K_{m+n}) ,
\eeqn
for all $K_1,...,K_{m+n} \in \underline{\R}^3 := \R^3 \times \Z_2$, and
using Fubini's theorem it is elementary to see that the vector valued map
 $(K_1,...,K_m) \mapsto a(K_1) \cdots a(K_m) \psi$ is
an element of $L^2((\underline{\R}^{3})^m; \FF)$. The following lemma states the well
known pull-through formula.
For a proof see for example \cite{BacFroSig98-2,HasHer11-1}.
\begin{lemma} \label{lem:pullthrough}
Let $f : \R_+ \to \C$ be a bounded measurable function. Then for all $K \in \R^3 \times \Z_2$
$$
f(H_{\rm f}) a^*(K) = a^*(K) f(H_{\rm f} + \omega(K) ) , \quad a(K) f(H_{\rm f}) = f(H_{\rm f} + \omega(K) ) a(K) .
$$
\end{lemma}
Let $w_{m,n}$ be function on $\R_+ \times ({\underline{\R}^{3})}^{n+m}$ with values
in the linear operators of $\HH_{\rm at}$ or with values in $\C$.
To such a function we associate the quadratic form
\begin{eqnarray*}
q_{w_{m,n}}(\varphi,\psi) := \int_{{(\underline{\R}^3)}^{m+n}} \frac{ dK^{(m,n)}}{|K^{(m,n)}|^{1/2}}
 \left\langle a(K^{(m)}) \varphi ,w_{m,n}(H_{\rm f}, K^{(m,n)}) a(\widetilde{K}^{(n)}) \psi \right \rangle ,
\end{eqnarray*}
defined  for all $\varphi$ and $\psi$ in $\HH_{\rm at} \otimes \FF$ or  $\FF$, respectively, for which the right hand side is defined as a complex number.
To associate an operator to the quadratic form we will use the following lemma.
\begin{lemma} \label{kernelopestimate} Let $\underline{X} = \R^3 \times \Z_2$. Then
\begin{eqnarray} \label{eq:defofH}
| q_{w_{m,n}}(\varphi,\psi) | \leq \| w_{m,n} \|_\sharp \| \varphi \| \| \psi \| ,
\end{eqnarray}
where
\begin{eqnarray*}
 \| w_{m,n} \|_\sharp^2 :=
\int_{\underline{X}^{m+n}} \frac{d K^{(m,n)}}{|K^{(m,n)}|^2} \sup_{r \geq 0} \left[ \|w_{m,n}(r,K^{(m,n)}) \|^2 \prod_{l=1}^m \left\{ r + \Sigma[K^{(l)}] \right\}
 \prod_{\widetilde{l}=1}^n \left\{ r + \Sigma[\widetilde{K}^{(\widetilde{l})}] \right\} \right] .
\end{eqnarray*}
\end{lemma}
\begin{proof} We set $P[K^{(n)}] := \prod_{l=1}^n ( H_{\rm f} + \Sigma[K^l])^{1/2}$ and insert 1's to obtain the trivial identity
\begin{align*}
 | q_{w_{m,n}}(\varphi,\psi) | &=
\Bigg| \int_{\underline{X}^{m+n}} \frac{d K^{(m,n)}}{|K^{(m,n)}|}
 \Big\langle
 P[K^{(m)}] P[K^{(m)}]^{-1} |K^{(m)}|^{1/2} a(K^{(m)}) \varphi , w_{m,n}(H_{\rm f} ,K^{(m,n)})
 \\
&
\times
P[\widetilde{K}^{(n)}] P[\widetilde{K}^{(n)}]^{-1} | \widetilde{K}^{(n)}|^{1/2} a(\widetilde{K}^{(n)}) \psi \Big\rangle \Bigg| .
\end{align*}
The lemma now follows using the Cauchy-Schwarz
inequality and the following well known identity for $n \geq 1$ and $\phi \in \FF$,
\begin{eqnarray}
\int_{\underline{X}^n}
d K^{(n)} | K^{(n)} | \left\| \prod_{l=1}^n \left[ H_{\rm f} + \Sigma[K^{(l)}] \right]^{-1/2} a(K^{(n)}) \phi \right\|^2 = \| 1_{N_{\rm op}  \geq n} \phi \|^2 \label{eq:trivialA}  ,
\end{eqnarray}
where
$N_{\rm op}$ is the number operator.
A proof of \eqref{eq:trivialA}  can for example be found in \cite{HasHer11-1} Appendix A.
\end{proof}

Provided the form $q_{w_{m,n}}$ is densely defined and $ \| w_{m,n} \|_\sharp$ is a finite real number,
then the form $q_{w_{m,n}}$ determines uniquely a bounded
linear operator  $\underline{H}_{m,n}(w_{m,n})$ such that
 $$
q_{w_{m,n}}(\varphi,\psi ) = \langle \varphi,\underline{H}_{m,n}(w_{m,n}) \psi \rangle ,
$$
for all $\varphi, \psi$ in the form domain of $q_{w_{m,n}}$. Moreover,
$\| \underline{H}_{m,n}(w_{m,n}) \|_{} \leq \| w_{m,n} \|_\sharp$.
Using the pull-through formula and Lemma \ref{kernelopestimate} it is easy to see
that for $w^{(I)}$, defined in \eqref{defofwI}, with $m+n=1,2$, the form
$$
q^{(I)}_{m,n}(\varphi, \psi) := q_{w_{m,n}^{(I)}}( \varphi , (H_{\rm f} + 1 )^{-\frac{1}{2}(m+n)} (-\Delta + 1 )^{-\frac{1}{2} \delta_{1,m+n}} \psi )
$$
is densely defined and bounded.
Thus we can associate a bounded linear operator $L_{m,n}^{(I)}$ such that
$q_{m,n}^{(I)}(\varphi, \psi) = \langle \varphi , L_{m,n}^{(I)} \psi \rangle$. This allows us to define
$$
\underline{H}_{m,n}(w_{m,n}^{(I)}) := L_{m,n}^{(I)} (H_{\rm f} + 1 )^{\frac{1}{2}(m+n)} (-\Delta + 1 )^{\frac{1}{2}\delta_{1,m+n}}
$$
as an operator in $\HH$.

\label{sec:appB}

In Theorem \ref{thm:wicktheorem}, below,  we state a generalized version of   Wick's  Theorem.
For $m,n \in \N_0$ let  $\underline{\mathcal{M}}_{m,n}$ denote the space of measurable functions on $\R_+ \times (\underline{\mathbb{R}}^3)^{m+n}$ with values
in the linear operators of $\HH_{\rm at}$. Let
$$
\underline{\mathcal{M}} = \bigoplus_{m+n=1,2} \underline{\mathcal{M}}_{m,n} .
$$
For  $w \in \underline{\mathcal{M}}$ we define
$$
\underline{W}[w] := \sum_{m+n=1,2} \underline{H}_{m,n}(w) .
$$
The following Theorem is from \cite[Theorem A4]{BacFroSig98-2} and \cite[Theorem 3.6]{BCFS}.

\begin{theorem} \label{thm:wicktheorem} Let $w \in \underline{\mathcal{M}}$
 and let $F_0,F_1,...,F_L \in   \underline{\mathcal{M}}_{0,0}     $. Then as a formal identity
$$
F_0(H_{\rm f}) \underline{W}[w] F_1(H_{\rm f}) \underline{W}[w] \cdots \underline{W}[w] F_{L-1}(H_{\rm f}) \underline{W}[w] F_L(H_{\rm f}) = \underline{H}( \widetilde{w}^{({\rm sym})} ) ,
$$
where
\begin{eqnarray}
\lefteqn{ \widetilde{w}_{M,N}(r;K^{(M,N)}) } \nonumber \\
& = &
\sum_{\substack{ m_1 + \cdots m_L = M \\ n_1+...n_L=N }} \sum_{\substack{ p_1, q_1,...,p_L,q_L: \\ m_l+p_l+n_l+q_l \geq 1 }} \prod_{l=1}^L
\left\{ \binom{ m_l + p_l}{ p_l} \binom{ n_l + q_l}{ q_l } \right\}
\nonumber \\
& & \times F_0(r + \tilde{r}_0)
\langle \Omega , \prod_{l=1}^{L-1} \left\{
\underline{W}_{p_l,q_l}^{m_l,n_l}[w]( r + r_l ; K_l^{(m_l,n_l)}) F_{l}(H_{\rm f} + r + \widetilde{r}_{l}) \right\} \nonumber \\
& &
\underline{W}_{p_L,q_L}^{m_L,n_L}[w]( r + r_L ; K_L^{(m_L,n_L)}) \Omega \rangle F_L(r + \widetilde{r}_L) , \label{eq:complicated}
\end{eqnarray}
with
\begin{align}
& K^{(M,N)} := (K_1^{(m_1,n_1)}, ... , K_L^{(m_L,n_L)}) , \quad K_l^{(m_l,n_l)} := (k_l^{(m_l)},\widetilde{k}_l^{(n_l)}) , \label{eq:KMNdef}
\\
& r_l := \Sigma[\widetilde{K}_1^{(n_1)}] + \cdots + \Sigma[\widetilde{K}_{l-1}^{(n_{l-1})}] + \Sigma[{K}_{l+1}^{(m_{l+1})}] + \cdots + \Sigma[{K}_L^{(m_L)}] , \label{eq:rldef}
\\
& \widetilde{r}_l := \Sigma[\widetilde{K}_1^{(n_1)}] + \cdots + \Sigma[\widetilde{K}_{l}^{(n_{l})}] + \Sigma[{K}_{l+1}^{(m_{l+1})}] + \cdots + \Sigma[{K}_L^{(m_L)}] . \label{eq:rltildedef}
\end{align}
\end{theorem}
A proof can be found in \cite{BacFroSig98-2} or 
\cite[Theorem 7.2]{HasHer11-1}.

\section{Analyticity Properties  for specific Operators}

\label{ana:con}

In this appendix we prove that specific operator families introduced in Section
\ref{sec:modalmainres}  are analytic.

\begin{lemma} \label{anatypeAtheta-12}
Let $V_C$ be given by  \eqref{dilatedhat2}. Then for any $Z\in \R$ and $e \in \R$  the following holds.
\begin{itemize}
\item[(a)] $V_C$ is invariant under rotations and permutations.
\item[(b)] $V_C$ is infinitesimally operator bounded with respect to $-\Delta$.
\item[(c)] If $N =  Z =1 $, then $\inf \sigma( -\Delta + V_C)$ is a non-degenerate isolated eigenvalue
of $-\Delta + V_C$.
\item[(d)]  For all $\theta \in \R$
\begin{align} \nonumber  \label{dilatedhat2333}
 V_{C,\theta}(x) &:= U_{\rm el}(\theta) V_C(x)  U_{\rm el}(\theta)^{-1} \\
& =   -  e^{-\theta} \sum_{j=1}^N \frac{Z e}{|x_j|}  +   e^{-\theta} \sum_{i < j } \frac{e^2}{|x_i - x_j| }  = e^{-\theta} V_C(x)
\end{align}
and $V_{C,\theta}$ is  operator bounded with respect to $-\Delta$. The map  $\theta \mapsto V_{C,\theta}(-\Delta+1)^{-1}$ has as a function
in the bounded operators on $\HH_{\rm at}$ an analytic extension to $\C$. For any $\theta_{\rm at} > 0$ and $a> 0$ there
exists a constant $C_{a,\theta}$ such that for all $\psi \in D(-\Delta)$
\begin{align} \label{eq:boundonc222}
 \left\| V_{C,\theta}(x)  \psi \right\| \leq a \| \Delta \psi \| + C_{a,\theta} \| \psi \| .
 \end{align}
\end{itemize}
In particular,   for $N=Z =1 $ Hypothesis \ref{potential}  holds for any $\theta_{\rm at} > 0$.
\end{lemma}
\begin{proof}  (a) is obvious to see. (b)
It follows from Kato's theorem, cf.  \cite[Thm X.16]{ReeSim2} and its   proof
that each term on the Coulomb potential is infinitesimally
bounded with respect to $-\Delta$. (c) The case $N=Z =1 $ is well known from
the  hydrogen atom. (d)  \eqref{dilatedhat2333} follows directly from  \eqref{trafoofxunidil}.
The analyticity statements  and the bound  \eqref{eq:boundonc222} now follow in view of   \eqref{dilatedhat2333} and (b).
\end{proof}

\begin{lemma} \label{anatypeAtheta} Suppose Hypothesis \ref{potential} holds for $\theta_{\rm at} > 0$.
 The map  $\theta \mapsto H_{\rm at}(\theta)$ with domain $\mathcal{D}(-\Delta)$ has an extension to   an analytic
family of type (A) on $\theta \in D_{\theta_{\rm at}}$ and satisfies $H_{\rm at}(\theta)^* = H_{\rm at}(\overline{\theta})$.
\end{lemma}

\begin{proof}
 Define for $\theta \in D_{\theta_{\rm at}}$
\begin{align} \label{dilatedhat0}
 \tilde{H}_{\rm at}(\theta) :=  e^{2\theta} H_{\rm at}(\theta) =  -\Delta + e^{2 \theta} V_\theta(x)
\end{align}
as an operator with domain  $\mathcal{D}(-\Delta)$.
It follows from \eqref{anabound(iv)} of  Hypothesis \ref{potential} that $\theta \mapsto \tilde{H}_{\rm at}(\theta) \psi $
is analytic on $D_{\theta_{\rm at}}$ for all $\psi \in \mathcal{D}(-\Delta)$.
Furthermore, it follows from  \eqref{eq:boundonc} of  Hypothesis \ref{potential}  that for any $a > 0$ there exists a $\tilde{C}_a$
such that
\begin{align} \label{eq:boundonc5555}
 \left\|e^{2 \theta}  V_\theta(x)  \psi \right\| \leq a \| \Delta \psi \| + \tilde{C}_a \| \psi \| , \quad \forall \psi \in \mathcal{D}(-\Delta) .
 \end{align}
Thus it follows  from the infinitesimal bound  \eqref{eq:boundonc5555} that for all $\theta \in D_{\theta_{\rm at}}$ the operator
$\tilde{H}_{\rm at}(\theta) $
is closed and has nonempty resolvent set (since the resolvent set of $-\Delta$ is nonempty). We conclude that  $\tilde{H}_{\rm at}(\theta)$
is an anlytic family of type (A), and hence so is $  H_{\rm at}(\theta)  = e^{-2\theta} \tilde{H}_{\rm at}(\theta)$.
Since $H_{\rm at}^*(\theta) = H_{\rm at}(\overline{\theta})$ holds for  all real $\theta \in D_{\theta_{\rm at}}$ it holds by analyticity
for  all  $\theta \in D_{\theta_{\rm at}}$.

\end{proof}

\begin{lemma} \label{anatypeAtheta2Field}
 Suppose Hypothesis \ref{potential} holds for $\theta_{\rm at} > 0$.
The map  $\theta \mapsto H_0(\theta)$ with domain $\mathcal{D}(-\Delta +H_{\rm f})$  is an analytic
family of type (A)   on  $D_{\theta_{\rm at}} \cap \{ \theta \in \C :  |{\rm Im} \theta |  < \pi/4\}$.
\end{lemma}
\begin{proof}
From  the infinitesimal bound \eqref{eq:boundonc}, we see   that  for $\psi \in \mathcal{D}(-\Delta + H_{\rm f})$
the map
\begin{align*}
\theta \mapsto H_0(\theta) \psi = e^{-2 \theta} (- \Delta)  \psi +  V_\theta \psi + e^{-\theta} H_{\rm f}\psi
\end{align*}
is analytic (since each summand is). Thus it remains to show closedness and existence of a  nonempty resolvent set.
To this end we consider first $ -\Delta + e^\theta H_{\rm f}$.
Observe that for $|{\rm Im} \theta| < \pi/4$ we have $c_\theta :=  \min\{ 1, e^{{\rm Re} \theta }  \cos({\rm Im} \theta ) \} > 0$ and
\begin{align}
{\rm Re} (  -\Delta + e^\theta H_{\rm f} ) = - \Delta + e^{{\rm Re} \theta} \cos({\rm Im} \theta) H_{\rm f}  \geq c_\theta  (-\Delta + H_{\rm f})   . \label{nonzerores}
\end{align}
This implies that for $\psi \in \mathcal{D}(-\Delta + H_{\rm f})$ we have
\begin{align} \label{boundonfree04}
 c_\theta^2 \| (-\Delta + H_{\rm f} ) \psi \|^2 \leq \| (-\Delta + e^{\theta}  H_{\rm f}  ) \psi \|^2    .
\end{align}
It follows from  \eqref{boundonfree04}  that for $|{\rm Im} \theta| < \pi/4$ the operator $-\Delta + e^\theta H_{\rm f}$
is a closed operator on $D(-\Delta + H_{\rm f})$. Furthermore  by   \eqref{nonzerores} its numerical range is in the right complex half plane, and therefore $-\Delta + e^\theta H_{\rm f}$   has nonempty resolvent set  if $|{\rm Im} \theta| < \pi/4$.
Since $e^{2 \theta} V_\theta$ is infinitesimally bounded with respect to $-\Delta$, cf. \eqref{eq:boundonc}and  \eqref{eq:boundonc5555}, it is infinitesimally small
with respect to  $(-\Delta + e^{\theta}  H_{\rm f}  )$  by  \eqref{boundonfree04} (for $|{\rm Im} \theta| < \pi/4$). Thus for $|{\rm Im} \theta| < \pi/4$
it follows 
that $(-\Delta + e^{2 \theta} V_\theta + e^{\theta}  H_{\rm f}  )$
is closed with nonempty resolvent set, and hence the same holds true for $H_0(\theta)$  (by multiplication with  the nonzero
number $e^{-2\theta}$).   This shows the lemma.
\end{proof}

\begin{lemma} \label{anaAalphathetaat} Suppose Hypothesis \ref{potential} holds for $\theta_{\rm at} > 0$.
The map  $(\theta,\alpha) \mapsto H_{\rm at}(\theta,\alpha)$ with domain $\mathcal{D}(-\Delta)$ is  an analytic
family of type (A) on $D_{\theta_{\rm at}} \times \C$ and satisfies $H_{\rm at}(\theta,\alpha)^* = H_{\rm at}(\overline{\theta},\overline{\alpha})$.
\end{lemma}

\begin{proof} First observe that  $p_j \cdot x_j \langle x \rangle^{-1}$,  $x_j \langle x \rangle^{-1} \cdot p_j $, $x_j^2 \langle x \rangle^{-2}$
and $e^{2 \theta} V_\theta(x)$ are infinitesimally small with respect to $-\Delta$ cf.  \eqref{eq:boundonc}.
Thus we can define for $(\theta,\alpha) \in D_{\rm at} \times \C$
\begin{align} 
  \tilde{H}_{\rm at}(\theta,\alpha) :=   e^{2 \theta} {H}_{\rm at}(\theta,\alpha)  =  \sum_{j=1}^N (p_j -  \alpha x_j \langle x \rangle^{-1} )^2    +  e^{2 \theta}  V_\theta(x)
\end{align}
as an operator with domain $\mathcal{D}(-\Delta)$. By  Part \eqref{anabound(iv)} of Hypothesis \ref{potential}
it follows that for every $\psi \in \mathcal{D}(-\Delta)$ the function $(\theta,\alpha) \mapsto   \tilde{H}_{\rm at}(\theta,\alpha)\psi$
is analytic. Furthermore since the terms involving $\theta$ or $\alpha$ are infinitesimally bounded with respect to $-\Delta$
the closedness and the nonempty resolvent property of   $\tilde{H}_{\rm at}(\theta,\alpha) $.
We conclude that $ H_{\rm at}(\theta,\alpha) = e^{-2 \theta}   \tilde{H}_{\rm at}(\theta,\alpha)  $
is an analytic family of type (A) on $D_{\theta_{\rm at}} \times \C$. The last statement follows from analytic continuation
and the fact that $H_{\rm at}(\theta,\alpha)^* = H_{\rm at}(\overline{\theta},\overline{\alpha})$ for real $(\theta,\alpha)  \in D_{\theta_{\rm at}} \times \C$.
\end{proof}

\begin{lemma} \label{anaAalphathetaField}
 Suppose Hypothesis \ref{potential} holds for $\theta_{\rm at} > 0$.
The map  $(\theta,\alpha) \mapsto H_0(\theta,\alpha)$ with domain $\mathcal{D}(-\Delta +H_{\rm f})$  is an analytic
family of type (A) on  $(D_{\theta_{\rm at}} \cap \{ \theta \in \C :  |{\rm Im} \theta |  < \pi/4\} )  \times \C$.

\end{lemma}

\begin{proof}
 From  the infinitesimal bound \eqref{eq:boundonc}, we see   that  for $\psi \in \mathcal{D}(-\Delta + H_{\rm f})$
the map
\begin{align*}
(\theta,\alpha) \mapsto H_0(\theta,\alpha) \psi = e^{-2 \theta}\sum_{j=1}^N (p_j - \alpha x_j \langle x \rangle^{-1} )^2  \psi +  V_\theta(x) \psi + e^{-\theta} H_{\rm f}\psi
\end{align*}
is analytic (since each summand is) on $D_{\theta_{\rm at}}$. Thus it remains to show closedness and existence of a  nonempty resolvent set.
As in the proof of Lemma \ref{anatypeAtheta2Field}   we consider first $ -\Delta + e^\theta H_{\rm f}$, where we showed
that this operator is closed with nonempty resolvent set, provided $|{\rm Im} \theta| < \pi/4$.
Since $p_j \cdot x_j \langle x \rangle^{-1}$,  $x_j \langle x \rangle^{-1} \cdot p_j $, $x_j^2 \langle x \rangle^{-2}$
and $e^{2 \theta} V_\theta(x)$ are infinitesimally bounded  with respect to $-\Delta$ cf.  \eqref{eq:boundonc}, it is infinitesimally
with respect to  $(-\Delta + e^{\theta}  H_{\rm f}  )$  by  \eqref{boundonfree04} (for $|{\rm Im} \theta| < \pi/4$). Thus for $|{\rm Im} \theta| < \pi/4$
it follows  that 
$$ \sum_{j=1}^N (p_j -  \alpha x_j \langle x \rangle^{-1} )^2  + e^{2\theta} V_\theta(x)  + e^{\theta} H_{\rm f}
$$
is closed with nonempty resolvent set for all $\alpha$.
Hence the same holds true for $H_0(\theta,\alpha)$  (by multiplication with  the nonzero
number $e^{-2\theta}$).  This shows the lemma.
\end{proof}

\begin{remark} \label{closedop} {\rm We note that  Lemmas \ref{anatypeAtheta2Field}  and  \ref{anaAalphathetaField}  imply that $H_g(\theta)$
and $H_g(\theta,\alpha)$ are closed  for $g$ in a neighborhood of zero, respectively.  This follows directly using    basic bounds
for the interaction, Lemma \ref{kernelopestimate},
 and abstract properties of operators \cite[Theorems 5.5 and 5.9]{Wei80}.  }
\end{remark}

\begin{lemma}  \label{anaint} Suppose Hypothesis  \ref{kappa} holds. Then for all $\psi_1= f_1 \otimes \cdots  \otimes f_n \otimes \varphi $ and
 $\psi_2= g_1 \otimes \cdots  \otimes g_m  \otimes \gamma$ with $f_s , g_l  \in \hh$  and $\varphi, \gamma \in \HH_{\rm at}$  the maps
\begin{align*}
 \theta \mapsto  \langle   \psi_1  ,  A_{\theta,j}(x)  \psi_2 \rangle   , \quad  \theta \mapsto  \langle   \psi_1  ,  A_{\theta}(x) \cdot A_{\theta}(x) \psi_2\rangle   ,
\end{align*}
are analytic on $D_{\theta_{\rm I}}$.
\end{lemma}
\begin{proof}
First observe that the first map is a finite linear combination of terms of the form
\begin{align}
&  \int \langle f_s ,  e^{-\theta}   \kappa_\theta  \omega^{-1/2}  e^{-i \beta ( \cdot )  \cdot x} \varepsilon_j \rangle_\hh
\overline{ \varphi}(x) \gamma(x) dx  \label{anafirsttermint}  , \\
&  \int \langle   e^{-\overline{\theta}}   \kappa_{\overline{\theta}}  \omega^{-1/2}  e^{-i \beta ( \cdot )  \cdot x} \varepsilon_j ,  {g}_l \rangle_\hh
\overline{ \varphi}(x) \gamma(x) dx  \label{anafirsttermint2}  .
\end{align}
For each fixed $x$ the expression in the inner product $\langle  \cdot , \cdot  \rangle_\hh$ is analytic in $\theta$  and uniformly bounded in $x$, by Hypothesis \ref{kappa} (i).
The analyticity of    \eqref{anafirsttermint} and  \eqref{anafirsttermint2}  thus follows by means of  dominated convergence
and using for example  Morera's theorem.
This shows the analyticity of the first map in the lemma. To show the analyticity of the second map observe that it is a finite  linear
combination of terms of the form
\begin{align*}
&  \int \sum_{j=1}^3 \langle f_{s_1} ,  e^{-\theta}   \kappa_\theta  \omega^{-1/2}  e^{-i \beta ( \cdot )  \cdot x} \varepsilon_j \rangle_\hh
 \langle f_{s_2} ,  e^{-\theta}   \kappa_\theta  \omega^{-1/2}  e^{-i \beta ( \cdot )  \cdot x} \varepsilon_j \rangle_\hh
\overline{ \varphi}(x) \gamma(x) dx , \\ 
&  \int  \sum_{j=1}^3  \langle_\hh   e^{-\overline{\theta}}   \kappa_{\overline{\theta}}  \omega^{-1/2}  e^{-i \beta ( \cdot )  \cdot x} \varepsilon_j ,  {g}_{l_1} \rangle_\hh
\langle   e^{-\overline{\theta}}   \kappa_{\overline{\theta}}  \omega^{-1/2}  e^{-i\beta  ( \cdot )  \cdot x} \varepsilon_j ,  {g}_{l_2} \rangle_\hh
\overline{ \varphi}(x) \gamma(x) dx  , \\
&  \int \sum_{j=1}^3
\langle f_{s} ,  e^{-\theta}   \kappa_\theta  \omega^{-1/2}  e^{-i \beta ( \cdot )  \cdot x} \varepsilon_j \rangle_\hh
 \langle   e^{-\overline{\theta}}   \kappa_{\overline{\theta}}  \omega^{-1/2}  e^{-i \beta  ( \cdot )  \cdot x} \varepsilon_j ,  {g}_{l} \rangle_\hh
\overline{ \varphi}(x) \gamma(x) dx   
, \\
&  \int  \sum_{j=1}^3
\langle   e^{-\overline{\theta}}   \kappa_{\overline{\theta}}  \omega^{-1/2}   \varepsilon_j ,
 e^{-\theta}   \kappa_\theta  \omega^{-1/2}   \varepsilon_j
 \rangle_\hh
\overline{ \varphi}(x) \gamma(x) dx  .  
\end{align*}
One now sees that the analyticity of the second map in the lemma follows analogously as the first.
Fir fixed $x$ the expression in the brackets $\langle  \cdots \rangle$ is analytic in $\theta$  and uniformly bounded in $x$, by Hypothesis \ref{kappa} (i). The analyticity then follows again by means of   dominated convergence and for example Morera's theorem. 
\end{proof}

\section{Smooth Feshbach Property}
\label{sec:smo}

In this appendix we follow \cite{BCFS,GriHas08}.
We introduce the Feshbach map and state basic isospectrality
properties.
Let $\chi$ and $\overline{\chi}$ be commuting, nonzero bounded operators, acting on a separable Hilbert space $\HH$
and satisfying $\chi^2 + \overline{\chi}^2=1$. A {\it Feshbach pair} $(H,T)$ for $\chi$ is a pair of
closed operators with the same domain,
$$
H,T : \mathcal{D}(H) = \mathcal{D}(T) \subset \HH \to \HH
$$
such that $H,T, W := H-T$, and the operators
\begin{align*}
&W_\chi := \chi W \chi , & &W_{\overline{\chi}} := \overline{\chi} W \chib \\
&H_\chi :=T + W_\chi , & &H_{\overline{\chi}} := T + W_{\chib} ,
\end{align*}
defined on $\mathcal{D}(T)$ satisfy the following assumptions:
\begin{itemize}
\item[(a)] $\chi T \subset T \chi$ and $\chib T \subset T \chib$,
\item[(b)] $T, H_{\chib} : \mathcal{D}(T) \cap \ran \chib \to \ran \chib$ are bijections with bounded inverse,
\item[(c)] $\chib H_{\chib}^{-1} \chib W \chi : \mathcal{D}(T) \subset \HH \to \HH$ is a bounded operator.
\end{itemize}
\begin{remark} \label{rem:abuse} {\em
By abuse of notation we write $ H_{\chib}^{-1} \chib$ for $ \left( H_{\chib} \upharpoonright \ran \chib \right)^{-1} \chib$ and
likewise $ T^{-1} \chib$ for $ \left( T \upharpoonright \ran \chib \right)^{-1} \chib$. }
\end{remark}
We call an operator $A: \mathcal{D}(A) \subset \HH \to \HH$ {\it bounded invertible} in a subspace $V \subset \HH$
($V$ not necessarily closed), if $A: \mathcal{D}(A) \cap V \to V$ is a bijection with bounded inverse.
Given a Feshbach pair $(H,T)$ for $\chi$, the operator
\begin{align} \label{eq:defoffesh}
&F_\chi(H,T) := H_\chi - \chi W \chib H_{\chib}^{-1} \chib W \chi
\end{align}
on $\mathcal{D}(T)$ is called the { \it Feshbach map of} $H$.
The auxiliary operator 
\begin{align} \label{eq:defofQ}
  Q_\chi := Q_\chi(H,T) := \chi - \chib H_{\chib}^{-1} \chib W \chi
\end{align}
is by conditions (a), (c), bounded, and $Q_\chi$ leaves $\mathcal{D}(T)$ invariant. The Feshbach map is
isospectral in the sense of the following theorem.
\begin{theorem} \label{thm:fesh}
Let $(H,T)$ be a Feshbach pair for $\chi$ on a Hilbert space $\HH$. Then the following holds.
$\chi \ker H \subset \ker F_\chi(H,T)$ and $Q_\chi \ker F_\chi(H,T) \subset \ker H$. The mappings
\begin{align*}
\chi : \ker H \to \ker F_\chi(H,T) , \quad Q_\chi : \ker F_\chi(H,T) \to \ker H ,
\end{align*}
are linear isomoporhisms and inverse to each other. $H$ is bounded invertible on $\HH$ if and only if $F_\chi(H,T)$ is
bounded invertible on $\ran \chi$.
\end{theorem}

The proof of Theorem \ref{thm:fesh} can be found in \cite{BCFS,GriHas08}. The next lemma
gives sufficient conditions for  two operators to be a Feshbach pair. It follows
from a Neumann expansion, \cite{GriHas08}.

\begin{lemma} \label{fesh:thm2}
Conditions {\rm (a), (b)}, and {\rm (c)} on Feshbach pairs are satisfied if:
\begin{itemize}
\item[(a')] $\chi T \subset T \chi$ and $\chib T \subset T \chib$,
\item[(b')] $T$ is bounded invertible in $\ran \chib$,
\item[(c')] $\| T^{-1} \chib W \chib \| < 1$, $\| \chib W T^{-1} \chib \| < 1$, and $T^{-1} \chib W \chi$ is a bounded operator.
\end{itemize}
\end{lemma}

\section{An Interpolation Result}

\begin{lemma} \label{intpol}  Let $A$ be a self-adjoint positive operator. Let $B$ be a bounded  operator with $\ran B \subset \mathcal{D}(A)$ such that for some constant $C$ we have
$$
 \| B  A \|  \leq C  , \quad  \| A  B  \|  \leq C .
$$
Then $ A^{1/2} B  A^{1/2} : \mathcal{D}(A) \to \HH$ is a  bounded operator with norm bounded by $C$.
\end{lemma}
\begin{proof}
We use interpolation.  Consider first  a regularization of $A$  as follows $A_n = \frac{n A}{n+ A}$, $n \in \N$. Then, clearly
\begin{align*}
 \| B A_n \| \leq  \| B  A \| \| (1 + A/n)^{-1} \| \leq  \| B  A \| \leq C \\
 \| A_n B   \| \leq   \| (1 + A/n)^{-1} \| \|  A  B  \| \leq  \| A B  \|  \leq C .
\end{align*}
It follows from interpolation, see for example \cite[Proposition 1 in  Appendix to IX.4]{ReeSim2} which  also holds for bounded operators,
that for all $n \in \N$
 $$\| A_n^{1/2}  B A_n^{1/2}  \| \leq C .  $$
Thus
$$  \inn{ f , A^{1/2}  B  A^{1/2} g } = \lim_{n \to \infty}  \inn{ f , A_n^{1/2} B A_n^{1/2} g } $$
for all $f, g \in \mathcal{D}(A)$. We conclude that $A^{1/2} B  A^{1/2}$ is bounded by $C$.
\end{proof}

\bibliographystyle{plain}
\bibliography{references}

\end{document}